\documentclass[reprint, aps, PRL, superscriptaddress, twocolumn, floatfix, preprintnumbers, pagewise, notitlepage, footinbib, longbibliography]{revtex4-1}

\usepackage{latexsym}
\usepackage{subcaption}
\usepackage{balance}
\usepackage{caption}
\usepackage{soul}
\usepackage[margin=0.75in]{geometry}
\usepackage[dvips]{color}
\usepackage[usenames,dvipsnames]{xcolor}
\usepackage[pdftex]{graphicx}
\usepackage{amsmath}
\usepackage{amssymb}
\usepackage{esdiff}
\usepackage{ragged2e}
\usepackage{lineno}
\usepackage{gensymb}
\usepackage{enumerate}
\usepackage[hidelinks]{hyperref}
\usepackage{mathrsfs}
\usepackage[version=4]{mhchem}
\usepackage{float}
\usepackage{subfiles}
\usepackage{bm}
\usepackage{physics}
\usepackage{soul}

\usepackage{tabularx}
\usepackage{enumitem}

\setlist{nolistsep}

\definecolor{bblue}{HTML}{000080}
\definecolor{mygray}{gray}{0.75}
\definecolor{lightBlue}{HTML}{e5f7ff}

\renewcommand{\arraystretch}{1.5}

\renewcommand{\thefigure}{\arabic{figure}}
\renewcommand{\thetable}{\Roman{table}}
\renewcommand{\theequation}{\arabic{equation}}

\def\bea{\begin{eqnarray}}
\def\eea{\end{eqnarray}}
\def\be{\begin{equation}}
\def\ee{\end{equation}}

\def\bal{\begin{aligned}}
\def\eal{\end{aligned}}

\usepackage{booktabs}
\usepackage{array}
\usepackage{colortbl}

\begin{document}
 
\title{\textbf{A tunable multi-gated platform for superlattice engineering in bilayer graphene}}
\author{Saisab Bhowmik}
\email{saisabb@iisc.ac.in}
\affiliation{\textit{Department of Instrumentation and Applied Physics, Indian Institute of Science, Bangalore, 560012, India}}
\author{Satyarth Srivastav}
\affiliation{\textit{Department of Instrumentation and Applied Physics, Indian Institute of Science, Bangalore, 560012, India}}
\author{Aditya Mehta}
\affiliation{\textit{Department of Physics, Indian Institute of Technology Bombay, Powai, Mumbai, 400076, India}}
\author{Robin Bajaj}
\affiliation{\textit{Department of Physics, Indian Institute of Science, Bangalore, 560012, India}}
\author{Suvronil Datta}
\affiliation{\textit{Department of Instrumentation and Applied Physics, Indian Institute of Science, Bangalore, 560012, India}}
\author{Rohit S Bhandari}
\affiliation{\textit{Department of Instrumentation and Applied Physics, Indian Institute of Science, Bangalore, 560012, India}}
\author{Nikhlesh S. Mehta}
\affiliation{\textit{Department of Instrumentation and Applied Physics, Indian Institute of Science, Bangalore, 560012, India}}
\author{Kenji Watanabe}
\affiliation{\textit{Research Center for Functional Materials, National Institute for Materials Science, Namiki 1-1, Tsukuba, Ibaraki 305-0044, Japan}}
\author{ Takashi Taniguchi}
\affiliation{\textit{International Center for Materials Nanoarchitectonics, National Institute for Materials Science, Namiki 1-1, Tsukuba, Ibaraki 305-0044, Japan}}

\author{Manish Jain}
\email{mjain@iisc.ac.in}
\affiliation{\textit{Department of Physics, Indian Institute of Science, Bangalore, 560012, India}}

\author{Hridis K. Pal}
\email{hridis.pal@iitb.ac.in}
\affiliation{\textit{Department of Physics, Indian Institute of Technology Bombay, Powai, Mumbai, 400076, India}}

\author{U. Chandni}
\email{chandniu@iisc.ac.in}
\affiliation{\textit{Department of Instrumentation and Applied Physics, Indian Institute of Science, Bangalore, 560012, India}}



\begin{abstract}

Periodic potentials with length scales much larger than crystal lattice constants can generate low-energy electronic minibands. While small-angle twisted van der Waals heterostructures have been widely used to realize such physics, correlated phases are often limited by twist-angle inhomogeneity, strain, and disorder. Here, we investigate high-quality Bernal bilayer graphene placed on a patterned graphite gate, featuring a triangular array of holes, forming a new class of `gate-engineered' superlattices tunable through three independent electrostatic gates. Near the charge neutrality point, we observe insulating states characteristic of miniband formation. Temperature-dependent transport reveals activated gaps consistent with theory, and Hall measurements indicate charge-carrier inversion. Notably, the superlattice can be reconfigured electrostatically using various gate combinations, enabling flexible band-structure design without mechanical twisting. The results highlight a new route to band engineering, potentially extendable to complex periodic structures that are untenable in natural materials.

\end{abstract}

\maketitle

\begin{figure}[t]
\centering
\includegraphics[width=\columnwidth]{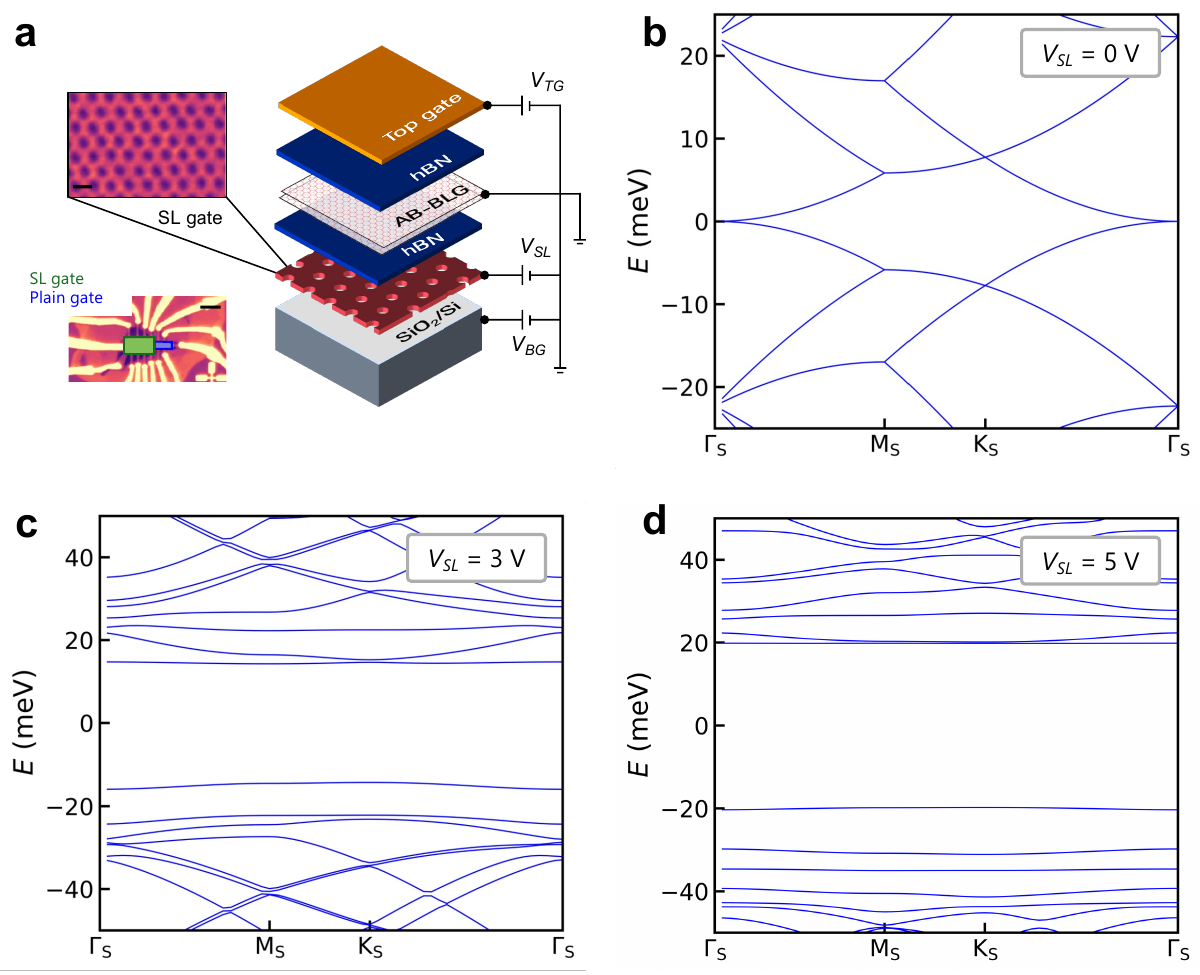}
\captionsetup{justification=raggedright,singlelinecheck=false}
\justify{FIG.~1.~Gate engineering and modifications in the band structure of BLG.}
(a)~Schematic of the triple-gated device consisting of hBN encapsulated Bernal BLG, where $V_{TG}$, $V_{SL}$, and $V_{BG}$ are the voltages applied to the top, superlattice, and Si bottom gate, respectively. The top left panel shows a false-colored scanning electron microscope image of the patterned graphite layer with a scale bar of 50 nm. The inset shows optical image of device D1 consisting of two distinct regions of SL gate (green) and plain gate (blue). The scale bar is 5 $\mu$m.(b)-(d)~Folded band structure of BLG at $V_{SL}=0$ V, 3 V, and 5 V along the high symmetry path $\mathrm{\Gamma_{S}-M_{S}-K_{S}-\Gamma_{S}}$ of the superlattice Brillouin zone.
\end{figure}

\begin{figure*}[bth]
\includegraphics[width=0.8\textwidth]{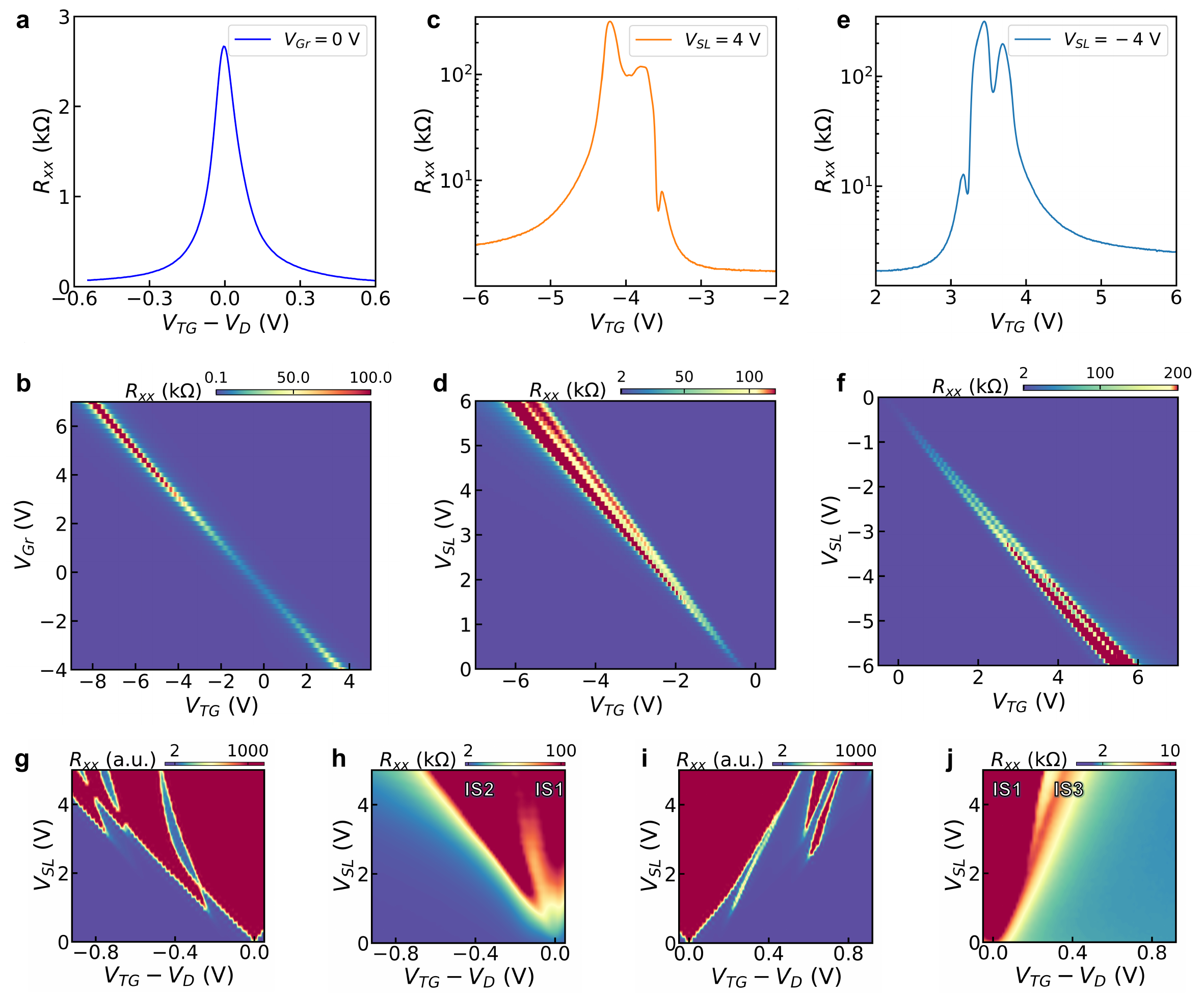}
\captionsetup{justification=raggedright,singlelinecheck=false}
\justify{FIG.~2.~Effect of superlattice gate on electrical transport}.~(a)~Four-probe longitudinal resistance $R_{xx}$ for device D1, as a function of $V_{TG}-V_{D}$ at plain graphite gate voltage $V_{Gr}=0$ V, where $V_{D}$ is the offset in the CNP due to residual charge carriers.~(b)~Color plot of $R_{xx}$ as a function of $V_{TG}$ and $V_{Gr}$ shows the magnitude of $R_{xx}$ increases at the CNP.~(c)~Line plot of $R_{xx}$ vs. $V_{TG}$ at $V_{SL}=4$ V shows multiple peaks.~(d)~The contour map of $R_{xx}$ vs. $V_{TG}$ at different $V_{SL}$ shows the evolution of these peaks.~(e)-(f)~Line plot of $R_{xx}$ vs. $V_{TG}$ at $V_{SL}=-4$ V and the contour plot show similar behaviour as panel c.-d.~(g),(i)~Resistivity calculated theoretically across the bands is plotted for positive $V_{SL}$ values.~(h),(j)~Experimentally obtained $R_{xx}$ in the same range. A qualitative comparison of the data reveals good agreement between theory and experiment, indicating the emergence of additional miniband features. In panel (h), the resistance peak around zero density ($V_{TG}-V_D = 0$) is marked as $IS1$, while the secondary peaks are marked as $IS2$ and $IS3$, respectively.
\end{figure*}

\begin{figure}[t]
\includegraphics[width=1.0\columnwidth]{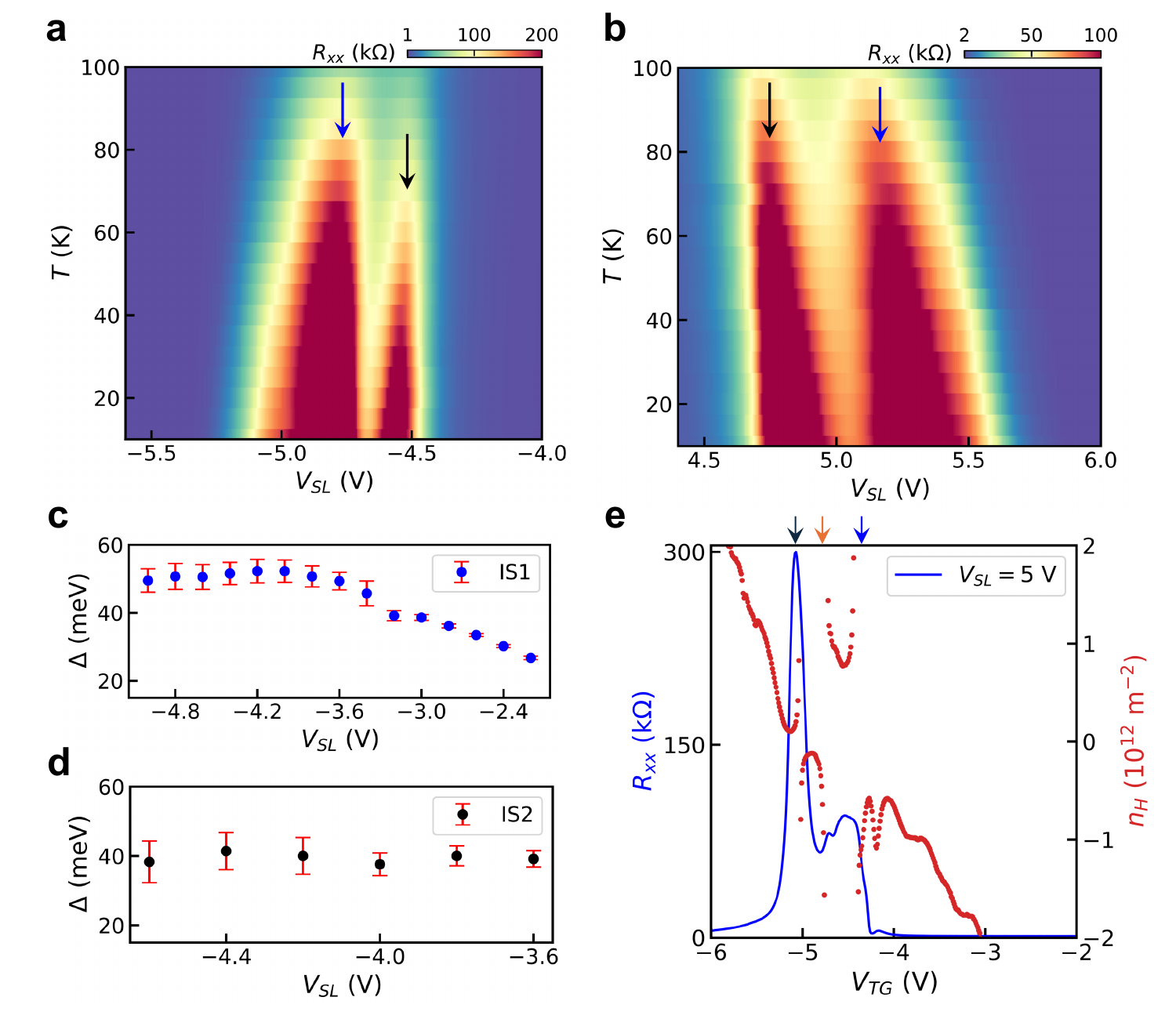}
\captionsetup{justification=raggedright,singlelinecheck=false}
\justify{FIG.~3.~Estimation of energy gaps.~(a)-(b)~Temperature $T$ dependence of $R_{xx}$ vs. $V_{SL}$ at $V_{TG}=\pm4.6$ V. The insulating states persist up to approximately $T=100$ K.~(c)-(d)~Thermally activated gaps $\Delta$ estimated for $IS1$ and $IS2$ are plotted as a function of $V_{SL}$.~(e) Hall density $n_{H}=-1/e(B/R_{xy})$, where $R_{xy}$ is Hall resistance $B$ is the magnetic field, and $e$ is the charge of an electron is plotted as a function of $V_{TG}$ at $V_{SL}=5$ V and $B=5$ T. The prominent sign changes observed in $n_H$ (red circles) are consistent with the position of the peaks $IS1$ and $IS2$ in $R_{xx}$ (blue trace), denoted by top arrows. An additional Lifshitz transition-like signature is observed in both $R_{xx}$ and $R_{xy}$, indicated by the dotted line.~All the data presented are for device D1.}
\end{figure}

\begin{figure*}[bth]
\includegraphics[width=0.9\textwidth]{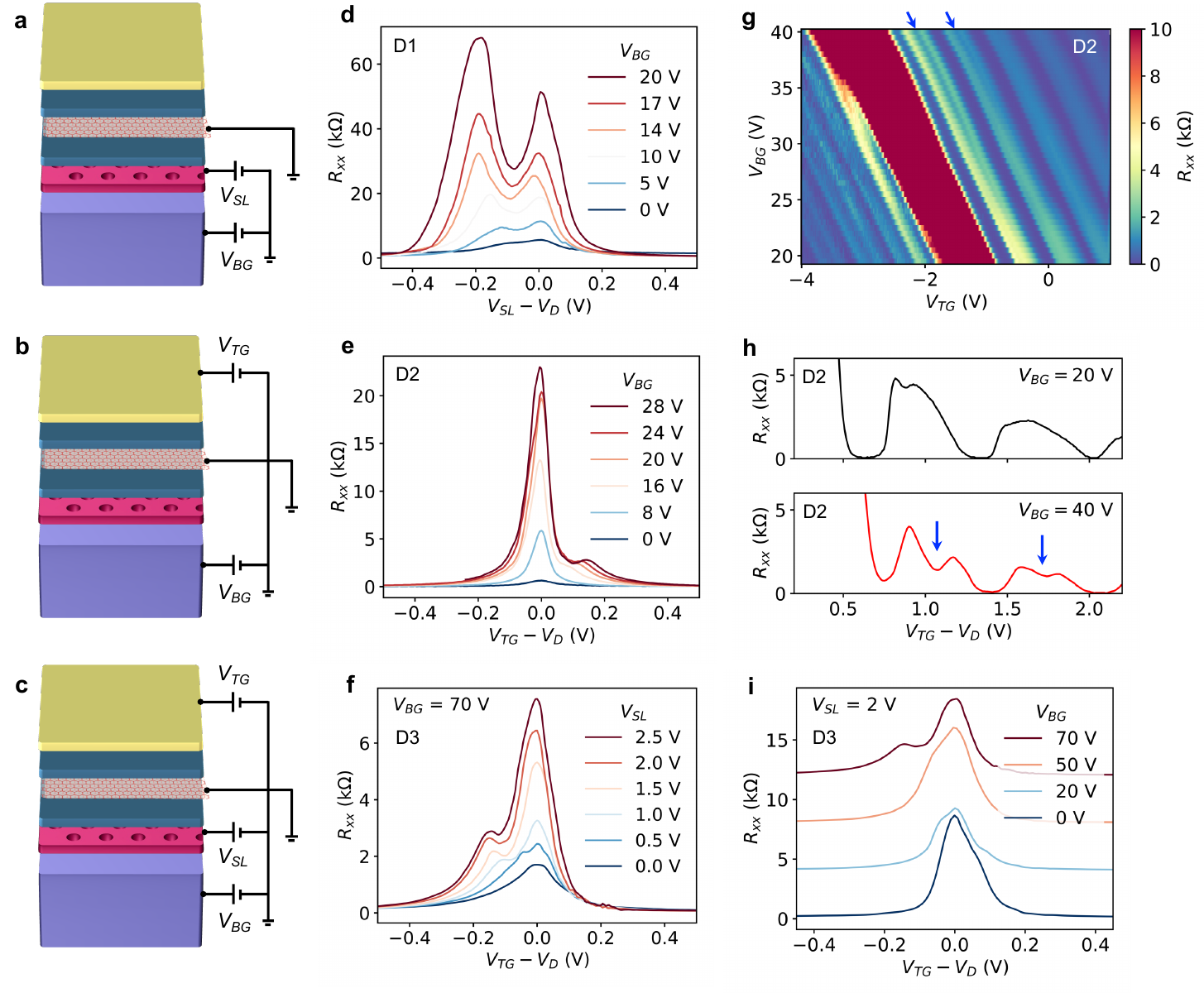}
\captionsetup{justification=raggedright,singlelinecheck=false}
\justify{FIG.~4.~Demonstration of multi-gate tunability.
(a-c)~Three different gate combinations used for devices D1, D2 and D3, respectively, that highlights the versatility of the multi-gated platform.
(d)~Line traces of $R_{xx}$ vs.\ $V_{SL}-V_D$ at several values of $V_{BG}$, where $V_D$ denotes the Dirac point offset.
(e)~Line traces of $R_{xx}$ vs.\ $V_{TG}-V_D$ at several values of $V_{BG}$.~Line traces of $R_{xx}$ vs.\ $V_{TG}-V_D$ at several values of $V_{SL}$, with $V_{BG}$ fixed at 70~V.
(g)~$R_{xx}$ as a function of $V_{TG}$ and $V_{BG}$ at $B=8$~T for device D2. Two additional minima emerge at higher $V_{BG}$, marked by blue arrows.
(h)~Line cuts at $V_{BG}=20$~V (black, top panel) and $40$~V (red, bottom panel) after subtracting $V_D$; the minima are marked by blue arrows in the bottom panel.
(i)~$R_{xx}$ vs.\ $V_{TG}-V_D$ at several values of $V_{BG}$ and fixed $V_{SL}=2$~V, highlighting the effect of a finite $V_{BG}$, for device D3. Curves are vertically offset by 4~k$\Omega$ for clarity.}
\end{figure*}

The discovery of strong correlations in flat bands of magic-angle twisted bilayer graphene~\cite{bistritzer2011moire, PhysRevB.82.121407, PhysRevB.86.155449} has driven the search for alternative systems with similar physics but fewer fabrication challenges~\cite{bistritzer2011moire, cao2018correlated, cao2018unconventional, park2021tunable, tang2020simulation}. While moiré materials are rapidly expanding through advanced stacking techniques, twist-angle control remains a major bottleneck $-$ introducing inhomogeneity, strain, and disorder that cause large sample-to-sample variations~\cite{uri2020mapping, pierce2021unconventional, lau2022reproducibility, bhowmik2024emergent}. Even devices with nominally identical twist angles can show vastly different properties. The quest for robust correlated physics has therefore led to explorations of alternate material platforms that host flatter bands, such as Bernal bilayer graphene (BLG)~\cite{zhou2022isospin} and rhombohedral multilayer graphene~\cite{zhou2021half, zhou2021superconductivity}. However, such samples are difficult to fabricate, demanding high levels of purity. In contrast, a distinct engineering approach would be to impose a smoothly varying electrostatic potential on a conventional two dimensional (2D) material such as graphene over length scales larger than its lattice constant, mimicking the moir\'{e} potential. This results in an effective superlattice with an enlarged real-space periodicity and a reduced Brillouin zone. Although implementing such artificial lattices in 2D materials remains experimentally challenging, recent advances have demonstrated promising strategies~\cite{dubey2013tunable, forsythe2018band, li2021anisotropic, sun2024signature, huber2020gate, PhysRevLett.133.066302, cai2026stack}.\\

For a triangular superlattice with lattice constant $\lambda_s$, the carrier density needed to fill each four-fold spin-valley degenerate miniband is given by $n_s=\frac{8}{\sqrt{3}\lambda_s^2}$. Experimental carrier densities are typically limited to $\sim 10^{12}-10^{13}$/cm$^2$, which requires superlattice periods of the order of tens of nanometers for the minibands to be experimentally accessible. Recent studies have explored two distinct approaches to enable this: dielectric patterning~\cite{forsythe2018band, li2021anisotropic} and the moiré proximity effect~\cite{gu2024remote, zhang2024engineering, wang2025moire}. The former involves patterning the bottom dielectric, typically SiO$_2$~\cite{forsythe2018band}, into a lattice of choice (triangular, one-dimensional, or square), thereby inducing resistance peaks, attributed to electron-hole miniband formation. Another approach is to use an external moir\'{e} material (such as twisted hBN or transition metal dichalcogenides) in close proximity to a target 2D layer, and open observable superlattice gaps~\cite{gu2024remote, wang2025moire}.~Despite their promise, both dielectric patterning and moiré-induced potentials face challenges: the former is limited by dielectric contrast and resolution, while the latter is highly sensitive to twist-angle disorder, limiting reproducibility and control.\\

In this work, we present gate engineering as an innovative approach to realize a robust and tunable superlattice potential in BLG, enabling programmable band structures with tailored functionalities~\cite{PhysRevLett.130.196201}. Fig.~1a shows the schematic of our hBN-encapsulated BLG device controlled by three independent gates: a top gate (TG), a patterned graphite superlattice gate (SL), and a global Si bottom gate (BG). While dual-gated BLG devices are well established for independent control of carrier density and displacement field, our device incorporates an artificially patterned graphite gate between BLG and the Si BG. The SEM image (Fig.~1a, top left) shows holes etched in graphite forming a triangular lattice with a period of ~50 nm.~A patterned gate enables direct electrostatic modulation of carrier density across the entire sample area. Moreover, the combination of the holey SL gate and global BG allows independent control of displacement fields in holey and non-holey regions. BLG is chosen for its tunable band gap via displacement fields and its rich correlated and topological phases near the band edge~\cite{zhou2022isospin, holleis2025nematicity, zhang2023enhanced, seiler2022quantum}. Recent theoretical investigations also highlight the promising emergence of topological and correlated phases in BLG through superlattice potential modulation~\cite{PhysRevLett.130.196201, PhysRevB.109.195406}.

In the absence of a superlattice gate voltage ($V_{SL}=0$ V), BLG exhibits the characteristic parabolic dispersion near the CNP, with conduction and valence bands touching at the K point ($\Gamma_S$ of the superlattice Brillouin zone) and preserving electron–hole symmetry, resulting in gapless semimetallic behavior (Fig.~1b).
Applying a periodic electric field via the uniform TG and patterned SL drastically modifies the band structure. As $V_{SL}$ increases (Fig.~1c–d), (i) a gap opens at the CNP, (ii) the bands flatten near high-symmetry points (M$_S$, K$_S$), indicating reduced group velocity and enhanced density of states, (iii) minibands and avoided crossings emerge with energy shifts, and (iv) reversing $V_{SL}$ interchanges conduction and valence bands, producing pronounced particle–hole asymmetry.
This asymmetry, absent in pristine BLG, originates from the asymmetric coupling of the periodic potential and demonstrates enhanced tunability compared to uniformly gated BLG.\\

Our devices consist of two functionally distinct regions: one with a plain graphite gate and another with a patterned superlattice graphite gate, enabling a direct comparison of the effect of the superlattice potential (see the optical image of a typical device in Fig.~1a). Our primary measurements are performed on device D1 with a superlattice periodicity of 50 nm, at $T=1.8$ K, except for the temperature-dependent data. ~The devices exhibit high mobility of $2\text{--}3 \times 10^5$ cm$^2$V$^{-1}$s$^{-1}$ (see Supplemental Material Fig. S2).~First, we present data for BLG with the plain gate, which produces a uniform electric field in the 2D plane. In Fig.~2a, we plot the longitudinal resistance $R_{xx}$ as a function of $V_{TG}-V_{D}$ at $V_{Gr}=0$ V, where $V_{TG}$ is the top-gate voltage, $V_{D}$ corresponds to the CNP offset, and $V_{Gr}$ is the plain graphite gate voltage. As expected, when a finite displacement field $D$ is applied, an increase in $R_{xx}$ (Fig.~2b) at CNP is observed due to band gap opening. We now present the primary data obtained by applying a finite $V_{SL}$.~Fig.~2c shows $R_{xx}$ as a function of negative $V_{TG}$ at $V_{SL}=4$ V. In contrast to standard BLG, which exhibits a single resistance peak, multiple peaks are clearly observed. 

To track the evolution of these features, we performed dual-gate sweeps. Fig.~2d shows a contour plot of $R_{xx}$ obtained by continuously sweeping $V_{TG}$ at discrete $V_{SL}$.~Similar to the standard BLG data in Fig.~2b, all the resistance peaks shift diagonally with gate voltages.~Near $V_{SL} = 0$ V, a single resistance peak is recovered, consistent with plain-gated BLG. Strikingly, as $V_{SL}$ increases, two distinct peaks emerge at intermediate voltages (dark red bands in Fig.~2d), followed by the appearance of a third peak at higher $V_{SL}$. Moreover, the magnitude of $R_{xx}$ increases with increasing $V_{SL}$. 
Similar features are observed when the polarities of both gates are reversed (Fig.~2e-f), although the magnitude of resistive peaks differs slightly.For clarity, we label the resistance peak at $V_{SL}=0$~V, corresponding to the CNP gap, as $IS1$. For $V_{SL}>0$~V, the secondary peaks are labeled $IS2$ (hole side, left), appearing at low/intermediate $V_{SL}$, and $IS3$ (electron side, right), appearing at higher $V_{SL}$ (Fig.~2h, 2j). Intriguingly, the secondary peak due to the band gap opening on the electron (hole) side is weaker for positive (negative) $V_{SL}$ values, hinting at electron-hole asymmetries that are in agreement with those seen in the band structure calculations (see Supplemental Material Fig.~S27). Consequently, for $V_{SL}<0$~V, the secondary peaks appear at mirrored positions (see Supplemental Material Fig.~S6).
The emergence of multiple resistive peaks marks a clear departure from the conventional BLG behaviour. 
Their appearance point to the emergence of multiple minibands separated by gaps in the band structure, as expected from theoretical calculations discussed earlier. To probe the robustness of these states and their dependence on the electric field, we varied the gate-sweep configuration.~Supplemental Material Fig.~S4 shows $R_{xx}$ measured by sweeping $V_{SL}$ continuously at discrete $V_{TG}$. Consistent with Fig.~2c-d, three resistance peaks appear for both polarities of the gate voltages, validating that the underlying mechanism is governed by the periodic electric field generated by the combined action of the two gates.~In both cases, the separation between the two dominant resistance peaks increases with the displacement field, suggesting that the bands shift as $V_{SL}$ increases. 

For a more direct comparison between experiment and theory, we calculated the longitudinal conductivity within the constant-relaxation-time approximation,  
$G_{xx} \propto \int \frac{d^2k}{(2\pi)^2} v_x^2 \left(-\frac{\partial f_0}{\partial E}\right)$,  
where $v_x$ is the $x$-component of the carrier velocity and $f_0$ is the Fermi--Dirac distribution. The electrostatically tuned band structures obtained from the model were used as input to this calculation. Importantly, both the band dispersion and the chemical potential evolve \emph{simultaneously} as $V_{TG}$ and $V_{SL}$ are varied; hence, they were self-consistently updated for each pair $(V_{TG}$, $V_{SL})$ (see Supplemental Material Section II B for details). The resulting resistance maps, $R_{xx} \propto 1/G_{xx}$ (Fig.~2g, 2i), capture the salient features of the experimental data (Fig.~2h, 2j)---notably the emergence of multiple gap regions separated by conducting minibands and a clear particle--hole asymmetry.In addition, we find that the first gap above the CNP is significantly smaller than the first gap below the CNP (Fig.~1c-d). The smaller resistive peak observed for $IS3$ on the electron side, compared to $IS2$ on the hole side, is a direct consequence of this asymmetry.

While a precise quantitative correspondence is limited by the sensitivity of the band structure to microscopic model parameters (see Supplemental Material Section II C), the principal qualitative features exhibit excellent agreement between theory and experiment.~We also highlight that the reported resistance - gate voltage maps are distinct from those previously reported for moir\'{e} materials~\cite{liu2020tunable, polshyn2020electrical, xu2021tunable}. Here, the highly resistive miniband features exhibit dispersive behavior in the gate voltage phase space (see Figs.~2h and 2j), governed by both $V_{TG}$ and $V_{SL}$. In contrast, in moir\'{e} systems~\cite{liu2020tunable}, the position of superlattice gaps is solely governed by the twist angle and remains unaffected by electric fields.\\



To gain further insights into the transport properties of the system, we perform temperature-dependent transport measurements which provide another powerful approach to probe the nature of ground states, whether metallic, semiconducting, or insulating.~We therefore measured $R_{xx}$ over a range of temperatures $T$ to estimate the corresponding energy gaps.~As discussed earlier, the increasing trend of $R_{xx}$ with $V_{SL}$ indicates the opening of a gap at the CNP in BLG.~Fig.~3a-b show $R_{xx}$ versus $V_{SL}$ for two different values of $V_{TG}$ ($\pm4.6$ V) across temperatures from 7.5 K to 100 K.~At low temperatures ($T \leq 30$ K), $R_{xx}$ remains nearly constant, whereas at higher temperatures it decreases, consistent with the insulating nature of these states.~Similar behaviour is observed when varying $V_{TG}$ continuously at constant $V_{SL}$ (see Supplemental Material Fig.~S7-8).


To estimate the energy gap, we plot the peak values of $R_{xx}$ as a function of $1/T$.~Prior studies have shown that at low temperatures ($T\leq30$ K, in our case), the peak resistance exhibits only a weak dependence on $1/T$, consistent with variable-range hopping (VRH) or a combination of VRH and nearest-neighbour hopping at low and intermediate $T$~\cite{PhysRevLett.105.166601, sui2015gate}.~At higher temperatures ($T\geq60$ K), however, $R_{xx}$ shows thermally activated behaviour, following: $R_{xx} \propto \exp(\Delta / 2k_BT)$, where $\Delta$ is the energy gap and $k_B$ is the Boltzmann constant.~Fitting this relation to the $R_{xx}$ versus $1/T$ data allows us to extract $\Delta$ (see Supplemental Material Fig. S8). To determine $\Delta$ at different $V_{SL}$, we performed temperature-dependent $R_{xx}$ versus $V_{TG}$ measurements across multiple $V_{SL}$ values.~Fig.~3c-d shows $\Delta$ as a function of $V_{SL}<0$~V for $IS1$ and $IS2$.~For $IS1$, $\Delta$ initially increases with $V_{SL}$, saturating for higher values. 
~By contrast, $IS2$, which emerges around $V_{SL} = -3.6$ V, shows only a weak dependence of $\Delta$ on $V_{SL}$.~These observations deviate from the conventional behaviour of gapped BLG, where $\Delta$ typically increases linearly with the displacement field~\cite{icking2022transport}.~Nonetheless, the magnitudes of $\Delta$ for both the peaks $IS1$ and $IS2$
are in order-of-magnitude agreement with our theoretical calculations (see Fig. 1c-d). However, the third peak $IS3$ was much weaker, and was therefore not ideal for the quantitative temperature dependent study. We note that although fitting the Arrhenius equation is a standard method for extracting energy gaps in thermally activated transport, the extracted values are highly sensitive to the temperature range used for the fit (see Supplemental Fig. S10).~Even small variations in the chosen $T$ window can yield significantly different $\Delta$.~Therefore, transport measurements alone cannot provide a highly accurate estimate of the gap.~In our analysis, we selected the fitting range that yields the least error in the higher-temperature regime (see Supplemental Material Section I B for details).\

Next, we present magnetotransport measurements.~In particular, measurement of the Hall resistance $R_{xy}$ can reveal signatures of charge carrier type, Fermi surface reconstructions and large density of states within minibands~\cite{kim2016charge, wu2021Chern, bhowmik2022broken, bhowmik2023spin}.~Fig.~3e shows $R_{xx}$ and the Hall density $n_H=-\frac{1}{e}\frac{B}{R_{xy}}$ as a function of $V_{TG}$ at $V_{SL} = 5$ V and a magnetic field $B = 5$ T.~ We observe clear sign changes in $n_H$ at the peak positions corresponding to $IS$1 and $IS$2, indicated by vertical arrows (blue and black), signifying a direct correlation between longitudinal and transverse transport signatures.~Sign changes in $n_H$ can occur when the Fermi energy $E_F$ reaches a band gap, where the band transitions take place and/or a van Hove singularity, where Lifshitz transitions may take place~\cite{wu2021Chern,kim2016charge}.~The highly resistive peaks in $R_{xx}$ in concurrence with the nature of sign changes in $n_H$ validates the presence of multiple energy gaps. We observe the emergence of an additional tiny peak in $R_{xx}$ (see Supplemental Material Fig. S3) and a corresponding sign change in $n_H$, indicated by the orange arrow. This additional peak along with the sign change in $n_H$ may be a correlation-induced phenomena, consistent with the high density of states in the minibands~\cite{kim2016charge, wu2021Chern, bhowmik2022broken, bhowmik2023spin}.
Taken together, the Hall responses align with the formation of minibands characterized by sizable band gaps and an enhanced density of states.\\

Finally, we demonstrate the multi-functionality of the periodic potential by systematically tuning different gate combinations in our device. While our measurements thus far have relied on two gates $V_{SL}$ and $V_{TG}$, our device architecture provides a third knob, namely, the bottom Si gate ($V_{BG}$), enabling multiple gate configurations to tailor the effective superlattice potential. For instance, while the potential along the non-holey regions is governed by the voltage applied to the graphite SL gate, the potential along the holey regions can be independently tuned via the Si BG~(see Supplemental Material Fig. S17). This design allows selective and continuous modulation of the SL potential strength in BLG, introducing a new degree of control that was not achievable in earlier superlattice-based studies~\cite{forsythe2018band, li2021anisotropic, sun2024signature, PhysRevLett.133.066302}. \

In the following, we demonstrate three configurations to illustrate the multi-gate tunability, as schematically described in Figs.~4a-c (see Supplemental Material Section I F for additional data and discussion.). We first demonstrate a new configuration using the gates $V_{BG}$ and $V_{SL}$, with $V_{TG}=0$ (see Fig.~4a). Line traces of $R_{xx}$ at different $V_{BG}$ reveal two distinct peaks emerging at finite $V_{BG}$ (see Fig.~4d). This finite $V_{BG}$ gives rise to a secondary peak, consistent with the features described in Figs.~2--3 (see Supplemental Material Fig.~S22-24 for device D3). For the second and third configurations, we examined two additional devices, D2 and D3, respectively. In Fig.~4b, the Si BG is used to induce a periodic potential in the BLG, with the graphite SL gate serving solely as a screening layer. This configuration can be approximated as a negative image of the potential profile generated by the graphite SL alone (see Supplemental Material Fig.~S13). We find a secondary peak by increasing $V_{BG}$, as shown in Fig.~4e.~Finally, in Fig.~4c, we demonstrate simultaneous operation of all three gates. The line plots show $R_{xx}$ vs.\ $V_{TG}-V_D$ for several values of $V_{SL}$ in the presence of a finite $V_{BG}$ (Fig.~4f). A clear trend emerges--increasing $V_{SL}$ leads to the emergence of a secondary peak. To isolate the role of $V_{BG}$, we additionally show data taken while varying $V_{BG}$ at fixed $V_{SL}$ (Fig.~4i). The system exhibits only a single peak at the CNP for low $V_{BG}$, whereas a pronounced secondary peak appears at $V_{BG}=70$~V---underscoring that the superlattice effect is substantially reinforced in the presence of a third gate.~In Fig.~4g-h, we present the magnetotransport data in device D2.~When $V_{BG}$ is finite, the system exhibits new states, manifested as additional resistance peaks in $R_{xx}$ (Fig.~4e) similar to the other configurations highlighting the versatility of gate engineering.~Interestingly, this configuration also revealed new states in the presence of magnetic fields. Fig.~4g shows a color map of $R_{xx}$ as a function of $V_{BG}$ and $V_{TG}$, at $B=8$ T, where Landau levels appear as minima.~Surprisingly, two new minima emerge at higher $V_{BG}$, which are absent at lower $V_{BG}$ values, indicated by blue arrows in the colour plot.~Fig.~4h shows line traces of $R_{xx}$ at $V_{BG}=20$ V and 40 V, where it is clear that the minima are better resolved at 40 V. These results suggest that quantum Hall states can also be manipulated by the multi-gate architecture.~However, whether these new states arise from spin/valley symmetry breaking at finite $V_{BG}$, or result from higher minibands, remains unclear and warrants further investigation. These results establish a realistic pathway for future investigations of higher-order topological bands, and fractional Chern insulators in artificially engineered architectures, where the three-gate electrostatic platform may enable  independent tuning of the carrier density, displacement field, and superlattice potential~\cite{PhysRevLett.130.196201}. A straightforward next step would be to implement this architecture with an enhanced periodic potential, achievable by reducing the hBN spacer thickness, and with a shorter superlattice periodicity comparable to that of twisted bilayer graphene-moiré lattices.\\

In summary, we have demonstrated the realization of minibands in Bernal-stacked BLG through the application of a SL gate potential. The periodic potential in our devices can be precisely tuned using various combinations of the three independently controlled gates, offering superior electrostatic control over the target 2D layer compared to previous architectures. Our key observation is the emergence of new insulating states near the CNP in BLG with large energy gaps, stabilized by the SL potential. Moreover, Hall measurements reveal signatures of charge carrier inversion, corroborating the formation of minibands. The inclusion of a third, independently tunable Si bottom gate further expands the accessible phase space, enabling dynamic modulation of the SL potential landscape. Theoretical calculations corroborate our experimental findings, demonstrating the formation of distinct minibands and the rapid evolution of the band structure with increasing SL strength. Overall, our results establish a versatile platform for engineering tunable miniband structures in BLG, paving the way for exploring a broad class of artificial lattice potentials and 2D material systems in the future. Whether the underlying physics of these artificially created bands is fundamentally similar to or different from that of twisted moir\'e bands remains an open question for further exploration. As the first demonstration of a three-gated architecture of this kind, our platform opens several promising directions for future work: \textit{(i)} stronger superlattice potentials via thinner dielectrics; \textit{(ii)} moir\'e-like superlattices with shorter periodicities; \textit{(iii)} engineered flat bands hosting complex topological phases and fractional Chern states, as proposed in Ref.~\cite{PhysRevLett.130.196201}; \textit{(iv)} extension to other material systems, such as oxides, enabling the study of correlation effects under moir\'e-like potentials without precise twist-angle engineering; and \textit{(v)} a versatile quantum simulation platform for nontrivial lattice geometries---Dice, Lieb, Kagome, and quasicrystalline structures---that are difficult to realize in naturally occurring materials.

\section*{Acknowledgements}

We thank A. Ghosh, V. Kochat and E. Tutuc for helpful discussions and comments. We gratefully acknowledge the usage of the MNCF and NNFC facilities at CeNSE, IISc. U.C. acknowledges funding from STARS (STARS-2/2023-0309) and Quantum Research Park, IQTI IISc. R.B. acknowledges funding from the Prime Minister’s Research Fellowship (PMRF), MHRD. M.J. acknowledges the Nano mission of the Department of Science and Technology, India, for financial support under Grant No. DST/NM/TUE/QM-10/2019. M.J. acknowledges the Nano mission of the Department of Science and Technology, India and the Anusandhan National Research Foundation (ANRF-ARG) for financial support under grant nos. DST/NM/TUE/QM-10/2019 and ANRF/ARG/2025/003370/PS, respectively. H.K.P. would like to thank DST SERB, India, for financial support via Grant No. CRG/2021/005453. 
K.W. and T.T. acknowledge support from the JSPS KAKENHI (Grant Numbers 21H05233 and 23H02052), the CREST (JPMJCR24A5), JST, and World Premier International Research Center Initiative (WPI), MEXT, Japan. This research also benefited from discussions at the International Centre for Theoretical Sciences (ICTS) through the program `Engineered 2D Quantum Materials' (code: ICTS/E2QM2024/07).

\section*{Methods}

\subsection*{Device fabrication}

Our patterned superlattice gate consists of a triangular network of holes with a period of $\sim50$ nm (D1, D3 and D4) and 100 nm (D2). The graphite flake to be patterned was first spin-coated with PMMA 950K A2 at 6000 rpm for 45 seconds, followed by baking at $180~^\circ\mathrm{C}$ for 1 minute. Electron-beam lithography was performed at an acceleration voltage of 30 kV, using a hole array of single-pixel dots. An optimized exposure dose of 0.00825 pC was used to obtain holes approximately 30 nm in diameter and with a periodicity of 50 nm. A 7.5 $\mu$m aperture was employed to minimize the beam diameter and reduce feature broadening on the sample. After exposure, the sample was developed in a MIBK:IPA (1:3) solution at $\sim5~^\circ\mathrm{C}$, followed by rinsing in IPA for 5 seconds at the same temperature to slow down the development process. The developed regions were then etched using reactive ion etching. The fabrication of the heterostructures began with a dry transfer process using a polycarbonate (PC) film. Each individual layer was sequentially picked up at a temperature of approximately $110~^\circ\mathrm{C}$ and finally released onto a target substrate of $\mathrm{SiO_2}$ ($285~\mathrm{nm}$)/Si at around $160~^\circ\mathrm{C}$. The assembled stack was shaped into a multi-terminal Hall bar geometry via reactive ion etching using $\mathrm{CHF_3/O_2}$ gas. Electron-beam lithography followed by thermal evaporation of $\mathrm{Cr/Au}$ ($5~\mathrm{nm}/60~\mathrm{nm}$) was employed to define Ohmic edge contacts and the top gate.\\

\subsection*{Transport Measurements} 
Electrical transport measurements were performed in a cryogen-free, pumped $\mathrm{^4He}$ cryostat equipped with a 9~T superconducting magnet and a base temperature of 1.6~K. The measurements were carried out using a bias current in the range of 10-100~nA, sourced through an SR830 lock-in amplifier operating at 17.777~Hz. Keithley~2450 source-measure units (SMUs) were used to apply the DC gate voltages. The voltage signal was amplified using a DL1201 preamplifier before lock-in detection.

\subsection*{Theory}

We investigate the low-energy electronic spectrum of Bernal-stacked bilayer graphene subjected to a displacement field and a superlattice potential using the continuum model\cite{sun2024signature, mccann2013electronic}. The total Hamiltonian is expressed as
\begin{equation*}
    H = H_{\mathrm{BLG}} + H_{V} + H_{SL},
\end{equation*}
where $H_{\mathrm{BLG}}$ represents the intrinsic bilayer graphene Hamiltonian, $H_{V}$ corresponds to the gate-induced mean potential, and $H_{SL}$ accounts for the superlattice modulation.

The mean potential term is given by
\begin{equation*}
    H_{V} =
    \begin{bmatrix}
    V_{0}^{(1)} \sigma_0 & 0_2 \\
    0_2 & V_{0}^{(2)} \sigma_0
    \end{bmatrix},
\end{equation*}
where $V_{0}^{(1)}$ and $V_{0}^{(2)}$ denote the average electrostatic potentials acting on layers 1 and 2, respectively.

The superlattice gate introduces a spatially varying two-dimensional cosine potential of the form
\begin{equation*}
    H_{SL} = \frac{1}{2}
            \begin{bmatrix}
            V_{SL}^{(1)} \sigma_0 & 0_2 \\
            0_2 & V_{SL}^{(2)} \sigma_0
            \end{bmatrix}
            \sum_{n=1}^{6} \cos(\mathbf{G}_n \cdot \mathbf{r}),
\end{equation*}
where $V_{SL}^{(1)}$ and $V_{SL}^{(2)}$ are the modulation amplitudes for layers 1 and 2. The reciprocal lattice vectors of the superlattice are defined as
$\mathbf{G}_n = G \bigg(\cos\Big(\frac{2\pi n}{6}\Big), \ \sin\Big(\frac{2\pi n}{6}\Big)\bigg)$. In the corresponding mini Brillouin zone, the high-symmetry points $\Gamma_\mathrm{S}$, $\mathrm{M}_\mathrm{S}$, and $\mathrm{K}_\mathrm{S}$ are located at $(0,0)$, $\frac{1}{2}\mathbf{G}_0$, and $\frac{1}{3}(\mathbf{G}_0 + \mathbf{G}_1)$, respectively.

The potential parameters, estimated from the carrier density data (see Supplemental Material Section III A), are given by
\begin{align*}
V_0^{(1)} &= 64.60(1-\alpha) V_{TG} + 64.60\alpha V_{SL} \ \mathrm{meV}, \\
V_0^{(2)} &= 64.60\alpha V_{TG} + 64.60(1-\alpha)V_{SL} \ \mathrm{meV}, \\
V_{SL}^{(1)} &= 4.54\alpha V_{SL} \ \mathrm{meV}, \\
V_{SL}^{(2)} &= 4.54(1 - \alpha) V_{SL} \ \mathrm{meV},
\end{align*}
where $\alpha$ is the screening parameter, $V_{TG}$ denotes the top-gate voltage, and $V_{SL}$ represents the applied superlattice-gate voltage.

Using this continuum model, the electronic band structures along the high-symmetry path of the mini Brillouin zone (see Supplemental Material Figs. S24-25) and the resistivity are computed for various values of $V_{SL}$ (see Fig.~2g,i).


\section*{Author Contributions}
S.S. fabricated the devices. S.B. assisted in device fabrication, performed the measurements, and analysed the data. A.M. and R.B. performed the theoretical calculations, supervised by H.K.P. and M.J., respectively. S.D. and R.S.B. assisted with measurements. K.W. and T.T. grew the hBN crystals.  S.B., A.M., R.B., M.J., H.K.P., and U.C. co-wrote the manuscript. U.C. supervised the project.

\section*{Competing interests}
The authors declare no competing interests.

\clearpage
\onecolumngrid
\setcounter{figure}{0}
\setcounter{table}{0}
\setcounter{equation}{0}
\renewcommand{\thefigure}{S\arabic{figure}}
\renewcommand{\thetable}{S\arabic{table}}
\renewcommand{\theequation}{S\arabic{equation}}
\section*{Supplemental Material: A tunable multi-gated platform for superlattice engineering in bilayer graphene}

\section{Experimental results}

\subsection{Device fabrication and basic characterization}

\begin{figure}[bth]
    \centering
    \includegraphics[width=0.8\textwidth]{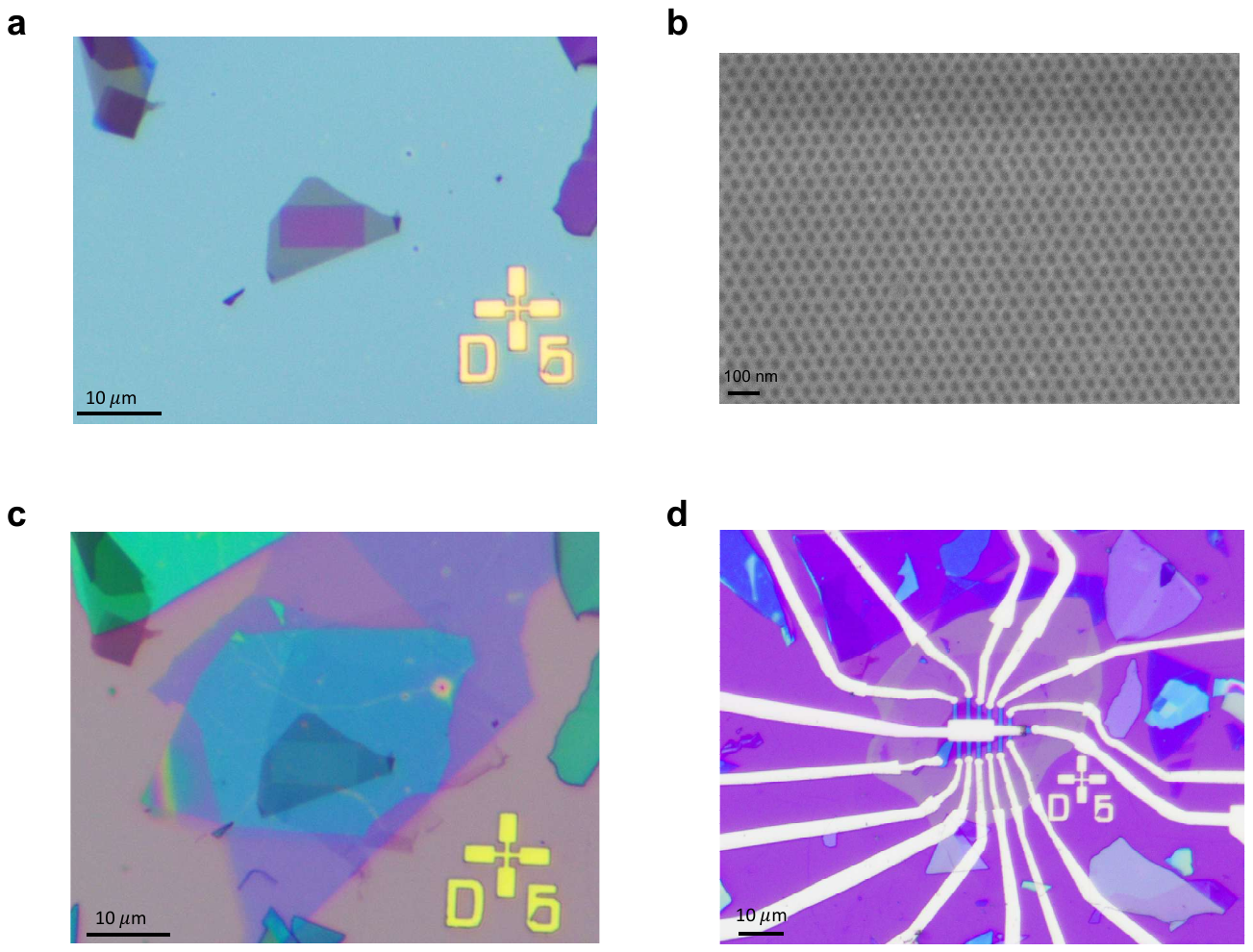}  \captionsetup{justification=raggedright,singlelinecheck=false}
    \caption{\justifying\textbf{Device fabrication.}~\textbf{a} Optical image of the patterned graphite flake transferred onto a SiO$_2$/Si substrate. The scale bar is 10 $\mu$m. \textbf{b} SEM image of the patterned graphite, showing the triangular periodic array of holes (dark circular features). The scale bar is 100 nm. \textbf{c} Optical image of the top hBN/bilayer graphene/bottom hBN heterostructure transferred onto the patterned graphite gate. \textbf{d} Optical image of the completed device with a scale bar of 10 $\mu$m. All images are for device D1.}
    \label{S1}
\end{figure}

\begin{figure}[bth]
\includegraphics[width=0.6\textwidth]{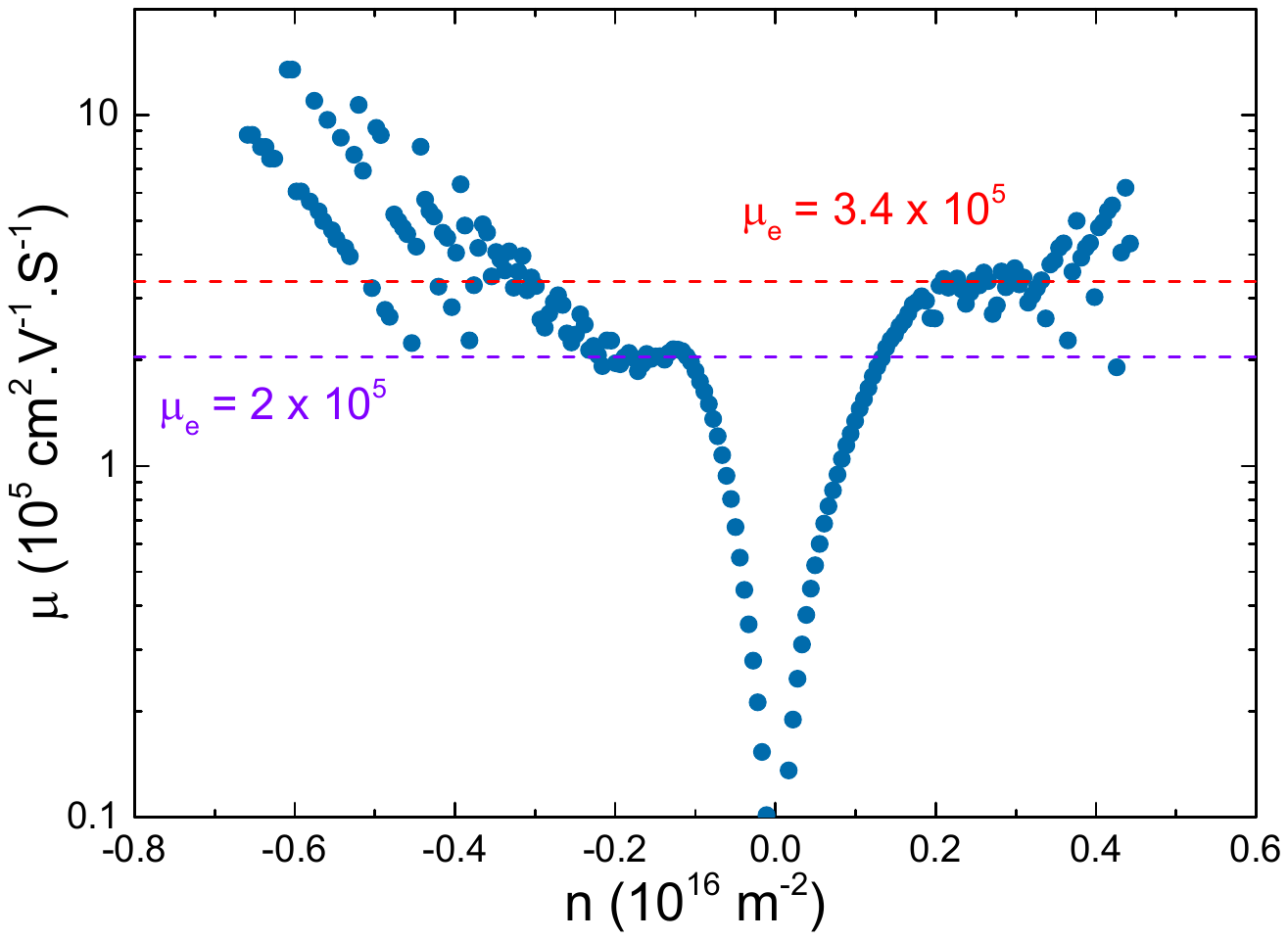}
\captionsetup{justification=raggedright,singlelinecheck=false}
\caption{\justifying\textbf{Mobility calculations}
  ~\textbf{a.}~$\frac{1}{e}\left| \frac{d\sigma}{dn} \right|$ is plotted as a function of $n$, where $\sigma$ is the electrical conductivity, and $e$ is the charge of an electron. The saturation of $\frac{1}{e}\left| \frac{d\sigma}{dn}\right|$ leads to a mobility of 340,000 cm$^2$V$^{-1}$s$^{-1}$ on the electron and 200,000 cm$^2$V$^{-1}$s$^{-1}$ on the hole side. Data shown for device D1. Devices D2 and D3 also showed similar values.}
  \label{S2}
\end{figure}

\begin{figure}[bth]
\includegraphics[width=0.8\textwidth]{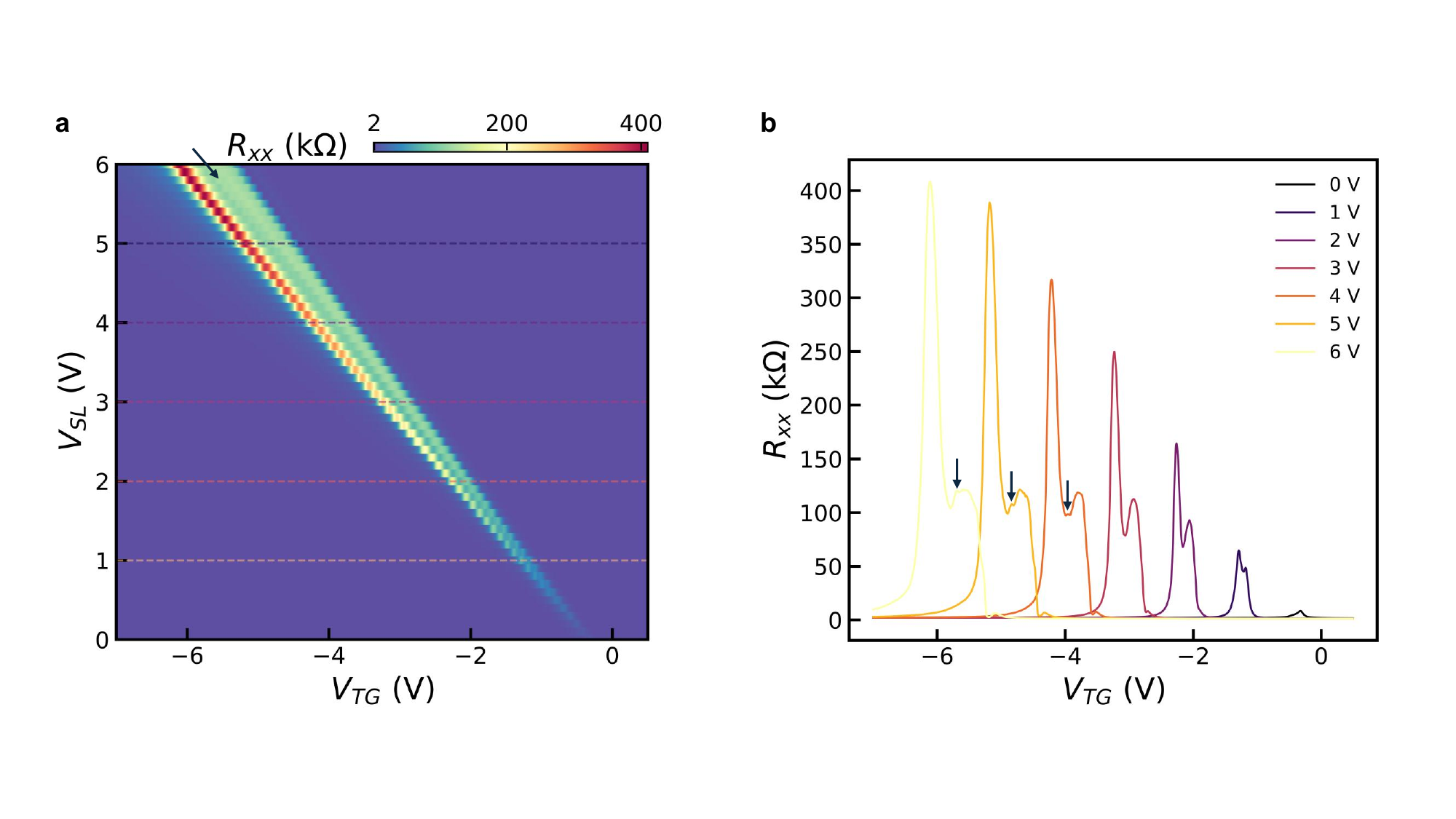}
\captionsetup{justification=raggedright,singlelinecheck=false}
\caption{\justifying\textbf{Effect of superlattice gate voltage}
  ~\textbf{a.} Color map of $R_{xx}$ as a function of top-gate voltage $V_{TG}$ and superlattice gate voltage $V_{SL}$.~\textbf{b.} Horizontal line cuts extracted from (a). At $V_{SL} > 3$ V, a tiny peak emerges between two prominent peaks, as indicated by the black arrow.}
    \label{S3}

\end{figure}

\begin{figure}[bth]
\includegraphics[width=0.98\linewidth]{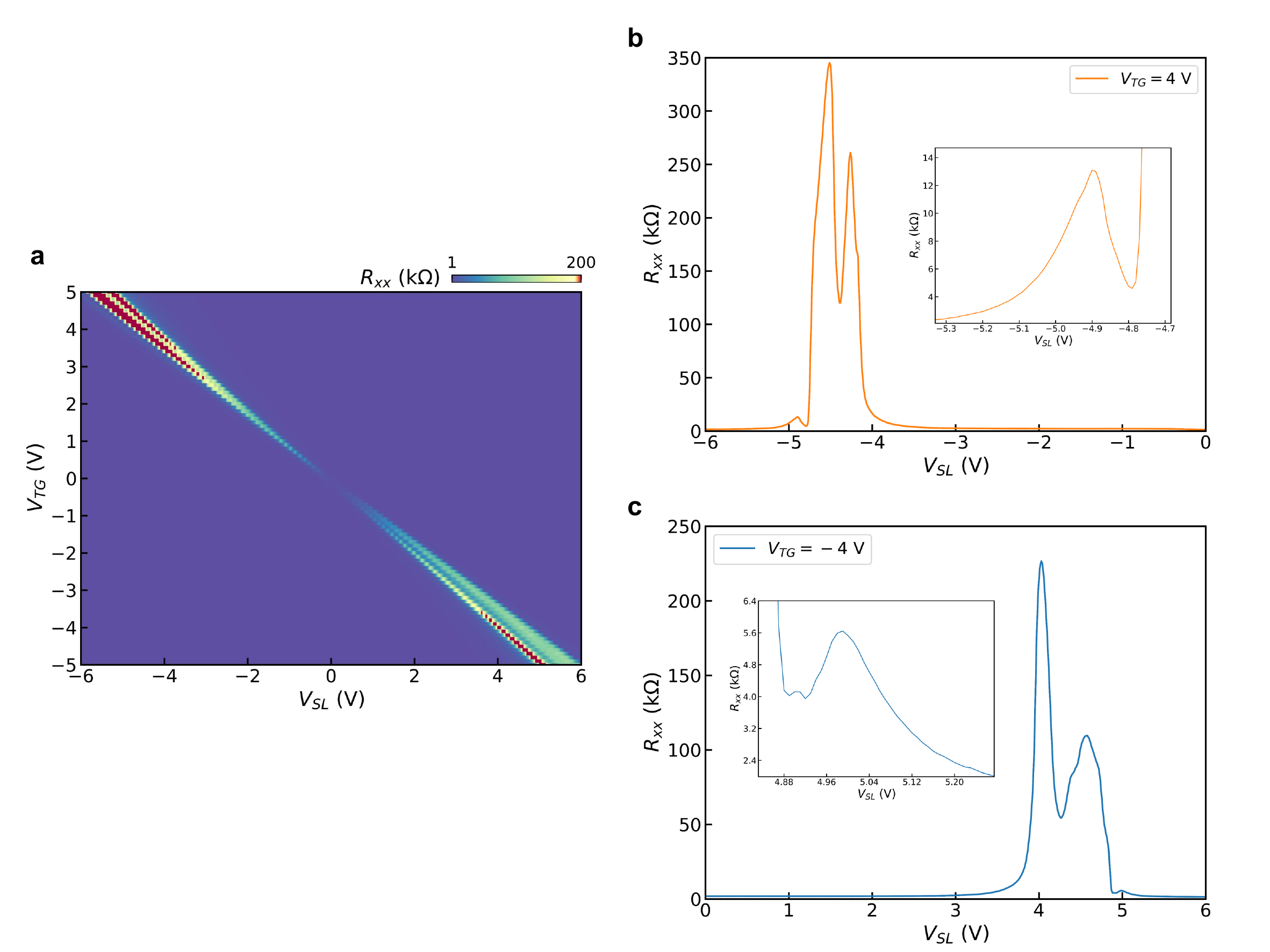}
\captionsetup{justification=raggedright,singlelinecheck=false}
\caption{\justifying\textbf{Phase space by sweeping the SL gate}
  ~\textbf{a.} $R_{xx}$ measured by sweeping $V_{SL}$ at discrete $V_{TG}$. Similar to Fig. 2 in the main text, three peaks are observed in $R_{xx}$ at finite $V_{TG}$ values.~\textbf{b.-c.} Line plots of $R_{xx}$ as a function of $V_{SL}$ at $V_{TG}=\pm4$ V. The insets show a closer view of the weaker peak. The measurements were performed at $T=1.8$ K.}
    \label{S4}

\end{figure}

\begin{figure}[bth]
\includegraphics[width=0.7\textwidth]{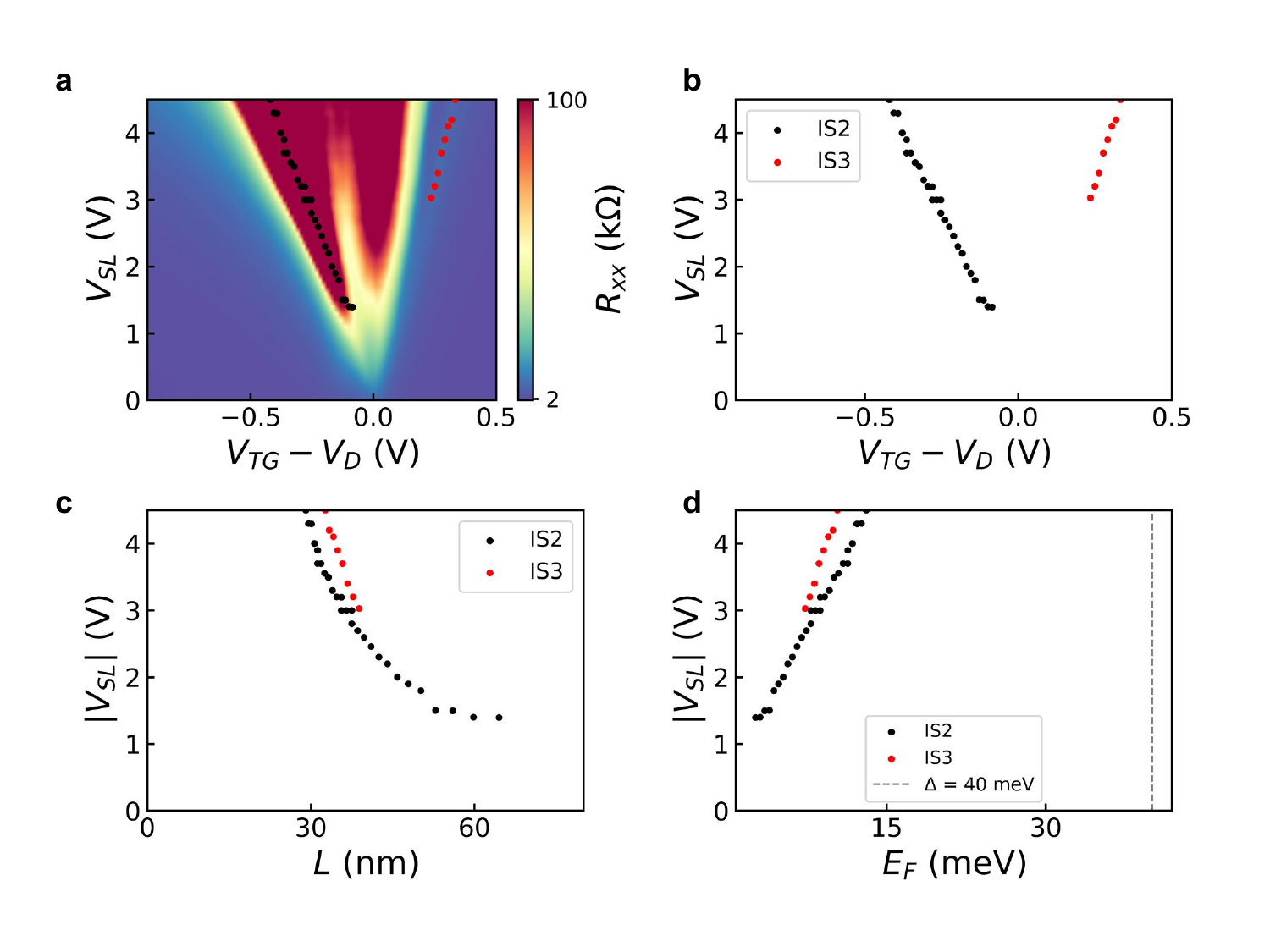}
\captionsetup{justification=raggedright,singlelinecheck=false}
\caption{\justifying\textbf{Estimation of the superlattice period and characteristic energy scale.}
\textbf{a.} Phase space of $R_{xx}$ as a function of $V_{TG}-V_D$ and $V_{SL}$. Black and red points mark the evolution of the resistance peaks $IS2$ and $IS3$, respectively.
\textbf{b.} Evolution of the $IS2$ and $IS3$ peak positions with $V_{SL}$, extracted from (a).
\textbf{c.} Estimated superlattice period $L$ obtained from the carrier density corresponding to the peak positions, assuming a triangular superlattice. The extracted values are of the same order as the nominal superlattice period of $50$ nm.
\textbf{d.} Estimated Fermi energy $E_F$ associated with the $IS2$ and $IS3$ peak positions as a function of $|V_{SL}|$, calculated assuming a parabolic dispersion with an effective mass $m^*=0.05m_e$. The gray dashed line marks the characteristic energy scale $\Delta=40$ meV. The extracted $E_F$ values remain below this energy scale over the measured range.}
    \label{lattice constant}

\end{figure}

\begin{figure}[bth]
\includegraphics[width=0.8\textwidth]{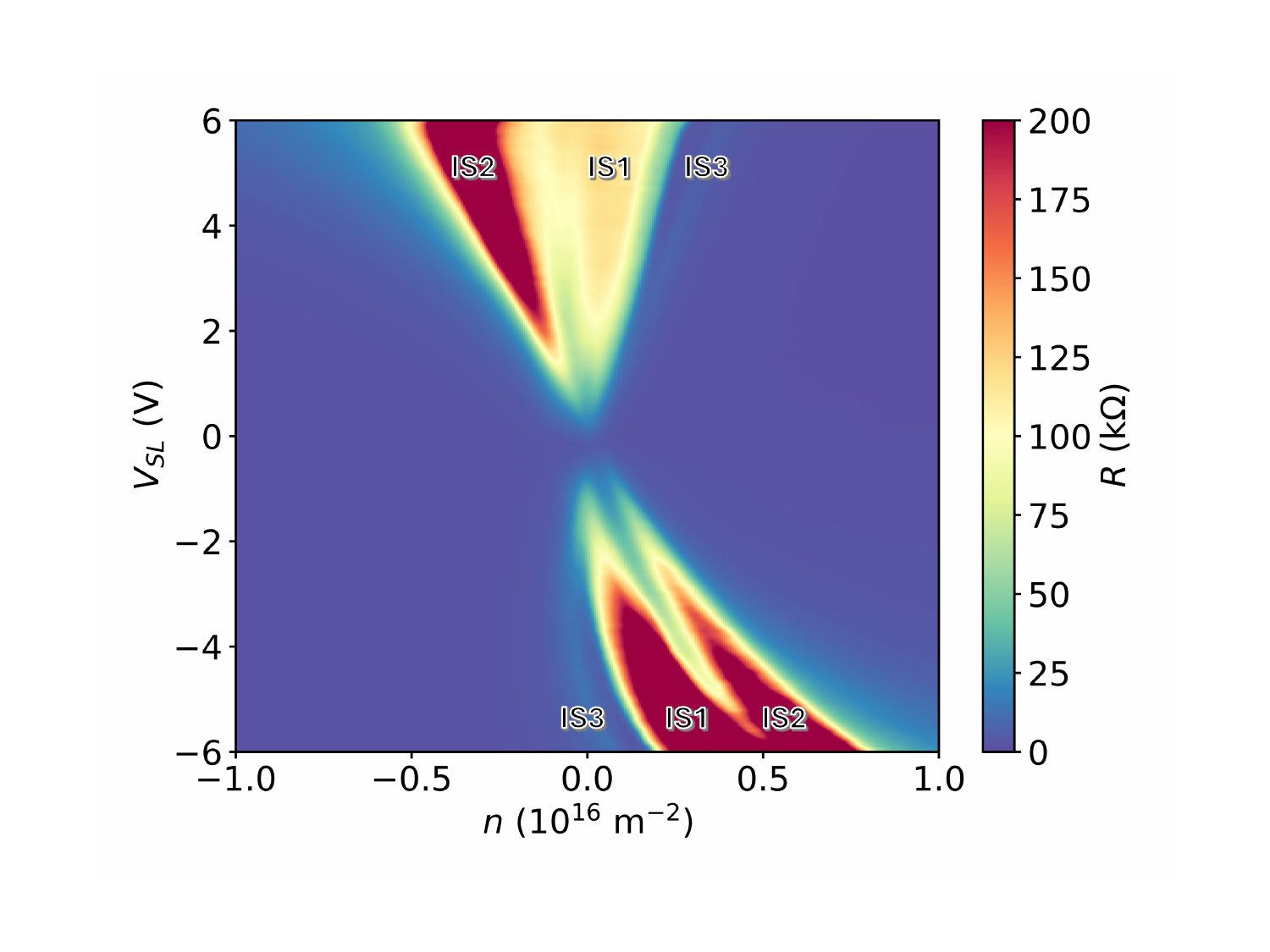}
\captionsetup{justification=raggedright,singlelinecheck=false}
\caption{\justifying\textbf{The whole phase of $R_{xx}$ vs. density $n$ and superlattice voltage $V_{SL}$} $-$
  $R_{xx}$ plotted as a function of total density $n$ and whole range of $V_{SL}$ values. Three primary resistance peaks are marked as $IS1, IS2$, and $IS3$. The position of $IS2$ and $IS3$ with respect to $IS1$ are reversed based on the sign of $V_{SL}.$ }
  \label{whole phase space}
\end{figure}

The device fabrication started with transferring the flakes in the required stacking order, followed by defining the Hall-bar geometry and making electrical contacts. First, a graphite flake was transferred onto a SiO$_2$/Si wafer using a PC polymer film on a PDMS stamp. The graphite was then patterned using the recipe described in the Methods section of the main text. Fig.~\ref{S1}a shows the optical image of the patterned graphite, and the corresponding SEM image in Fig~\ref{S1}b. Fig.~\ref{S1}b highlights the triangular periodic array of holes (dark circular regions). A heterostructure consisting of hBN/bilayer graphene/hBN was subsequently transferred onto the patterned gate (Fig.~\ref{S1}c). The stack was processed in multiple steps: depositing a metal top gate, etching the Hall-bar geometry, and finally forming 1D edge contacts. We used Cr/Au (5 nm/50 nm) for the top gate and Cr/Au (10 nm/100 nm) for the edge contacts. The optical image of the final device is presented in Fig.~\ref{S1}d.\\

In Fig.\ref{S2}, we show the estimation of mobility $\mu$ in our device. $\mu$ is calculated using Drude conductivity $\sigma=ne\mu$, where $n$ is the number density, $e$ is the charge of an electron. However, fitting a straight line in $\sigma$ vs. $n$ plot often leads to an error in the estimation of $\mu$ as it depends on the range of fitting. Therefore, we have plotted $\frac{1}{e}\left| \frac{d\sigma}{dn} \right|$ which saturates at finite $n$ on both electron and hole side, indicating $\sigma$ is linear in $n$. The $\mu$ values are found to be 340,000 cm$^2$V$^{-1}$s$^{-1}$ on the electron and 200,000 cm$^2$V$^{-1}$s$^{-1}$ on the hole side. Device D2 and D3 also exhibit high mobility of $150,000-200,000$ cm$^2$V$^{-1}$s$^{-1}$ and $90,000-100,000$ cm$^2$V$^{-1}$s$^{-1}$, respectively.\\

Fig.\ref{S3}a displays the color map of the longitudinal resistance, $R_{xx}$, as a function of top gate (TG) voltage $V_{TG}$ and superlattice (SL) gate voltage $V_{SL}$, with corresponding horizontal line cuts shown in Fig.\ref{S3}b. The charge neutrality point (CNP) shifts along a diagonal trajectory in the color plot, accompanied by an increase in the magnitude of resistance. Multiple prominent peaks emerge at higher $V_{SL}$. Notably, for $V_{SL} > 3$ V, a tiny peak is observed situated between two dominant peaks (indicated by the black arrow).\\

Fig.~2 in the main text shows the phase space obtained by sweeping the TG at a discrete SL gate voltage. The features are found to be stabilized by the periodic displacement field $D$ by the combination of both gates. In Fig.\ref{S4}, we show the reverse configuration obtained by sweeping the SL gate at discrete TG voltages. Multiple peaks in $R_{xx}$ appear at a finite periodic $D$ field. Fig.\ref{S4}a shows the color plot. The line plots at $V_{TG}=\pm4$ V are shown in Fig.\ref{S4}b-c. The insets show a closer view of the third weaker peak that appears at a high electric field. These results are identical to those shown in Fig. 2 of the main text.\\

In Fig.~\ref{lattice constant}, we show the estimation of the superlattice period from the densities corresponding to the secondary peaks. Fig.~\ref{lattice constant}a shows the phase space of $R_{xx}$ vs $V_{TG}-V_D$ for different values of $V_{SL}$. The black and red points show the evolution of $IS2$ and $IS3$, respectively. Fig.~\ref{lattice constant}b shows this more clearly, where the $V_{SL}$ values corresponding to the peaks are plotted. As can be seen, the peak positions shift toward higher $V_{TG}-V_D$, and hence higher density, as $V_{SL}$ increases. In Fig.~\ref{lattice constant}c, we estimate the lattice constant using the density $n$ obtained from the electrostatic capacitance; assuming a triangular lattice structure, we find a period $L$ between $30$-$60$ nm, close to the lithographically defined period of $\sim50$ nm. The effective reduction of the period is a consequence of the electrostatics of the patterned gate voltage when a finite voltage is applied.\\

We further estimate the Fermi energy $E_F$ associated with the IS2 and IS3 minibands as a function of $V_{\mathrm{SL}}$, using the carrier densities extracted from the peak positions and an effective mass of $m^*=0.05m_e$ (Fig.~\ref{lattice constant}d). Across the entire range of superlattice potential investigated, the estimated $E_F$ remains below $\sim 15$~meV. In comparison, the experimentally extracted gap of $\sim 40$~meV is approximately two to three times larger than $E_F$ at the highest $V_{SL}$.\\

In Fig.~\ref{whole phase space}, the three insulating states are labelled $IS1, IS2$, and $IS3$. For $V_\mathrm{SL}>0$ V, $IS1$ corresponds to the original charge neutrality point of bilayer graphene at $V_\mathrm{SL}=0$ V. $IS2$ denotes the first secondary peak at finite $V_\mathrm{SL}$ and is associated with the first miniband gap on the hole side ($V_\mathrm{TG}-V_\mathrm{D}<0$ V), whereas $IS3$ corresponds to the higher-$V_\mathrm{SL}$ secondary peak associated with the electron side ($V_\mathrm{TG}-V_\mathrm{D}>0$ V). For $V_\mathrm{SL}<0$ V, the features are approximately mirror symmetric, and the labels of IS2 and IS3 are reversed accordingly.

\clearpage

\subsection{Temperature dependence}

\begin{figure}[bth]
\includegraphics[width=0.98\linewidth]{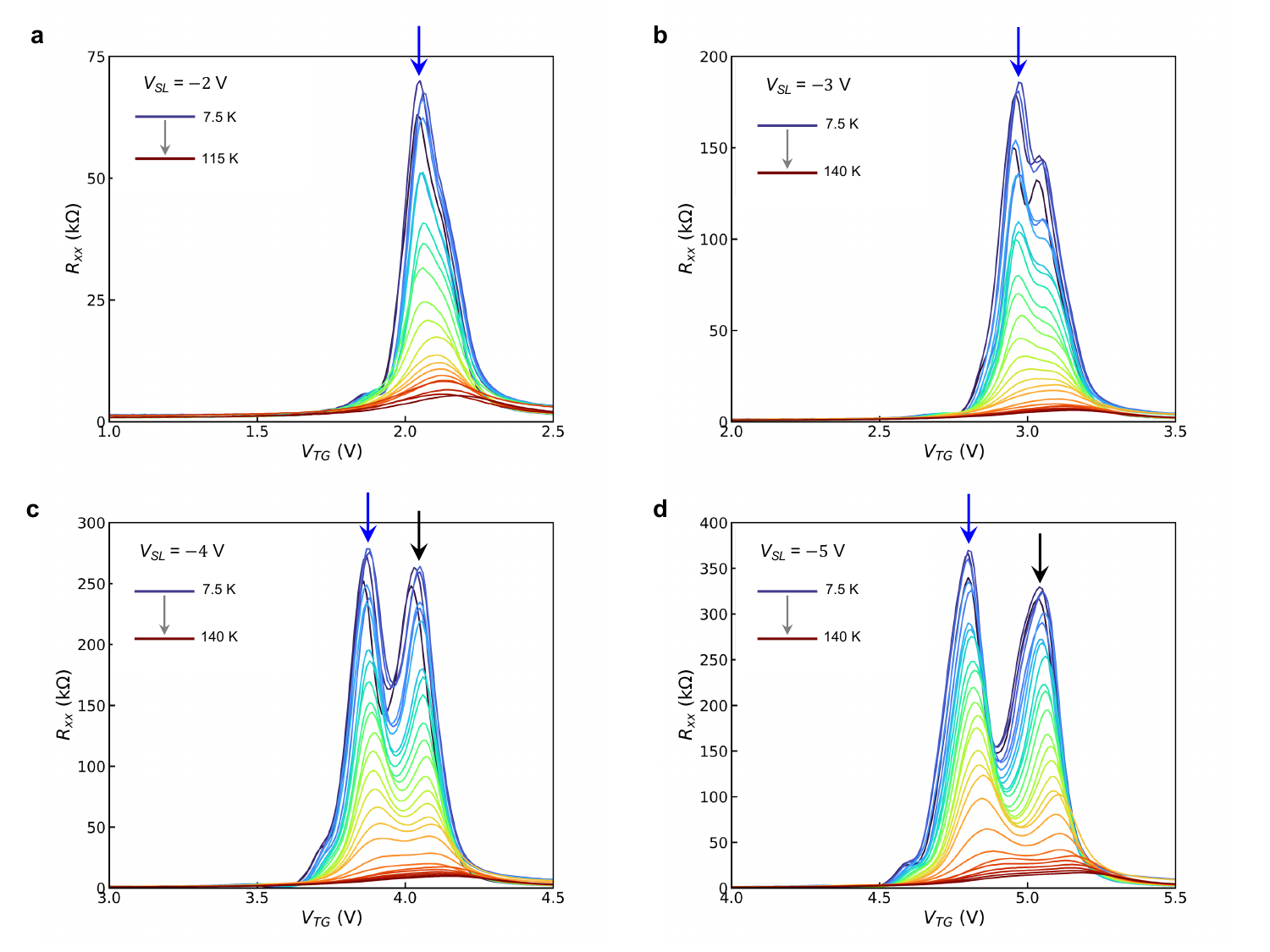}
\captionsetup{justification=raggedright,singlelinecheck=false}
\caption{\justifying\textbf{Temperature dependence of $R_{xx}$ at different $V_{SL}$}
  ~\textbf{a.} $R_{xx}$ as a function of $V_{TG}$ at $V_{SL}=-2$ V taken for different temperatures $T$ ranging from 7.5 K to 115 K.~\textbf{b.} $V_{SL}=-3 V$ up to $T=140$ K~\textbf{c.}~$V_{SL}=-4$ V.~\textbf{d.}~$V_{SL}=-5$ V. The blue and black arrows denote two insulating states $IS1$ and $IS2$ as defined in the text. Data for device D1.}
  \label{S5}
\end{figure}

The temperature dependence of $R_{xx}$ can give information about the band gap for insulating states. In Fig.\ref{S5}, we show $T$ dependence of $R_{xx}$ vs. $V_{TG}$ at different $V_{SL}$. All four panels (four SL voltages) show the insulating nature of the peaks in $R_{xx}$ up to a highest $T=140$ K. In Fig.\ref{S5}a-b, $R_{xx}$ is plotted for $V_{SL}=-2$ V and $-3$ V, where a single peak is observed, marked by the blue arrow ($IS1$ as defined in the main text). Fig.\ref{S5}c-d shows $R_{xx}$ vs. $V_{TG}$ at $V_{SL}=-4$ V, $-5$ V, which clearly show two peaks with the secondary peak ($IS2$ as defined in the main text) marked by a black arrow. Both the peaks remain insulating up to a high $T=140$ K.\\

\begin{figure}[bth]
\includegraphics[width=0.98\linewidth]{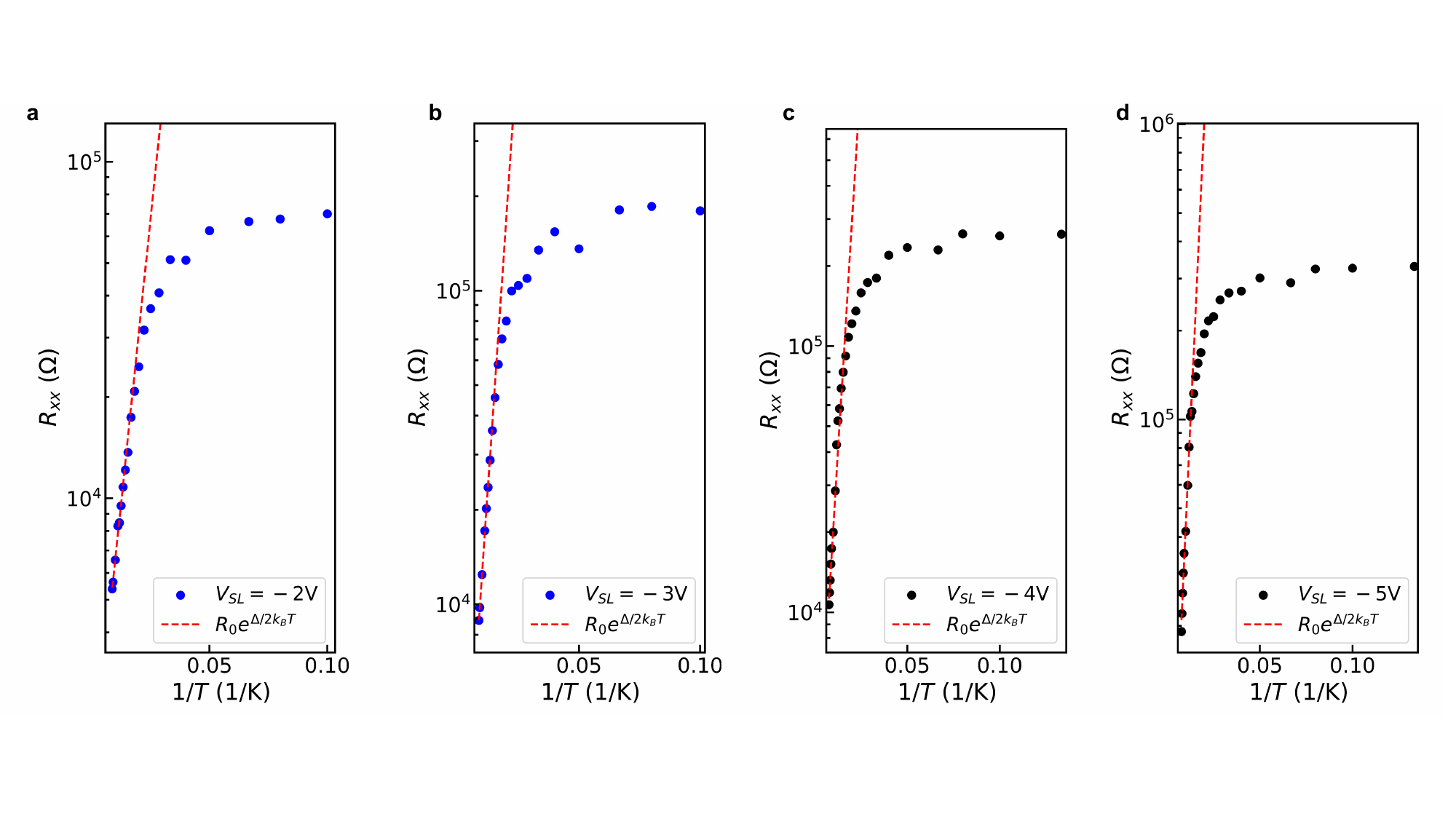}
\captionsetup{justification=raggedright,singlelinecheck=false}
\caption{\justifying\textbf{Fitting Arrhenius equation for thermally activated behavior.}
  ~\textbf{a.-b.} The maximum resistance of $IS1$, indicated by the blue arrow, is plotted as a function of $1/T$. The Arrhenius equation $R_{xx}=R_{0}$exp$(\Delta/2k_BT)$ is fitted for the high $T$ range that means lower values of $1/T$. The slope of these red dashed lines is $\Delta/2k_B$.~\textbf{c.-d.} Similar fitting done for the state $IS2$. Data for device D1.}
    \label{S6}

\end{figure}

To estimate the energy gap, we plot the peak values of $R_{xx}$ as a function of $1/T$. Prior studies have shown that at low temperatures ($T\leq30$ K in our case), the peak resistance weakly depends on $1/T$, consistent with variable range hopping (VRH) or a combination of VRH and nearest-neighbor hopping at low and intermediate $T$. At higher temperatures ($T\geq60$ K), however, $R_{xx}$ exhibits thermally activated behavior, following:

\[
R_{xx} \propto \exp(\Delta / 2k_BT),
\]

where $\Delta$ is the energy gap and $k_B$ is the Boltzmann constant. Fitting this relation to the $R_{xx}$ versus $1/T$ data allows us to extract $\Delta$. Fig.\ref{S6}a-b shows the peak $R_{xx}$ of $IS1$ as a function of $1/T$ for $V_{SL} = -2$ V and $-3$ V. From the slope of the dashed red fit lines, the extracted gaps are found to be on the order of tens of meV. A similar activated behavior is observed for $IS2$ at higher $V_{SL}$ values of $-4$ V and $-5$ V, as shown in Fig.\ref{S6}c-d. Repeating this process for each $V_{SL}$, we extract $\Delta$ and study its evolution with $V_{SL}$.~Although fitting the Arrhenius equation is a standard method for extracting energy gaps in the case of thermally activated transport, the results are highly sensitive to the temperature range chosen for the fit. Even small variations in the selected $T$ range can lead to significantly different energy gap values. Therefore, transport measurements alone are not sufficient to determine the energy gaps with high accuracy. In our analysis, we have used the fit that yields the lowest error in the higher temperature regime.\\

\begin{figure}[bth]
\includegraphics[width=0.98\linewidth]{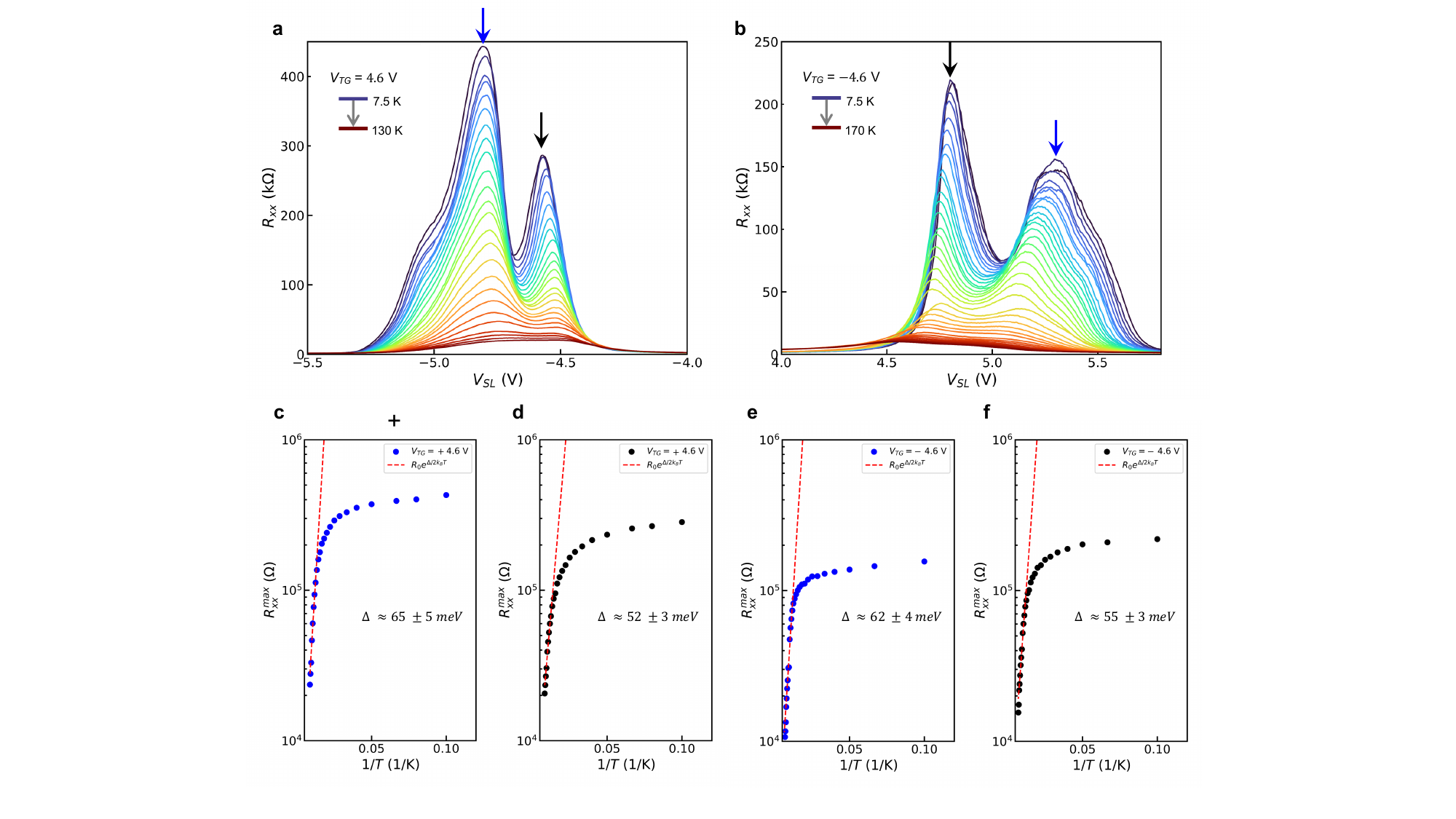}
\captionsetup{justification=raggedright,singlelinecheck=false}
\caption{\justifying\textbf{Activated energy gaps obtained from the SL gate sweep.}
  ~\textbf{a.} $R_{xx}$ vs. $V_{SL}$ at $V_{TG}=4.6$ V V plotted in a range of $T=7.5$ K to $130$ K. The blue and black arrow denotes the states $IS1$ and $IS2$, respectively.~\textbf {b.}~$R_{xx}$ vs. $V_{SL}$ at $V_{TG}=-4.6$ V plotted in a range of $T=5$ K to $170$ K.~\textbf {c.-d.}~At $V_{TG}=4.6$ V Peak resistance for $IS1$ (blue circles) and $IS2$ (black circles) plotted as a function of $1/T$. The slope of the dashed line results in $\Delta$ of $65\pm5$ meV for $IS1$ and $52\pm3$ meV for $IS2$.~\textbf {e.-f.}~Similar fitting is done for $V_{TG}=-4.6$ V for $IS1$ and $IS2$. The obtained $\Delta$ values are $62\pm4$ meV for $IS1$ and $55\pm3$ meV for $IS2$. Data for device D1.}
    \label{S7}

\end{figure}

Fig.\ref{S7}a-b presents the temperature dependence of $R_{xx}$ measured by sweeping $V_{SL}$ at fixed $V_{TG} = \pm 4.6$ V, up to 130 K and 170 K, respectively. Both plots exhibit two prominent peaks ($IS1$ and $IS2$), which are indicated by blue and black arrows, respectively. In Fig.\ref{S7}c-f, we have plotted the maximum resistance of $IS1$ and $IS2$ as a function of $1/T$. Fitting the data with the red dashed lines gives gap values of approximately in the range of $\approx52-65$ meV, which is slightly larger than the gaps extracted in Fig.~3e-f. This difference may arise from a pristine gap resulting from reduced charge disorder when using a graphite gate instead of a metal gate, as previously reported~\cite{icking2022transport}.\\

\begin{figure}[bth]
\includegraphics[width=0.98\linewidth]{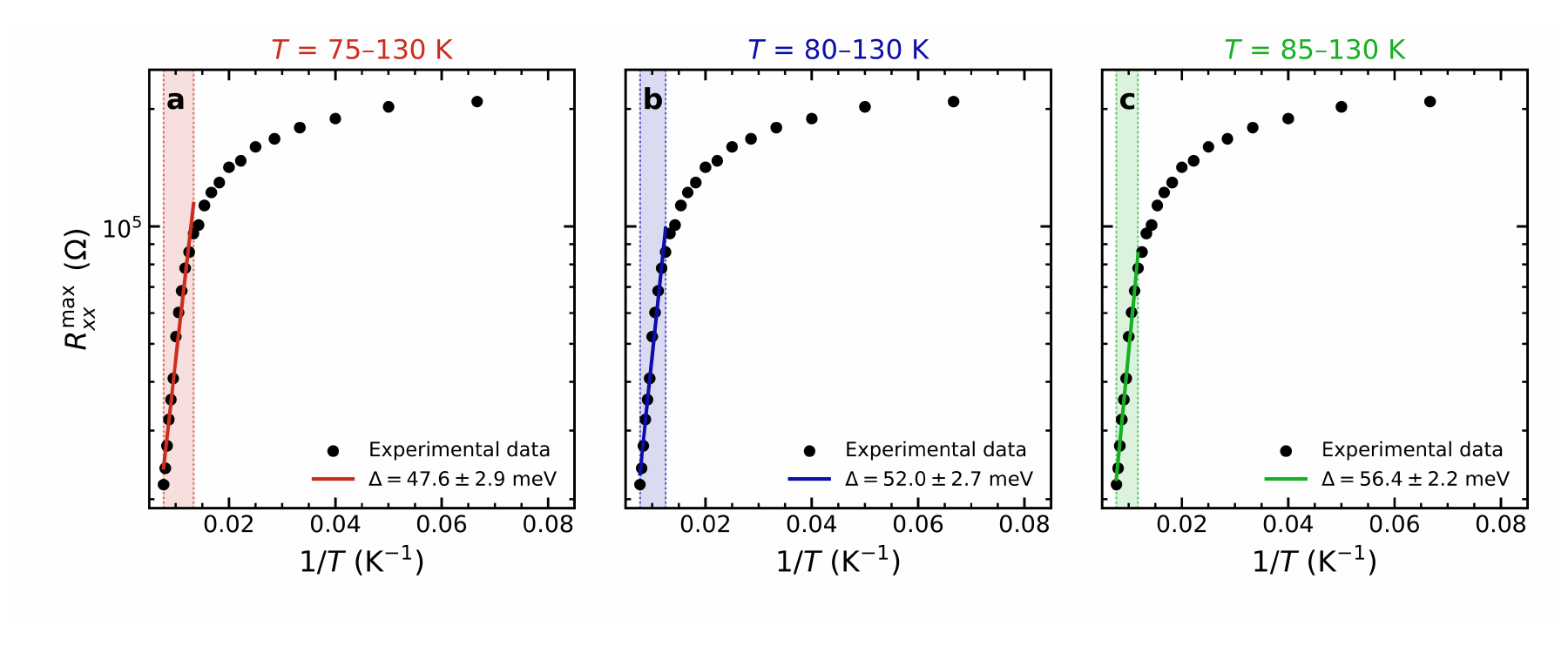}
\captionsetup{justification=raggedright,singlelinecheck=false}
\caption{\justifying\textbf{Sensitivity of the extracted activation gap to the fitting range.}
\textbf{a.-c.} Arrhenius plots of $R_{xx}^{\mathrm{max}}$ for different temperature fitting ranges at $V_{TG}=-4.6$ V. The extracted activation gap varies by approximately 10 meV when the lower bound of the fitting temperature range is shifted by 10 K.
}
 \label{delta fitting_sl}

\end{figure}

Analysis of thermally activated transport does not necessarily provide a robust estimate of the energy gap~\cite{icking2022transport, PhysRevB.82.081407}. To assess the reliability of the extracted activation gap, we fit the temperature-dependent resistance using the thermally activated form $R_{xx}^{\mathrm{max}} = R_0 \exp(\Delta/2k_{\mathrm{B}}T)$ over three different $1/T$ intervals. As shown in Figs.~\ref{delta fitting_sl}a--c, different temperature ranges of the data can be reasonably described by thermally activated behavior. However, the corresponding extracted values of $\Delta$ depend strongly on the selected fitting range, with a small shift of approximately 5 K in the lower temperature bound resulting in a change of approximately 5 meV in $\Delta$. These results demonstrate that the extracted activation energy is highly sensitive to the temperature interval chosen for fitting, indicating that activated-transport analysis may not always provide a reliable estimate of the underlying energy gap.

\subsection{Summary of transport data in Sections A and B}

As discussed above, the additional insulating states, $IS2$ and $IS3$, emerge only upon applying a finite superlattice potential, whereas conventional BLG behavior with a single CNP feature ($IS1$) is recovered in its absence. This, together with the emergence of multiple resistance peaks at finite periodic displacement field, supports their origin from a superlattice-induced reconstruction of the BLG band structure rather than from conventional BLG physics alone. A particularly important signature is the pronounced electron-hole asymmetry of these states: upon reversing the sign of $V_{SL}$, the relative positions of $IS2$ and $IS3$ are reversed. This behavior is reproduced by our band-structure calculations, which show that reversing $V_{SL}$ inverts the electron-hole asymmetry, providing qualitative agreement between experiment and theory. We further characterize these insulating states through their temperature dependence. At high temperatures, where thermally activated transport dominates, Arrhenius analysis provides an order-of-magnitude estimate of the associated energy gaps. Although the extracted values are sensitive to the selected fitting window and should therefore not be regarded as precise measurements of the band gaps, their overall energy scales are broadly consistent with the theoretical calculations. This combined agreement in the emergence, gate-polarity dependence, and characteristic energy scales of the insulating states provides evidence that $IS2$ and $IS3$ originate from the superlattice-engineered miniband structure.

\clearpage
\subsection{Transport data in the presence of a magnetic field}

\begin{figure}[bth]
\includegraphics[width=0.98\linewidth]{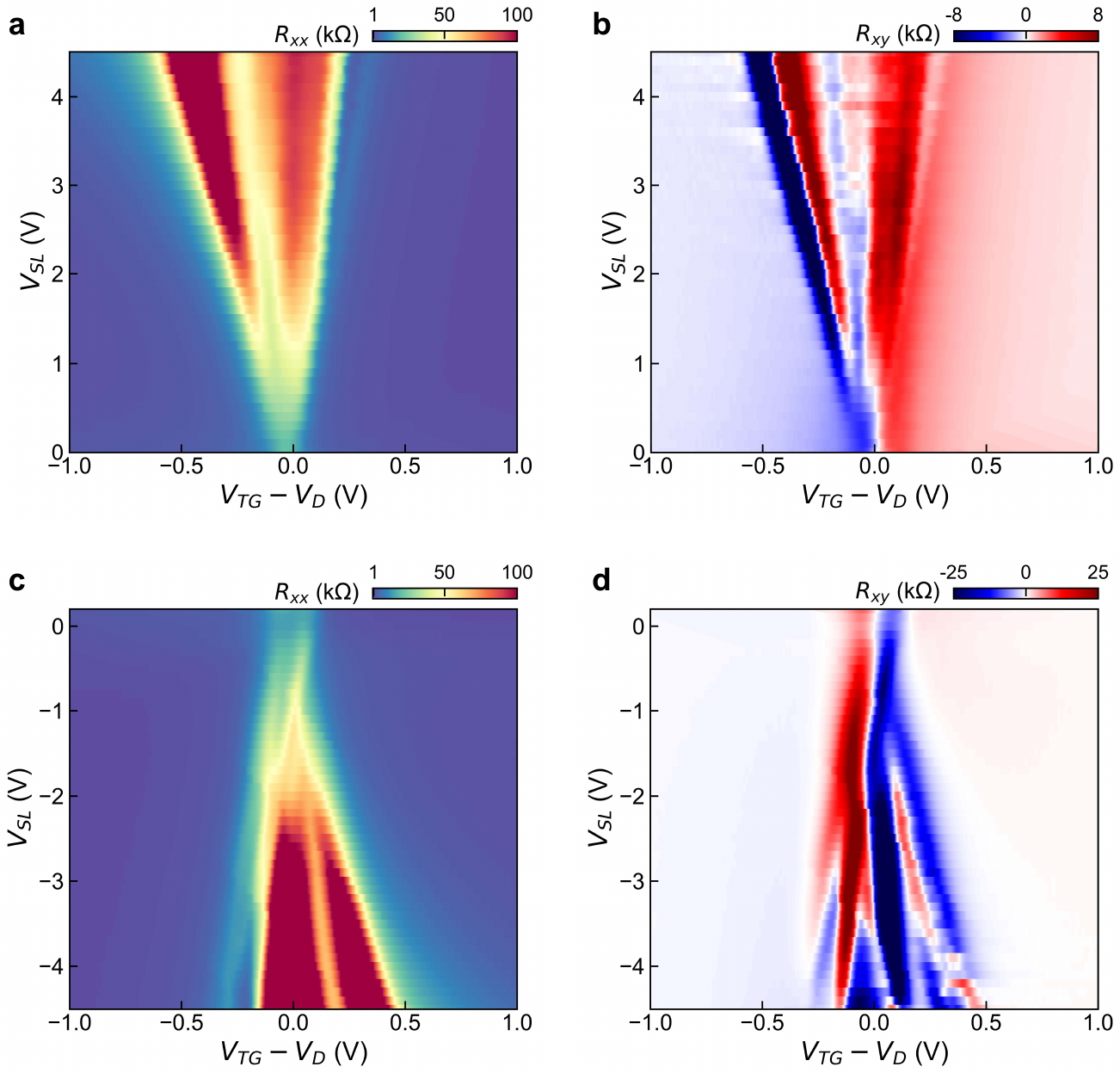}
\captionsetup{justification=raggedright,singlelinecheck=false}
\caption{\justifying\textbf{Phase space of $R_{xx}$ and $R_{xy}$ at $B=1$ T and $V_{BG}=-20$ V.}
  \textbf{a.}~$R_{xx}$ plotted as a function of positive SL voltage $V_{SL}$ and $V_{TG}-V_D$, where $V_{TG}$ is the top gate voltage and $V_{D}$ is the CNP offset. Two peaks are clearly observed. \textbf{b.}~$R_{xy}$ is plotted in the same range of gate voltages. Sign changes are observed for the peaks in $R_{xx}$. \textbf{c.}~$R_{xx}$ plotted as a function of negative SL voltage $V_{SL}$ and $V_{TG}-V_D$. \textbf{d.} $R_{xy}$ is plotted in the same range of gate voltages. Similar to panel b, multiple sign changes are observed for the peaks in $R_{xx}$. Data for device D1.}
    \label{S8}

\end{figure}

In Fig.\ref{S8}, we present contour plots of $R_{xx}$ and $R_{xy}$ as a function of $V_{TG}-V_D$ for positive and negative $V_{SL}$ values at $B = 1$ T and $V_{BG}=-20$ V. At $V_{SL} = 0$ V, a single sign change is observed, corresponding to carrier inversion at the CNP of bilayer graphene. However, with increasing $V_{SL}$, new resistive peaks and corresponding sign changes in $R_{xy}$ emerge. The sign changes in $R_{xy}$ somewhat match with the peaks in $R_{xx}$ and may be associated with Fermi surface reconstructions and charge carrier inversion upon crossing band gaps.\\

\subsection{COMSOL simulations}

\begin{figure}[bth]
\includegraphics[width=0.98\linewidth]{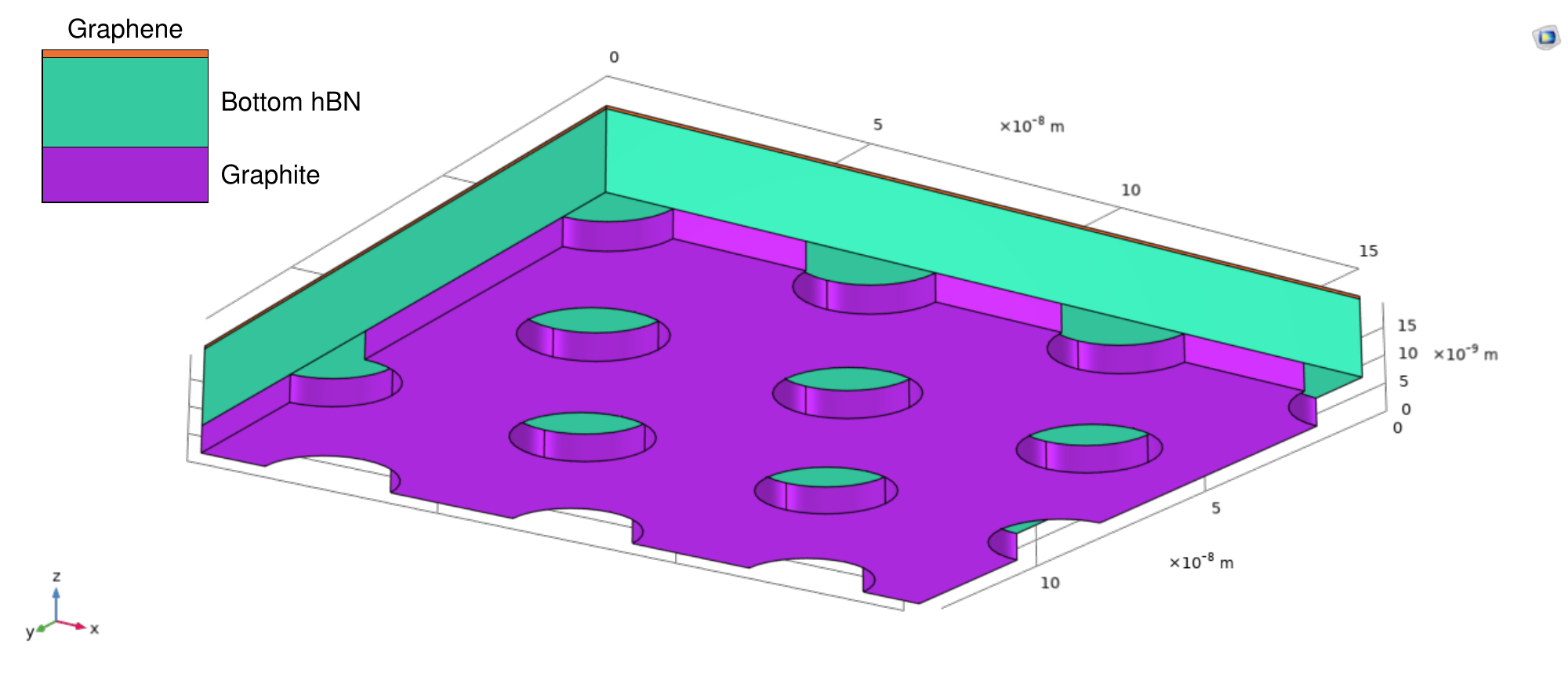}
\captionsetup{justification=raggedright,singlelinecheck=false}
\caption{\justifying\textbf{COMSOL model.}~Three vertically stacked blocks are made in COMSOL. Graphene and bottom hBN are denoted by the orange and cyan colors, respectively. The holes with a diameter of 25 nm are made in the graphite layer (purple color). A lattice periodicity of 50 nm was chosen for the simulation. The total dimension of the model system is 150 nm $\times$ 129 nm.}
  \label{S9}

\end{figure}

In order to find the charge density profile on BLG due to the patterned graphite gate, we have used the COMSOL Multiphysics simulation. The geometry was constructed using COMSOL Multiphysics 6.0. Fig.\ref{S9} shows the model, where holes with a diameter of 25 nm are patterned in a graphite layer (shown in purple). The superlattice period is fixed at 50 nm, matching that of the patterned graphite gate in our device. The bilayer graphene (BLG, shown in orange) is separated from the graphite gate by a 14 nm-thick bottom hBN dielectric layer (cyan). This hBN thickness was estimated from low-field Hall measurements. In the simulation, the BLG is modeled as a grounded metallic sheet with its upper boundary set at $V_{BLG} = 0$ V. A finite voltage $V_{SL}$ is applied to the graphite gate to determine the induced gate voltage profile in the device. Since the BLG is grounded, the top hBN layer is excluded from the model, as it does not affect the induced charge density from the patterned graphite gate. These details are shown in the table~\ref{tab:model_parameters} \\

\begin{table}[htbp]
\centering
\caption{Device parameters}
\begin{tabular}{lll}
\hline
\textbf{Parameter} & \textbf{Value} & \textbf{Description} \\
\hline
Hole diameter 
& $25~\mathrm{nm}$ 
& Diameter of holes patterned in the graphite gate \\

Superlattice period 
& $50~\mathrm{nm}$ 
& Periodicity of the patterned graphite gate \\

Bottom hBN thickness 
& $14~\mathrm{nm}$ 
& Separation between BLG and patterned graphite gate \\

Active layer 
& BLG 
& Bilayer graphene \\

Patterned gate 
& Graphite 
& Graphite layer containing the periodic hole array \\

Dielectric 
& hBN 
& Bottom dielectric separating BLG from graphite
\\
\hline
\end{tabular}
\label{tab:model_parameters}
\end{table}

\begin{figure}[bth]
\includegraphics[width=0.98\linewidth]{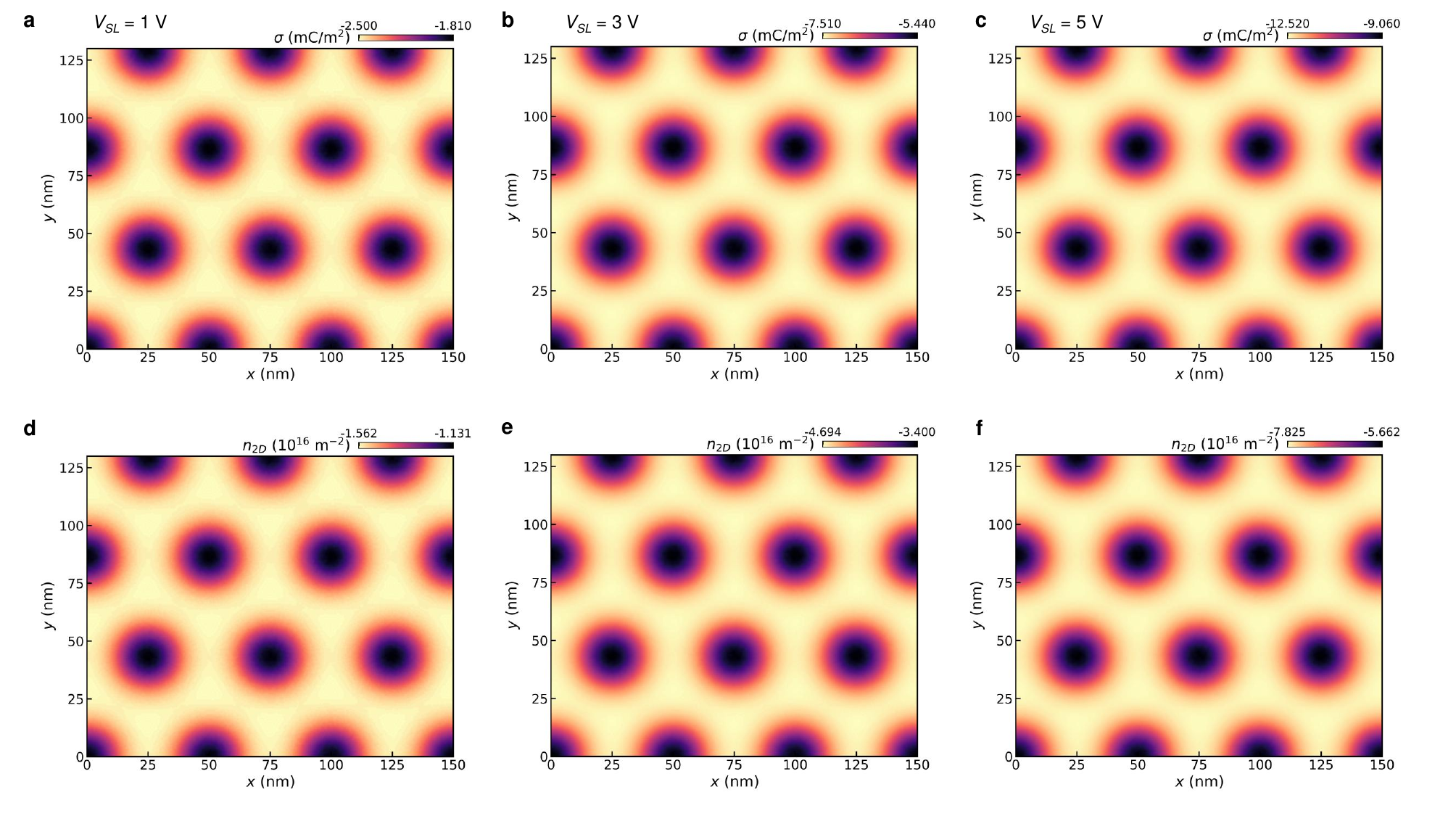}
\captionsetup{justification=raggedright,singlelinecheck=false}
\caption{\justifying\textbf{Simulation of charge density.}~\textbf{a.-c.}~The 2D profile of surface charge density $\sigma$ on BLG when a constant dc voltage of $V_{SL}=1$, 3, and 5 V is applied to the graphite gate. The density in the hole region is non-zero because of the fringing effect of the sub-nm pattern. \textbf{b.}~Charge carrier density $n_{\mathrm{2D}}=\sigma/e$ shown three superlattice gate voltages shown in \textbf{a}. The density is minimum inside the holes and maximum outside the holes.~Both $\sigma$ and $n_{\mathrm{2D}}$ increase with increasing $V_{SL}$ as shown by the minima and maxima of the colour bars.}
\label{S10}

\end{figure}

\begin{figure}[bth]
\includegraphics[width=0.98\linewidth]{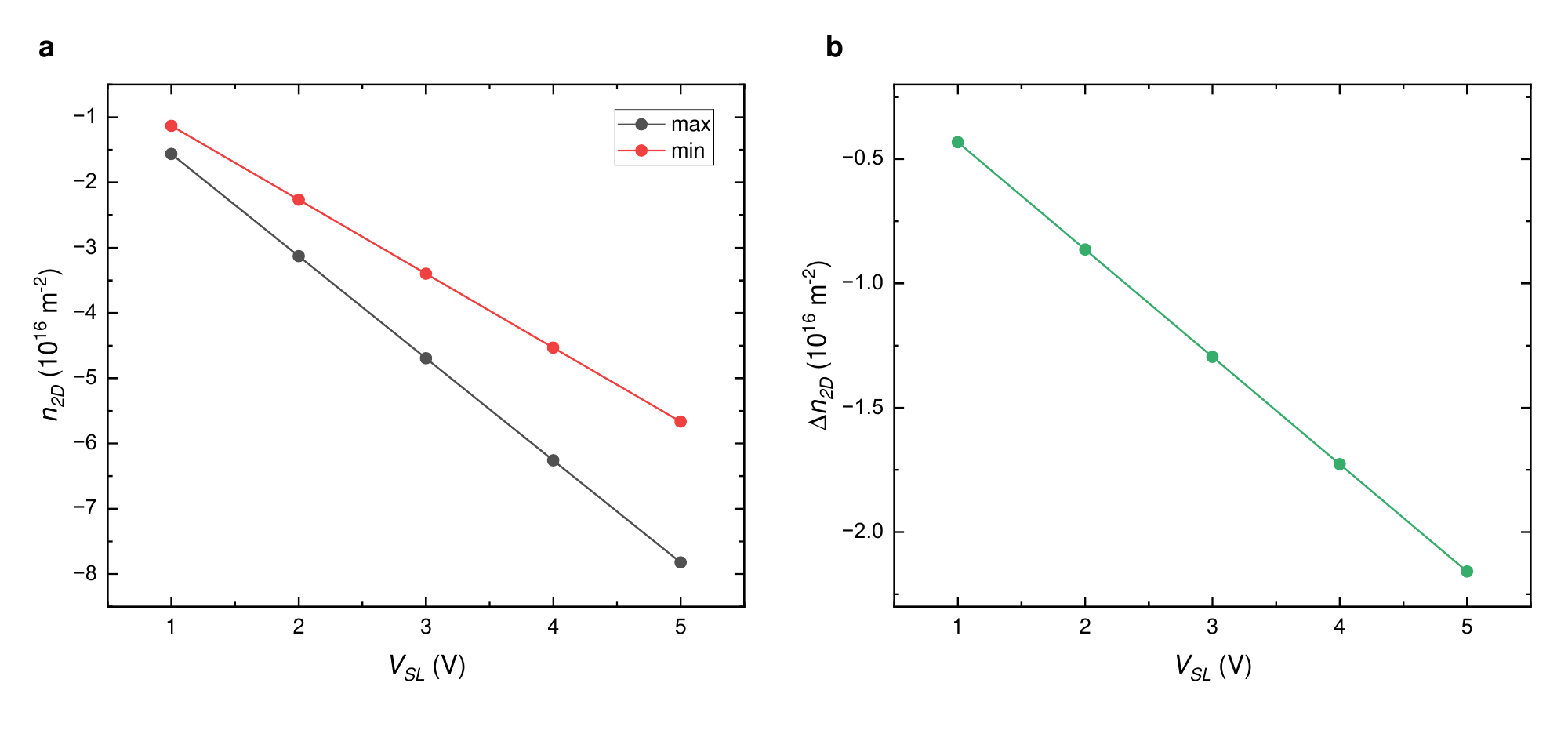}
\captionsetup{justification=raggedright,singlelinecheck=false}
\caption{\justifying\textbf{Estimation of maximum and minimum density from COMSOL.}~\textbf{a.}~Maximum and minimum density $n_{\mathrm{max}}$ and $n_{\mathrm{min}}$, respectively obtained from the oscillating profile of $n_{2D}$ is plotted as function of $V_{SL}$. The magnitude of both the quantities increases with increasing $V_{SL}$.~\textbf{b.}~Difference $\Delta n=n_{\mathrm{max}}-n_{\mathrm{min}}$ plotted as a function $V_{SL}$ also shows a increasing trend with increasing $V_{SL}$.}
  \label{S11}

\end{figure}

The dielectric constant of hBN was explicitly defined within the material properties. The graphite gate was modeled as a standard metallic electrode, similar to gold, to simplify the boundary conditions. We employed the electrostatics module in COMSOL, where charge conservation is inherently enforced. First, simulations were carried out using a plain graphite gate configuration, and the resulting charge density profiles showed good agreement with estimates derived from electrostatic capacitance calculations. This consistency validates the reliability of our modeling approach for capturing the spatial charge distribution in the system. The surface charge density $\sigma(\vec{r})$ on the BLG is obtained from the displacement field $\vec{D}$ via $\sigma(\vec{r}) = \hat{n} \cdot \vec{D}$, where $\hat{n}$ is the unit normal vector to the BLG plane. Fig.\ref{S10}a-c shows the spatial distribution of $\sigma(\vec{r})$ in the $x$–$y$ plane at $V_{SL} = 1, 3$ and 5 V, in units of mC/m$^2$. It is observed that $\sigma(\vec{r})$ is minimum inside the holes and maximum outside. The negative sign of $\sigma(\vec{r})$ indicates electron doping. Fig.\ref{S10}d-f presents the carrier density distribution $n_{2D} = \sigma/e$ in the BLG plane at $V_{SL} = 1, 3$ and 5 V, revealing clear density modulation. In Fig.\ref{S11}a, we show the maximum and minimum in the oscillating profile of $n_{\mathrm{2D}}$ for $V_{SL}$ values from 1 to 5 V. The value of $n_{\mathrm{max}}$ matches with the carrier density expected from a uniform gate voltage at the corresponding $V_{SL}$. We define the amplitude of density modulation as $\Delta n = n_{\mathrm{max}} - n_{\mathrm{min}}$.~Notably, $\Delta n$ also increases linearly with increasing $V_{SL}$ (Fig.\ref{S11}b).\\

\begin{figure}[bth]
\includegraphics[width=0.98\linewidth]{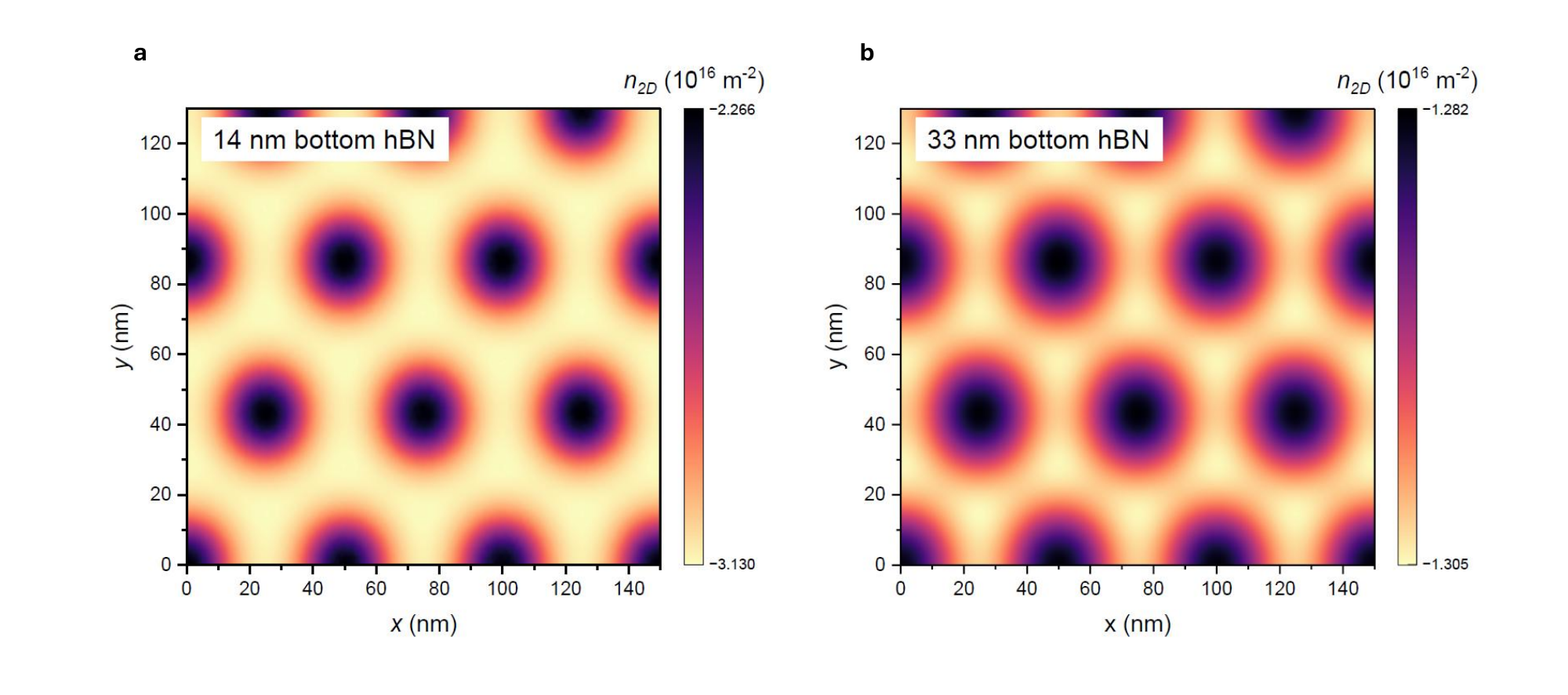}
\captionsetup{justification=raggedright,singlelinecheck=false}
\caption{\justifying\textbf{hBN thickness dependence of number density.}~\textbf{a.-b.}~Number density from the superlattice gate $n_\mathrm{2D}$ plotted as a function of $x$ and $y$ at a superlattice gate voltage $V_{SL}=2$ V for a bottom hBN thickness of 14 nm and 33 nm, respectively. The modulation of density is extremely weak, and the potential variation is found to be almost uniform for the 33 nm hBN, with a variation of $\Delta n=-0.02\times10^{16}$ m$^{-2}$. The simulation for a bottom hBN thickness of 14 nm shows an enhanced density modulation with a $\Delta n=-0.86\times10^{16}$ m$^{-2}$ at $V_{SL}=2$ V.}
  \label{S12}

\end{figure}

The thickness of the bottom hBN plays a crucial role in determining the modulation of density $\Delta n$. If the thickness of the bottom hBN becomes comparable to the period of the superlattice, then it gives rise to a smooth density profile on BLG. Fig.\ref{S12} shows the spatial distribution of $n_{\mathrm{2D}}$ at $V_{SL}=2$ V for two thicknesses of hBN of 14 nm (Fig.\ref{S12}a) and 33 nm (Fig.\ref{S12}b). We note that $\Delta n$ is significantly reduced for a 33 nm-thick hBN layer compared to 14 nm. This indicates that the influence of the superlattice gate becomes weaker with increasing bottom hBN thickness. We have estimated the total density $n$ as a function of the density from the top gate $n_{\mathrm{TG}}$ and the superlattice gate $n_{\mathrm{SL}}$ and expressed as $n=n_{TG}+n_{SL}+n_0$, where $n_0=-0.25\times10^{16}$ is the residual carrier density. $n_{\mathrm{SL}}$ is an average density of $n_{\mathrm{max}}$ and $n_{\mathrm{min}}$ and can be expressed as $n_{SL}=\alpha n_{\mathrm{max}}+\beta n_{\mathrm{min}}$ where $\alpha$ and $\beta$ are the fraction of area of the total area for plain graphite and holes in the graphite.\\

\begin{figure}[bth]
\includegraphics[width=0.98\linewidth]{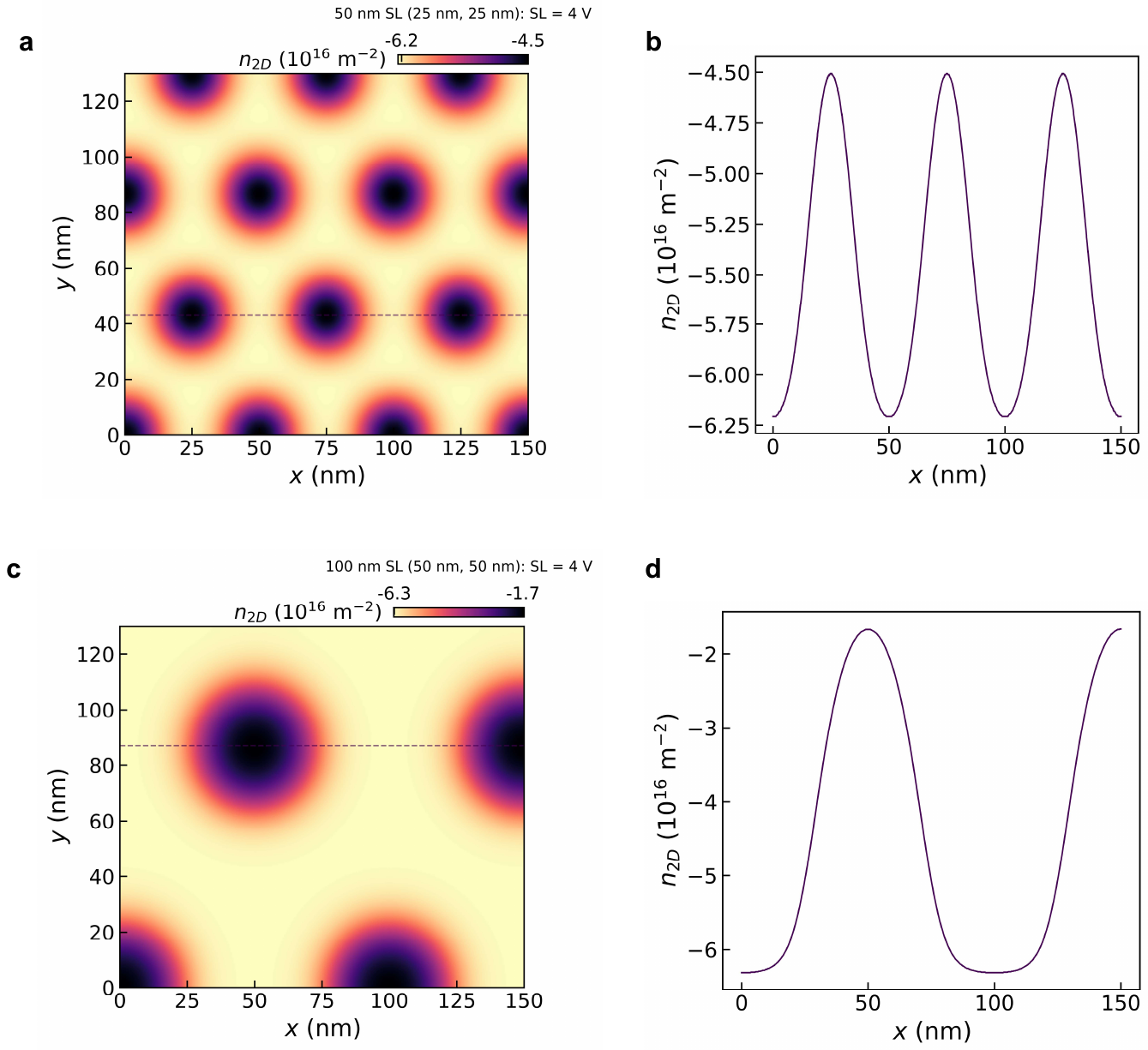}
\captionsetup{justification=raggedright,singlelinecheck=false}
\caption{\justifying\textbf{Periodicity dependence of the charge density ($n_{2D}$).}~\textbf{a.-b.}~Charge density distribution for a SL with a 50 nm period (25 nm diameter holes separated by 25 nm spacing). The line cut is taken along the horizontal dashed line in a.~\textbf{c.-d.} Charge density for a 100 nm period SL (50 nm diameter holes separated by 50 nm spacing). Calculations are shown for a superlattice gate voltage $V_{SL}=4$ V. The density modulation is observed to be stronger for the 100 nm period due to the relationship between hole diameter and dielectric layer thickness.}
  \label{S12_2}

\end{figure}

We now calculate the periodicity dependence of the charge density in BLG. Fig.\ref{S12_2} illustrates the charge density, $n_{2D}$, for superlattice periods of 50 nm and 100 nm. The 50 nm period consists of 25 nm diameter holes separated by 25 nm, while the 100 nm period features 50 nm holes with 50 nm spacing. At $V_{SL}=4$ V, the density modulation is notably stronger for the 100 nm SL compared to the 50 nm case (compare Fig.\ref{S12_2}a-b and Fig.\ref{S12_2}c-d). This amplitude is determined by the interplay between the hole diameter and the thickness of the dielectric layer.\

\begin{figure}[bth]
\includegraphics[width=0.98\linewidth]{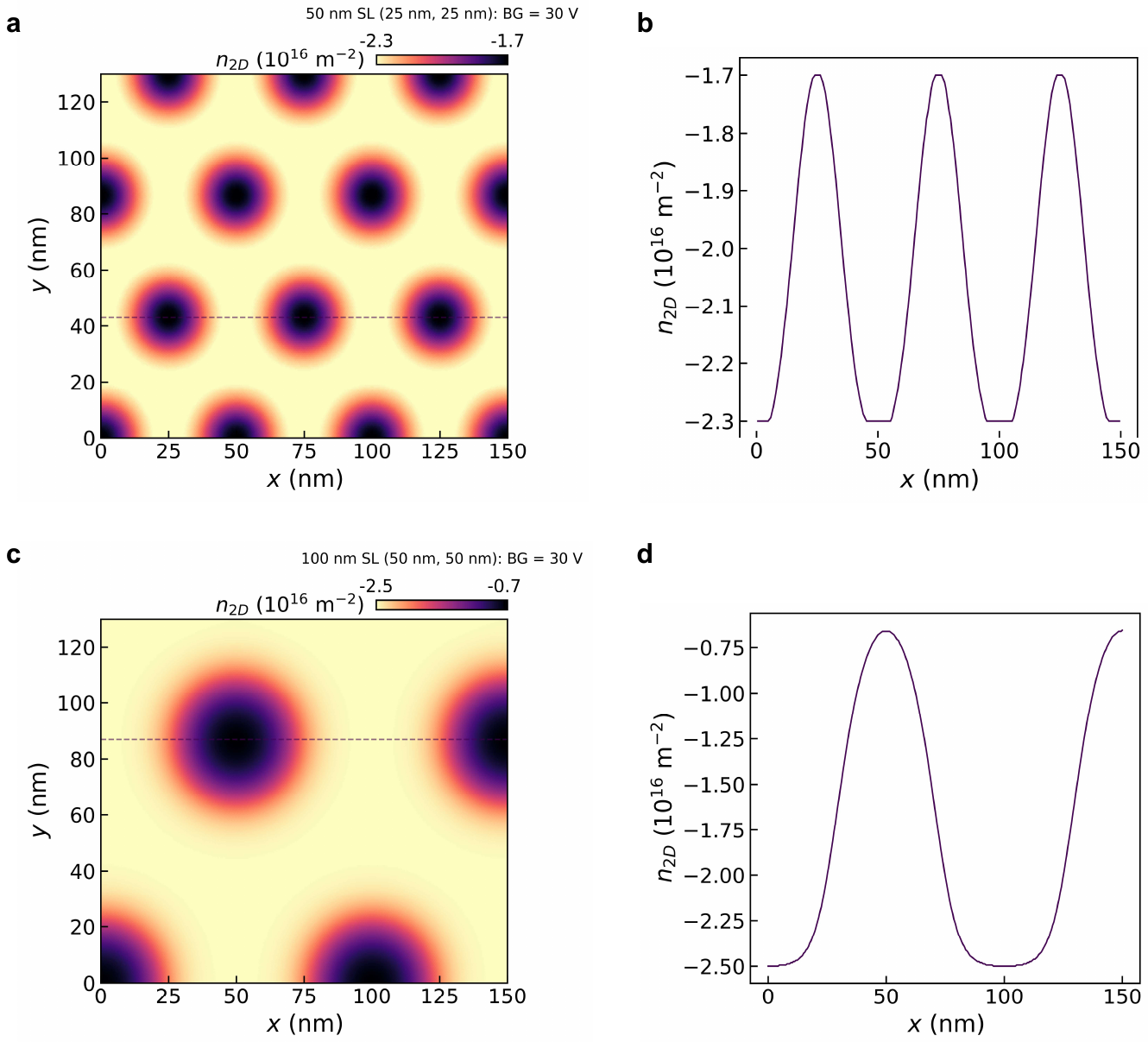}
\captionsetup{justification=raggedright,singlelinecheck=false}
\caption{\justifying\textbf{Charge density modulation induced by the global Si bottom gate.}~\textbf{a.} Spatial $x-y$ map of $n_{2D}$ for the 50 nm period SL at a back gate voltage $V_{BG}=30$ V. \textbf{b.} Corresponding line plot of $n_{2D}$ along the horizontal dashed line in a. \textbf{c.} Spatial $x-y$ map of $n_{2D}$ and \textbf{d.} corresponding line plot for the 100 nm period SL at $V_{BG}=30$ V. The bottom gate induces a periodic modulation dependent on dielectric thickness, with higher density observed for the 100 nm SL.}
  \label{S13}

\end{figure}

The application of a global Si bottom gate (BG) also induces a periodic charge density, as discussed in the Supplemental Material and Fig. 4 of the main text. Fig.\ref{S13}a and c display spatial maps of $n_{2D}$ at $V_{BG}=30$ V for 50 nm and 100 nm SLs, respectively. Analogous to the SL gate, the BG induces a finite modulation where the magnitude depends on $V_{BG}$ and the bottom dielectric thickness. Notably, the density is observed to be higher for the 100 nm SL (see line plots in Fig.\ref{S13}b and d).\

\clearpage
\subsection{Additional Transport Data}

\begin{figure}[bth]
\includegraphics[width=0.98\linewidth]{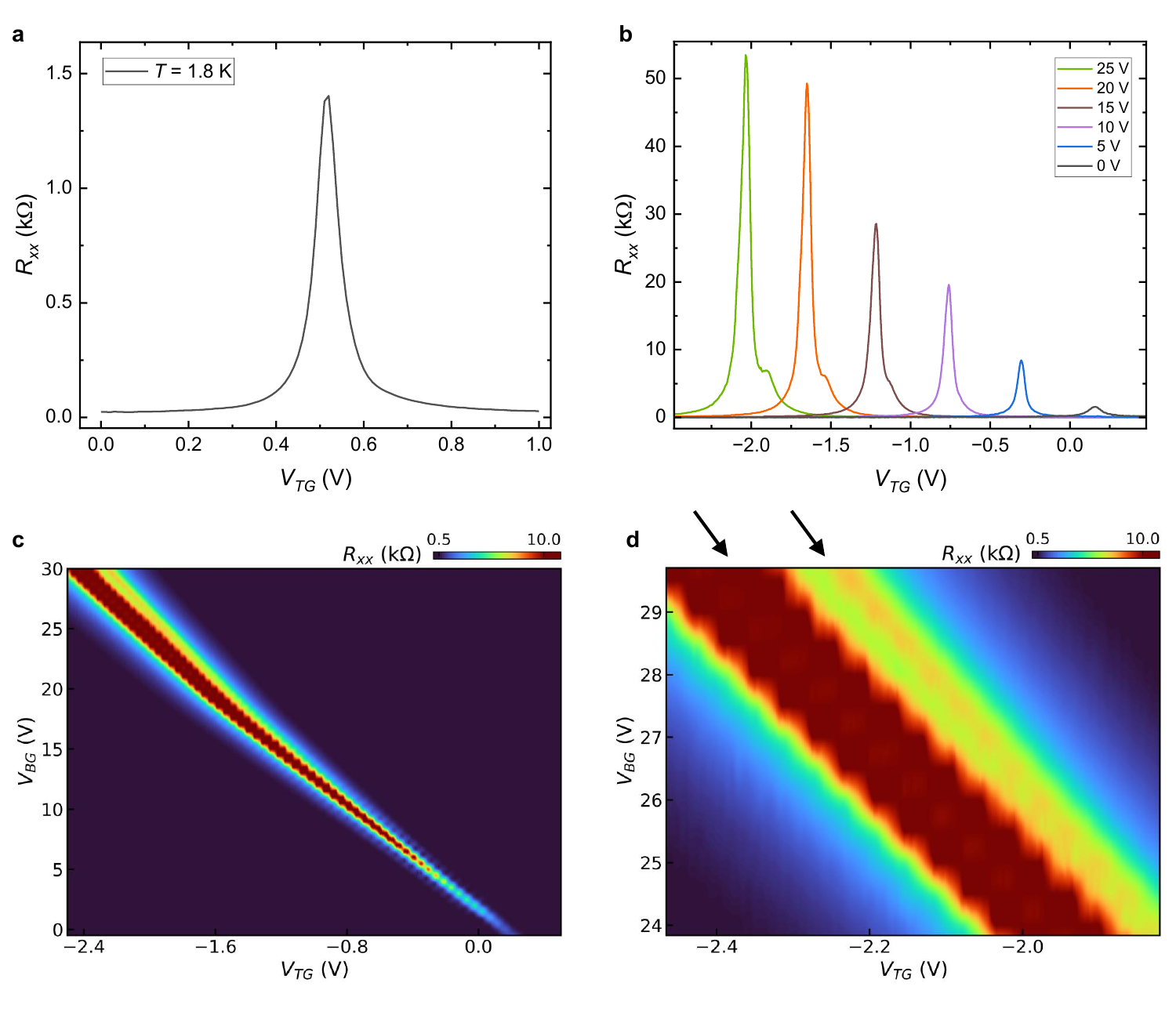}
\captionsetup{justification=raggedright,singlelinecheck=false}
\caption{\justifying\textbf{Transport characteristics of device D2.}~\textbf{a.}~Longitudinal resistance $R_{xx}$ as a function of top gate voltage $V_{TG}$ at $T = 1.8~\mathrm{K}$, showing a single peak at the CNP.~\textbf{b.}~Application of a finite $V_{BG}$ gives rise to an additional shoulder peak.~\textbf{c.}~Color map of $R_{xx}$ as a function of $V_{TG}$ and $V_{BG}$, illustrating the evolution of the peaks, where two distinct resistance maxima emerge at higher $V_{BG}$ (indicated by black arrows) in the zoomed-in panel in~\textbf{d.}}
  \label{S14}

\end{figure}

In the following we discuss various transport measurements performed on three additional devices D2, D3 and D4, demonstrating consistency and reproducibility of our gate-engineering approach, while also investigating factors such as dielectric thickness dependence and gate tunability. In Fig.\ref{S14}, we show the transport data of device D2, in which a global Si BG was used to apply a periodic potential on BLG, given that the SL graphite gate acts as a screening layer. The hBN layer separating BLG from SiO$_2$ has a thickness of 33 nm. Fig.\ref{S14}a shows $R_{xx}$ as a function of $V_{TG}$ at $T=1.8$ K, where a single peak at the CNP is observed. The application of $V_{BG}$ leads to a secondary shoulder peak (Fig.\ref{S14}b). The color map of $R_{xx}$ vs. $V_{TG}$ and $V_{BG}$ illustrates the evolution of the peaks (Fig.\ref{S14}c). A closer look at the higher $V_{BG}$ clearly shows two peaks in $R_{xx}$, as indicated by two black arrows.\\

\begin{figure}[bth]
\includegraphics[width=0.98\linewidth]{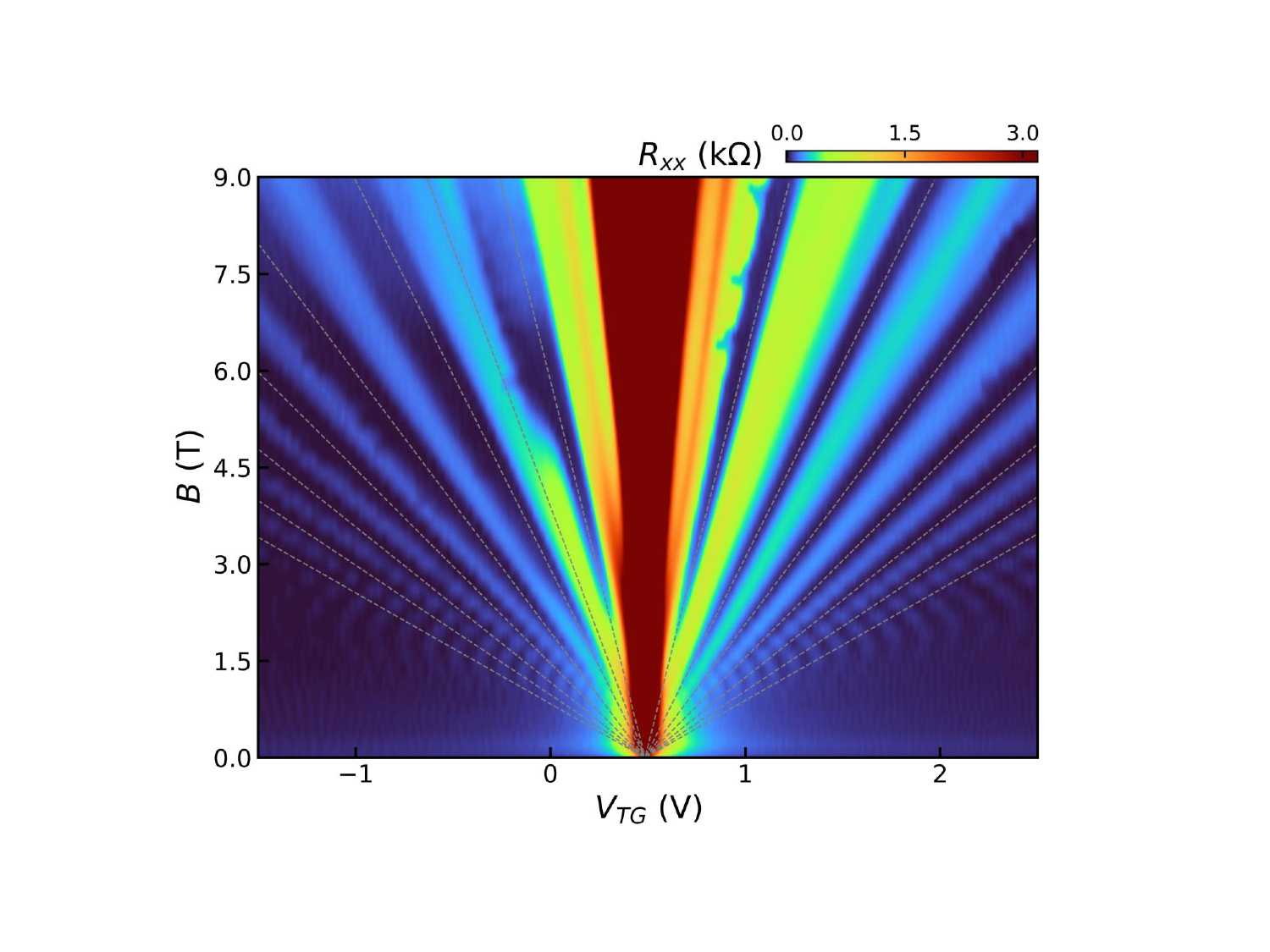}
\captionsetup{justification=raggedright,singlelinecheck=false}
\caption{\justifying\textbf{Landau fan diagram of BLG at $V_{BG}=0$ V; Device D2.}~Longitudinal resistance $R_{xx}$ plotted as a function of top gate voltage $V_{TG}$ and magnetic field $B$ at $V_{BG}=0$~V.~A series of minima emerge from the CNP with increasing $B$, corresponding to the formation of Landau levels. We observe four-fold degenerate Landau levels with $\nu=\pm4,\pm8,\pm12,\ldots$ emanating from the CNP at $V_D\sim0.5$~V.}
  \label{S15}

\end{figure}

\begin{figure}[bth]
\includegraphics[width=0.98\linewidth]{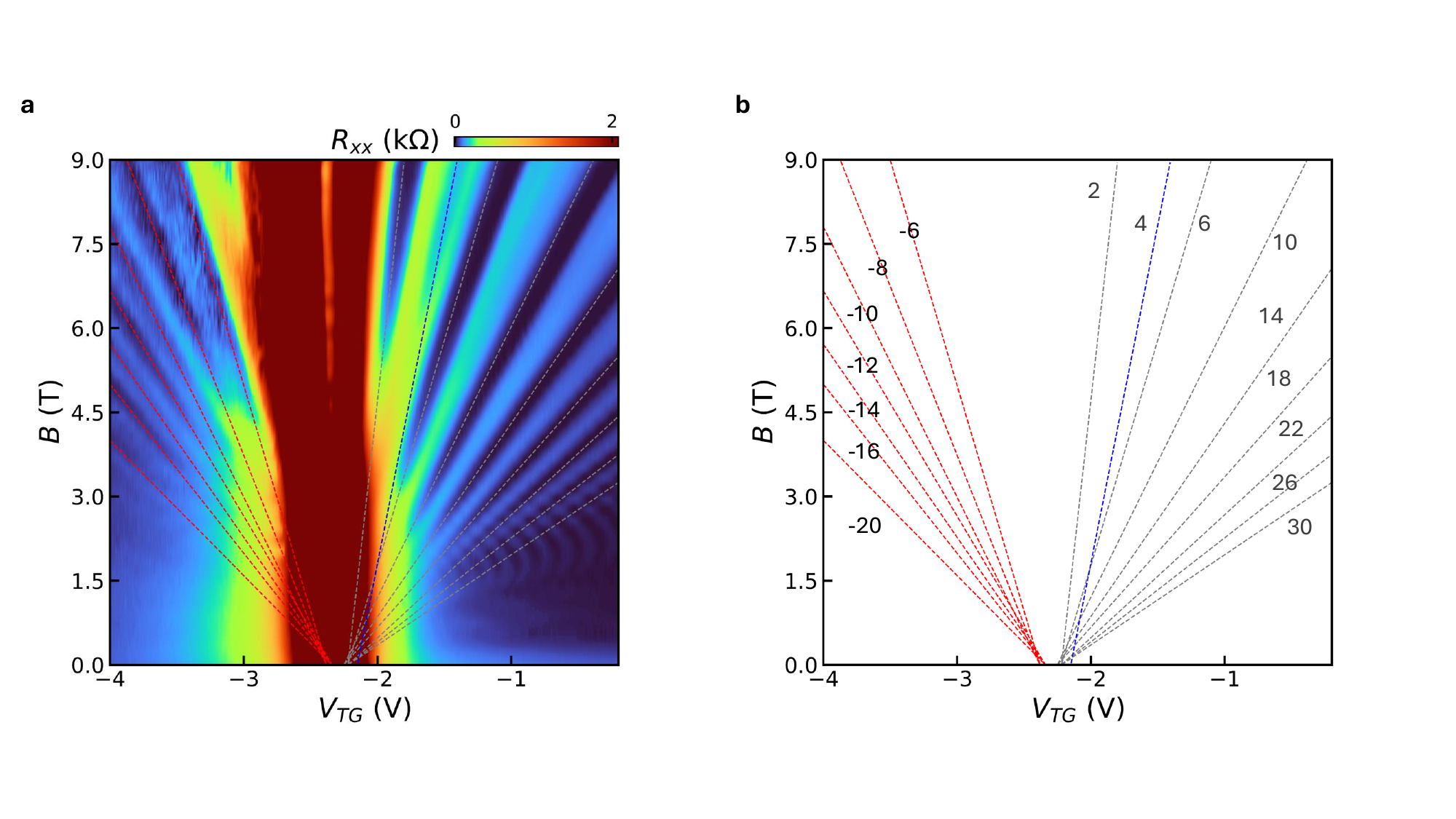}
\captionsetup{justification=raggedright,singlelinecheck=false}
\caption{\justifying\textbf{Landau fan diagram of BLG at $V_{BG}=30$ V; Device D2}~\textbf{a.}~Longitudinal resistance $R_{xx}$ plotted as a function of top gate voltage $V_{TG}$ and magnetic field $B$ at $V_{BG}=30$~V.~A clear departure from the standard BLG-like filling factor $\nu$ is observed. The red dashed lines originate from the CNP peak, while the grey dashed lines emerge from the secondary peak.~A pronounced asymmetry in $\nu$ values for these two sets suggests spin-valley symmetry breaking.~Additionally, a state with $\nu=4$ (blue dashed line) appears from a higher carrier density at larger $B$, likely associated with remote bands.~The slopes of these lines are shown in $\textbf{b.}$}
  \label{S16}

\end{figure}

\begin{figure}[bth]
\includegraphics[width=0.98\linewidth]{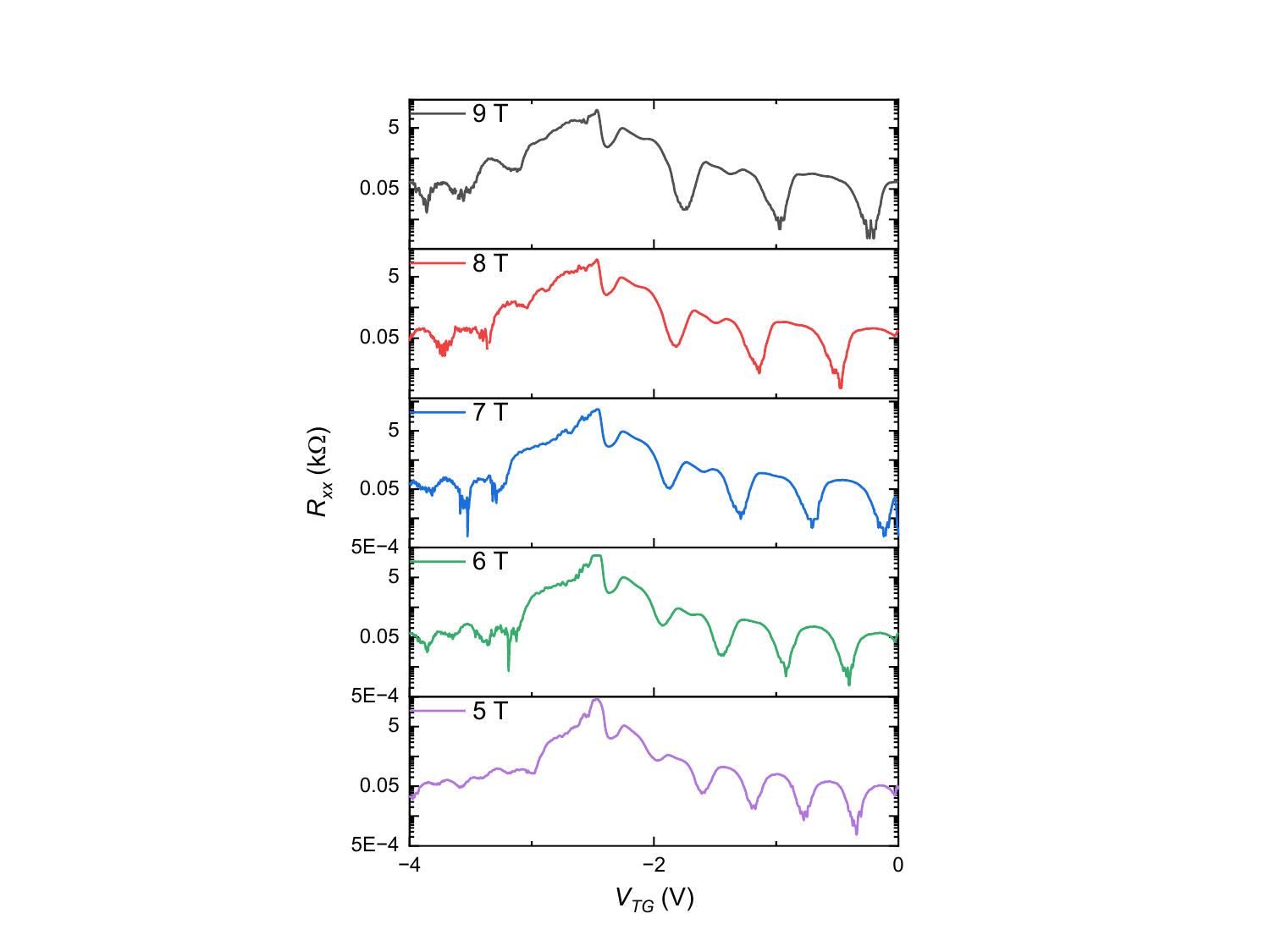}
\captionsetup{justification=raggedright,singlelinecheck=false}
\caption{\justifying\textbf{Line traces of longitudinal resistance at finite magnetic fields for device D2.}~Longitudinal resistance $R_{xx}$ as a function of top gate voltage $V_{TG}$ at $V_{BG}=30$~V for magnetic fields $B=5-9$~T.~These line traces, extracted from Fig.~S14, clearly show the evolution and splitting of Landau levels with varying magnetic field.}
  \label{S17}

\end{figure}

Fig.\ref{S15} shows the Landau fan diagram where $R_{xx}$ is plotted as a function of $V_{TG}$ and $B$ at $V_{BG}= 0$ V. As a finite $B$ field, we find a series of minima emerging from the CNP. These minima are attributed to the Landau levels, and the filling factor can be estimated by fitting the Diophantine equation, $B = (h/\nu e) (n-s)$, where $n$ is the charge carrier density, $\nu$ is the Landau filling factor, $B$ is the magnetic field, $h$ is Planck's constant, $e$ is the charge of an electron, and $s$ is charge density at $B = 0$ T. In our analysis, $n=C_{TG}(V_{TG}-V_{D})/e$, where $C_{TG}$ is the top gate capacitance per area and $V_{D}$ is the residual gate voltage at the CNP. The slope of the dashed lines (Diophantine equation) gives the filling factor $\nu$. In the absence of $V_{BG}$, we find four-fold degenerate Landau levels with $\nu=\pm4,\pm8,\pm12.....$, appearing from the CNP at $V_D\sim0.5$ V~\cite{novoselov2006unconventional}. These results are consistent with the conventional behavior of BLG without a periodic potential. In Fig.\ref{S16}a, we present the data for $V_{BG}=30$ V. We find a departure of $\nu$ from the standard BLG-like $\nu$ (Fig.\ref{S16}b). Interestingly, two primary sets of lines with opposite signs of $\nu$ appear from two different densities. The red dashed lines appear from the CNP peak, and the grey dashed lines appear from the secondary peak. We also find a significant asymmetry in $\nu$ values for these two sets, possibly determined by the spin valley symmetry breaking. In addition, we find a state with $\nu=4$ (blue dashed line) that appears from a higher density at higher $B$ and may be attributed to remote bands. Fig.\ref{S17}, we show the line traces of $R_{xx}$ vs. $V_{TG}$ at $V_{BG}=30$ V for $B=5-9$ T, where the evolution of the Landau levels is clearly visible. These line plots are obtained from Fig.\ref{S16}.


\begin{figure}[bth]
\includegraphics[width=0.98\linewidth]{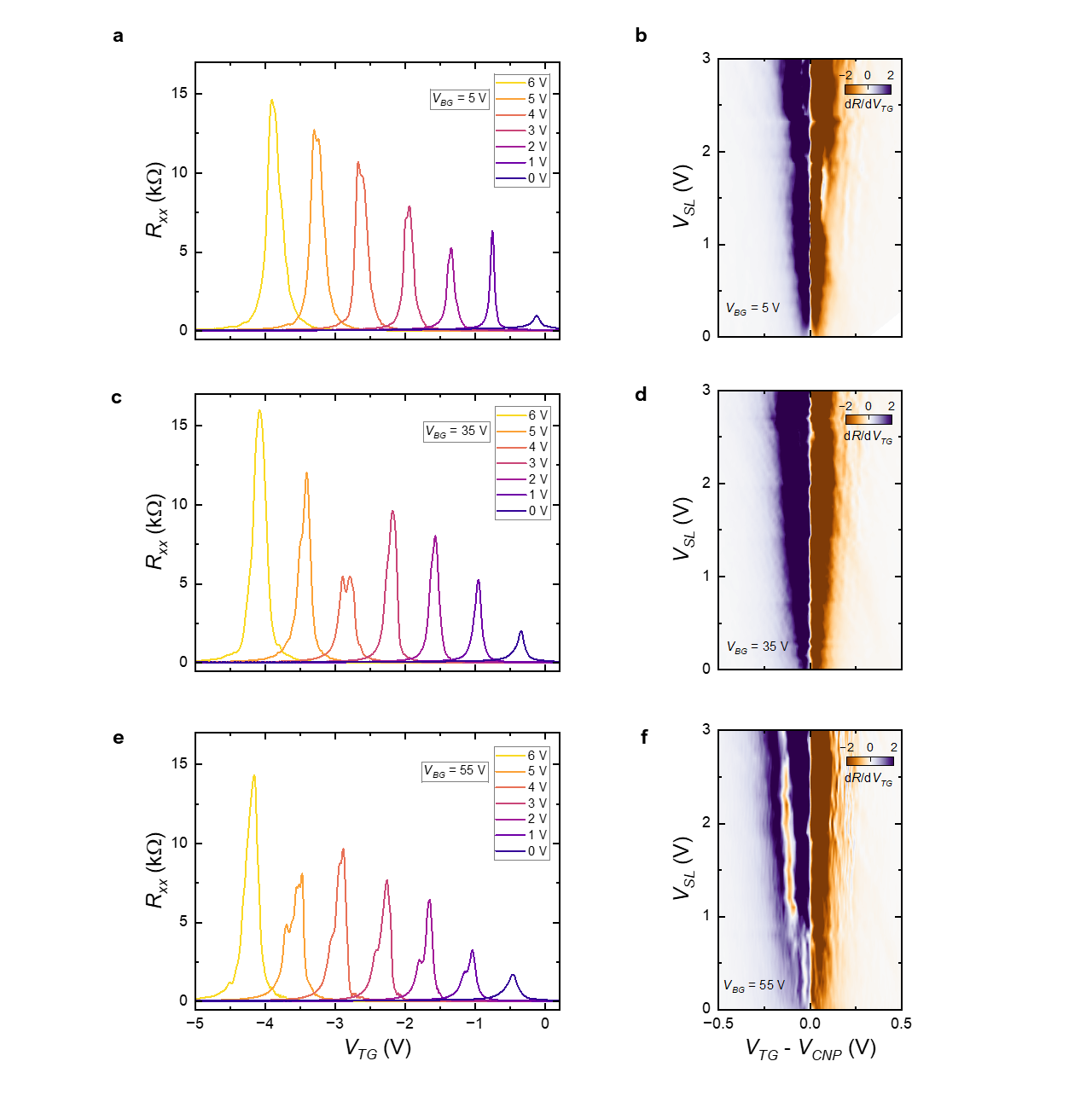}
\captionsetup{justification=raggedright,singlelinecheck=false}
\caption{\justifying\textbf{Transport characteristics of device D3.}~\textbf{a., c., e.} Longitudinal resistance $R_{xx}$ as a function of top-gate voltage $V_{TG}$ for $V_{SL}=0-6$ V at fixed back-gate voltages of $V_{BG}=5$ V, 35 V, and 55 V, respectively. A distinct secondary peak emerges near the charge neutrality point (CNP) at $V_{BG}=55$ V. \textbf{b, d, f.} Colour maps of $dR_{xx}/dV_{TG}$ plotted against $V_{SL}$ and the normalized gate voltage $V_{TG}-V_{CNP}$, where $V_{CNP}$ is the Dirac point offset. While the maps for lower $V_{BG}$ display a single sign change at the CNP, two sign changes are visible at $V_{BG}=55$ V, confirming the presence of the secondary peak.}
  \label{S18}

\end{figure}

\begin{figure}[bth]
\includegraphics[width=0.98\linewidth]{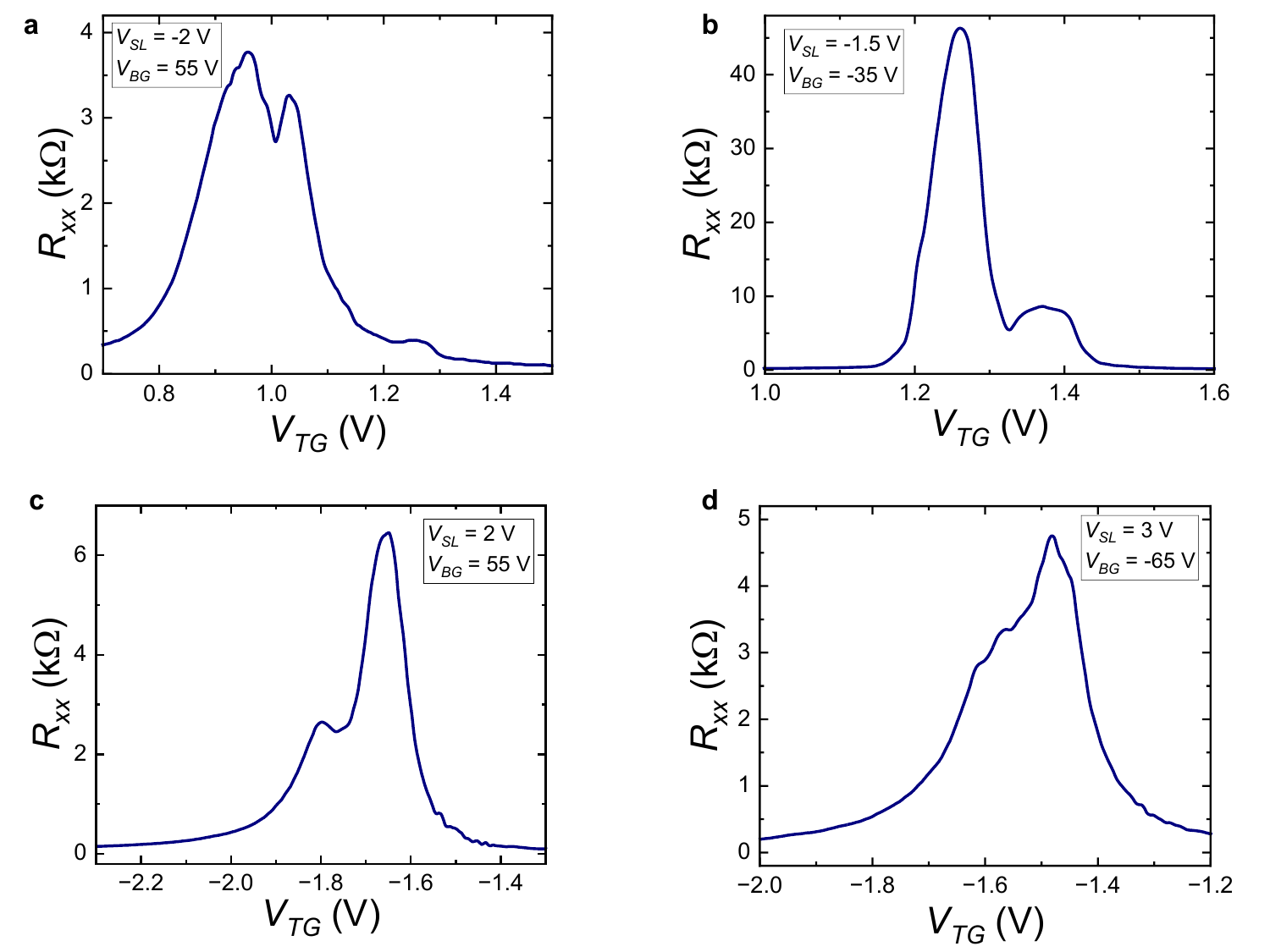}
\captionsetup{justification=raggedright,singlelinecheck=false}
\caption{\justifying\textbf{Various configuration of SL potential for device D3.}~$R_{xx}$ vs. $V_{TG}$ for $(V_{SL}, V_{BG})=$ $\textbf{a.}~(-2~\mathrm{V}, 55~\mathrm{V})$, \textbf{b.}~$(-1.5~\mathrm{V}, -35~\mathrm{V})$, $\textbf{c.}~(2~\mathrm{V}, 55~\mathrm{V})$, $\textbf{d.}~(3~\mathrm{V}, -65~\mathrm{V})$. All four configurations show multiple peaks.}
  \label{S19}

\end{figure}

Fig.\ref{S18} presents data for a third device, D3, which has an SL period of $\sim50$ nm. The effect of $V_{SL}$ on this device is relatively weak, attributed to the thicker bottom hBN layer ($\sim33$ nm; see Fig.\ref{S12}). However, applying a Si $V_{BG}$ enhances the modulation. Fig.\ref{S18}a displays $R_{xx}$ vs. $V_{TG}$ traces for $V_{SL}=0-6$ V at a fixed $V_{BG}=5$ V. While the magnitude of $R_{xx}$ increases with increasing $V_{SL}$, the secondary peak remains notably weaker than in the other two devices. The color map of $dR_{xx}/dV_{TG}$ (Fig.\ref{S18}b), corrected for the $V_{CNP}$ shift, reveals a single sign change, confirming the presence of only a single peak. Notably, increasing $V_{BG}$ induces a qualitative difference: at $V_{BG}=55$ V, a distinct secondary peak emerges within the same $V_{SL}$ range (see Fig.\ref{S18}e and f).\

Fig.\ref{S19}a-d display $R_{xx}$ vs. $V_{TG}$ for four different combinations of $V_{SL}$ and $V_{BG}$. In all cases, we observe multiple peaks near the CNP, suggesting the presence of a periodic potential that modulates the band structure on the length scale of the superlattice.\\

\begin{figure}[bth]
\includegraphics[width=0.98\linewidth]{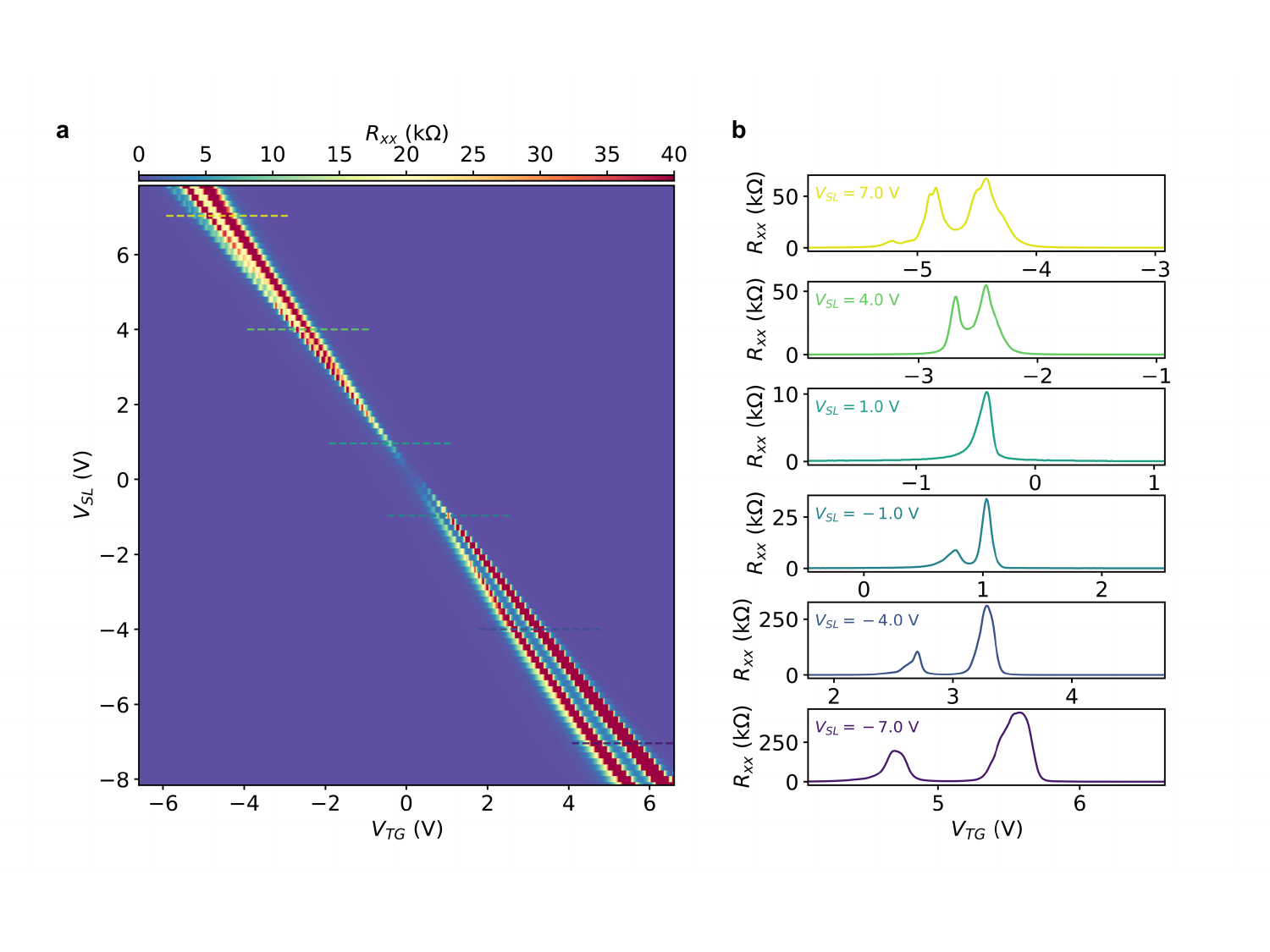}
\captionsetup{justification=raggedright,singlelinecheck=false}
\caption{\justifying\textbf{Phase space of D3.} (a) Colorplot of $R_{xx}$ vs. $V_{TG}$ for different values of $V_{SL}$. (b) Line cuts along the horizontal dashed line in (a), showing multiple peaks emerging as $V_{SL}$ increases on both the positive and negative sides, consistent with the behavior of device D1 discussed in the main text.}
  \label{new_data}

\end{figure}

\begin{figure}[bth]
\includegraphics[width=0.98\linewidth]{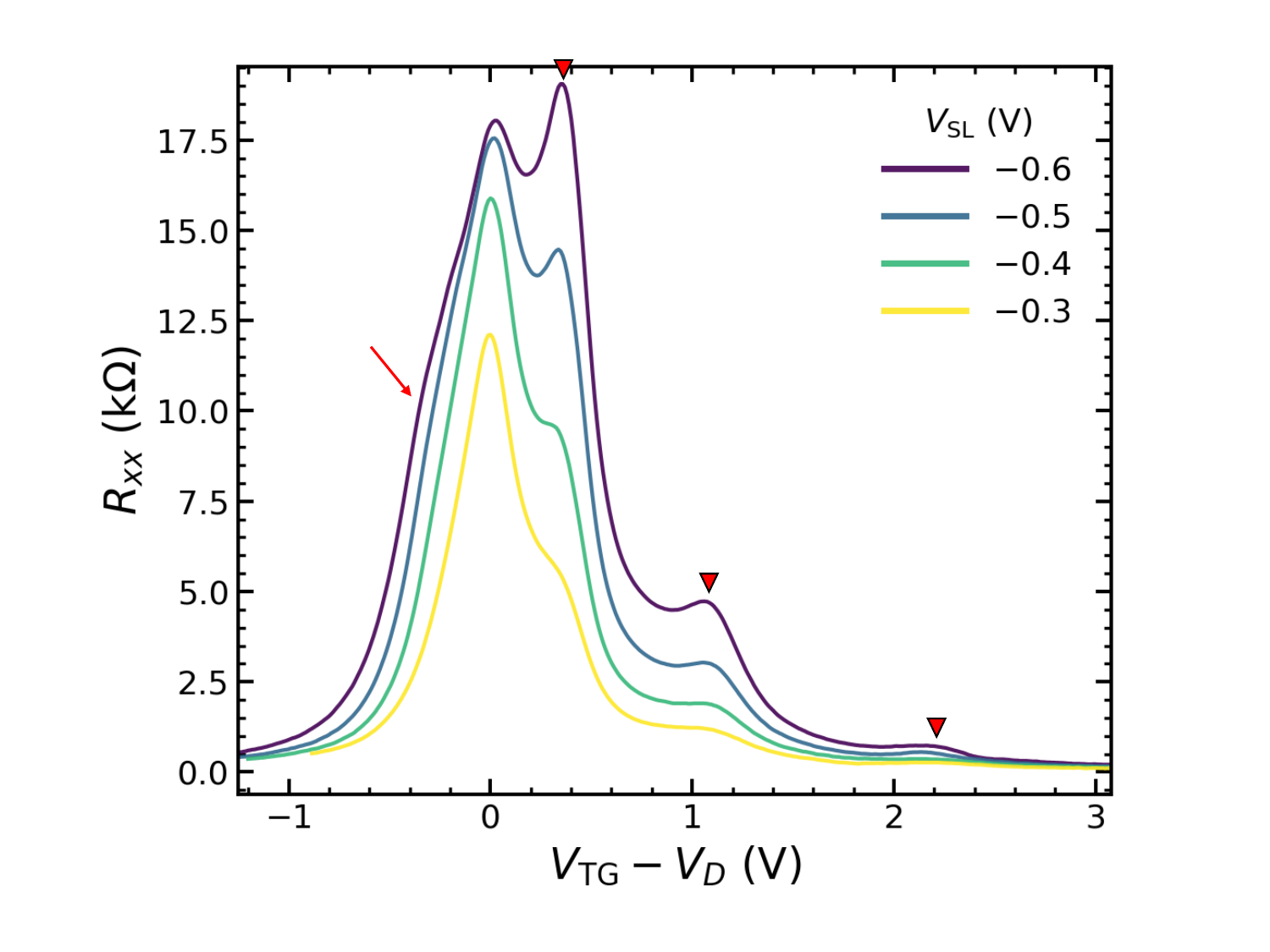}
\captionsetup{justification=raggedright,singlelinecheck=false}
\caption{\justifying\textbf{Additional data for Device D4.} $R_{xx}$ vs. $(V_{TG}-V_D)$ at different $V_{SL}$ showing multiple peaks marked by the red triangles and arrow are consistent with superlattice-induced band engineering.}
  \label{new_data_2}

\end{figure}

In Fig.~\ref{new_data}a-b, we show $R_{xx}$ as a function of $V_{TG}-V_D$ for several values of $V_{SL}$. The data reveal multiple peaks, which we attribute to band gaps emerging at finite $V_{SL}$. These observations are consistent with those from device D1, presented in the main text.\\

Lastly, in order to strengthen the claim that the observed features are from superlattice minibands and not due to inherent Dirac point inhomogenieties in the holey and non-holey regions,  we present data from an additional device (D4), with a thinner hBN ($\sim 7$~nm) between the BLG and SL gate.  This device exhibited multiple resistance peaks in the presence of finite superlattice potentials. Fig.~\ref{new_data_2} shows the $R_{xx}$ vs. $(V_{TG}-V_D)$ plot for varying $V_{SL}$, clearly demonstrating the emergence of three additional peaks on the electron side. A faint feature is observed on the hole side, signifying electron-hole asymmetry, similar to that observed in all other devices.
\\

In summary, the usage of the multi-gated platform for superlattice engineering offers the following distinct advantages:
\begin{enumerate}
    \item 
The combination of holey and non-holey potentials can be used to enhance the effective superlattice strength. As demonstrated for device D2 (see Fig. 4a,d, Fig.\ref{S14}), while the additional miniband peaks are hardly observable at a back gate voltage of $V_{BG}=0$, the strength is highly enhanced for a combination of negative and positive $V_{SL}$ and $V_{BG}$, respectively.

\item The superlattice potential can be accessed even without an explicit connection to the SL gate, via the non-holey regions and a back gate voltage, as demonstrated for device D3. While this combination reasserts the uniformity of the potential profile across the device, it also offers a very useful experimental method that can be extended to various architectures. For example, this approach allows a graphite gate to be placed very close to the 2D material under study (BLG or other materials, including oxides), while avoiding the fabrication complexity and potential shorting risks associated with integrating a gate electrode directly into the device. 

\item A full combination using all three gates show the expected miniband features, while also offering a rich voltage space to observe the associated features. This effect is exemplified in Figs. \ref{S18} and \ref{S19}. While the superlattice effect is minimal owing to a thicker bottom hBN (33 nm), they are enhanced on adding a back gate.

\item The three-gate configuration provides independent control over the bulk bandgap and the superlattice potential in an electrically tunable material such as BLG. For example, in Figs. S23a and S23b, by changing the sign and magnitude of the electric field in the patterned hole regions while keeping the electric field and electrostatic potential in the non-holey regions nearly unchanged, we observe an approximately order-of-magnitude difference in resistance between the two configurations, along with a change in the number of resistance peaks. These variations suggest changes in the local bandgap within the hole regions and the overall strength of the resulting superlattice potential, highlighting the additional electrostatic tunability enabled by the three-gate configuration.
\end{enumerate}

\clearpage

\definecolor{headerblue}{RGB}{173,216,230}
\definecolor{rowgray}{RGB}{242,242,242}

\renewcommand{\arraystretch}{1.3}

\begin{table}[p]
\centering
\caption{Summary of papers on dielectric and gate patterning in 2D materials.}
\label{tab:gate_patterning_refs}
\renewcommand{\arraystretch}{1.3}
\small
\resizebox{\linewidth}{!}{%
\begin{tabular}{|c|c|l|l|c|l|}
\hline
\rowcolor{headerblue}
\textbf{Refs} & \textbf{Material} & \textbf{Substrate} & \textbf{Lattice structure} & \textbf{Period} & \textbf{Gates} \\
\hline
\rowcolor{rowgray}
\cite{dubey2013tunable} & SLG & Patterned metal gate & 1D lattice & 100 nm & Dual-gate controlled \\
\hline
\cite{forsythe2018band} & SLG & Patterned SiO$_2$ & 2D triangular lattice & 40 nm & Dual-gate controlled \\
\hline
\rowcolor{rowgray}
\cite{li2021anisotropic} & SLG & Patterned SiO$_2$ & 1D lattice & 60 nm & Dual-gate controlled \\
\hline
\cite{sun2024signature} & BLG & Patterned SiO$_2$ & 2D triangular lattice & 60 nm & Dual-gate controlled \\
\hline
\rowcolor{rowgray}
\cite{huber2020gate} & SLG & Patterned graphite gate & 2D triangular lattice & 40 nm & Dual-gate controlled \\
\hline
\cite{PhysRevLett.133.066302} & SLG & Patterned graphite gate & 2D Kagome lattice & 80 nm & Dual-gate controlled \\
\hline
\rowcolor{rowgray}
\textbf{This work} & \textbf{BLG} & \textbf{Patterned graphite gate} & \textbf{2D triangular lattice} & \textbf{50 nm} & \textbf{Triple-gate controlled} \\
\hline

\end{tabular}%
}
\end{table}

We have summarized prior work on engineering the band structure of single- and bilayer graphene via periodic potentials imposed through dielectric or gate patterning (Table~\ref{tab:gate_patterning_refs}). We also note reports that mimic such a potential by placing a moiré lattice in proximity to the target layer\cite{gu2024remote, zhang2024engineering, wang2025moire}.\\

In a dual-gated device with one plain gate and one patterned gate, the tunability of the periodic potential is limited by the physical lattice constant of the patterned structure. For dielectric patterning, the amplitude of the potential is determined mainly by the difference in thickness between the holey and non-holey regions, whereas for gate patterning, the amplitude is primarily governed by the lattice constant. A triple-gated structure, with the patterned gate sandwiched between two plain gates, offers greater tunability, since the potential in the holey and non-holey regions can now be tuned independently, enabling the imposition of a complex potential landscape on the target layer. Such a three-gated platform is expected to realize intriguing topological signatures in Bernal bilayer graphene~\cite{PhysRevLett.130.196201}, though no previous experimental study has attempted to implement this architecture. Although the predicted topological signatures are not observed in the present work, realizing high-quality electrostatically defined superlattices remains a significant experimental challenge. We anticipate that future studies, with further optimization of the superlattice potential, will enable the observation of the predicted correlated and topological phenomena. Furthermore, the three-gated architecture also ensures that the miniband landscape is accessible even without explicitly contacting the sandwiched graphite gate, ensuring versatility in device fabrication - a combination of top and back gates still provides access to the miniband physics. This could be an important criterion to access miniband physics in engineered heterostructures such as complex oxides where contacting a sandwiched gate layer may pose difficulties.

\clearpage

\section{Theoretical details}

\subsection{Parameter Calculations}

The number density depends on both the top gate and superlattice gate voltage. It can be calculated using
    \begin{equation}
        n = \frac{c_{TG}V_{TG} + c_{SL}V_{SL}}{e}
    \end{equation}
where $c_{TG} = 2.36 \times 10^{-3} \ C/Vm^2$ and $c_{SL} = 2.36 \times 10^{-3} \ C/Vm^2$ are the capacitances of the top gate and superlattice gate, respectively. The capacitance values are determined using COMSOL simulations. This gives the total charge induced on both layers. We take the effective mass to be $0.05 \text{m}_e$, and the density of states is approximated as $D = m^*/\pi \hbar^2$ to estimate the energy.

For $V_{SL} = 1 \ V$, the magnitude of the constant charge density induced due to the superlattice gate is $1.348 \times 10^{12} \text{cm}^{-2}/\text{V}$, and the amplitude of charge density is $2.131 \times 10^{11} \text{cm}^{-2}/\text{V}$. The constant charge density arises due to the mean potential generated by the combination of both gate voltages, and the charge density modulation occurs due to superlattice formation. The ratio is 0.158. Converting the charge density modulation to energy gives the amplitude of superlattice modulation to be $10.206V_{SL}$ meV. The superlattice Hamiltonian is of the form~\cite{sun2024signature}

    \begin{equation}
        H_{SL} = \frac{1}{2}
        \begin{bmatrix}
            V_{SL}^1 \sigma^0 & 0_2 \\
            0_2 & V_{SL}^2 \sigma^0
        \end{bmatrix}
        \sum_{n=1}^{6} \cos(\mathbf{G}_n \cdot \mathbf{r})
    \end{equation}
    
The amplitude of $1/2 \sum_{n=1}^{6} \cos(\mathbf{G}_n \cdot \mathbf{r})$ is 2.25, which gives the coefficient of $V_{SL}$ to be 4.536 meV. Using the ratio calculated above, we get the coefficient of mean potential to be 64.595 meV. The parameters are then given by 

    \begin{align}
        V_0^{(1)} &= 64.595(1-\alpha) \ V_{TG} + 64.595\alpha \ V_{SL} \ meV \\ \nonumber
        V_0^{(2)} &= 64.595\alpha \ V_{TG} + 64.595(1-\alpha) \ V_{SL} \ meV \\ \nonumber
        V_{SL}^{(1)} &= 4.536\alpha \ V_{SL} \ meV \\ \nonumber
        V_{SL}^{(2)} &= 4.536(1 - \alpha) \ V_{SL} \ meV \\ \nonumber
    \end{align}
where $\alpha$ is a screening parameter. It accounts for the asymmetry in the charge induced on each layer due to the gate voltages. 

\subsection{Band structure}

\begin{figure}[bth]
    \centering
    \includegraphics[width=0.98\linewidth]{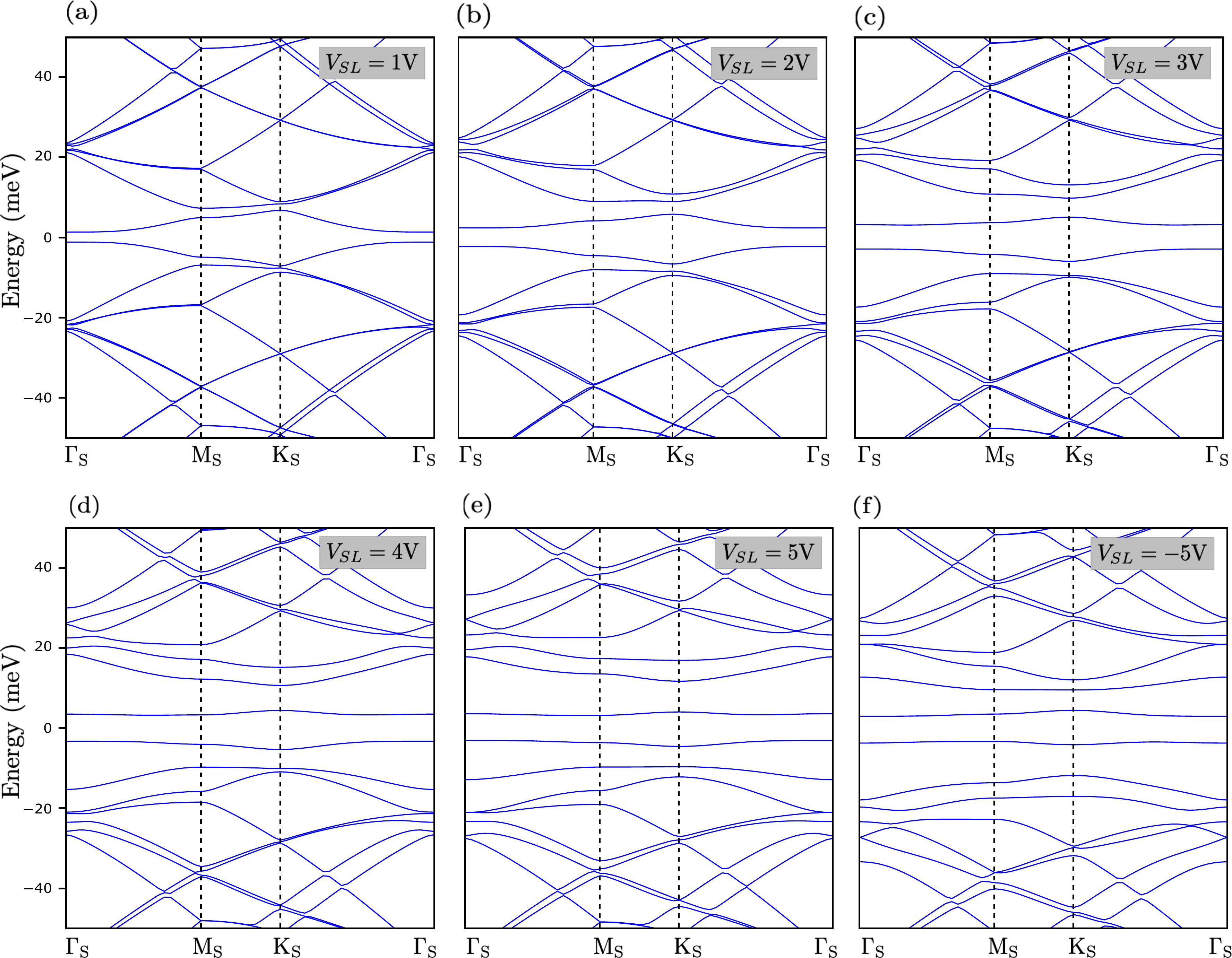}  \captionsetup{justification=raggedright,singlelinecheck=false}
    \caption{
        \justifying
        \textbf{Electronic band structures of bilayer graphene (BLG) under different superlattice potentials}. \textbf{(a)–(e)} show the band structures for $V_{SL} = 1$–$5$ V, and \textbf{(f)} corresponds to $V_{SL} = -5$ V, all calculated with the screening parameter $\alpha = 0.51$.
    }
    \label{SM_fig1}
\end{figure}

\begin{figure}[bth]
    \centering
    \includegraphics[width=0.98\linewidth]{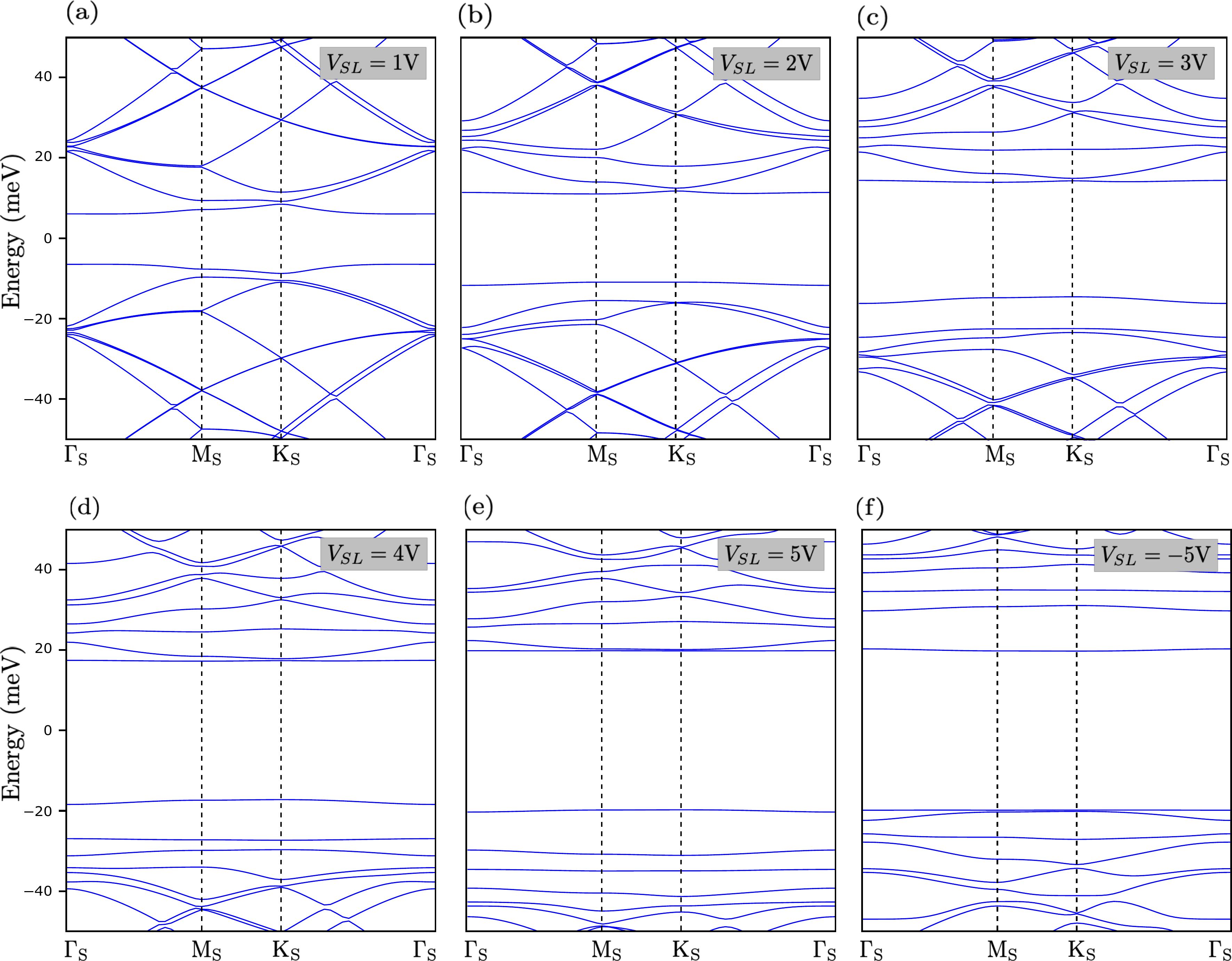}  \captionsetup{justification=raggedright,singlelinecheck=false}
    \caption{
        \justifying
        \textbf{Electronic band structures of bilayer graphene (BLG) under different superlattice potentials}. \textbf{(a)–(e)} show the band structures for $V_{SL} = 1$–$5$ V, and \textbf{(f)} corresponds to $V_{SL} = -5$ V, all calculated with the screening parameter $\alpha = 0.55$.
    }
    \label{SM_fig2}
\end{figure}

\begin{figure}[bth]
    \centering
    \includegraphics[width=0.98\linewidth]{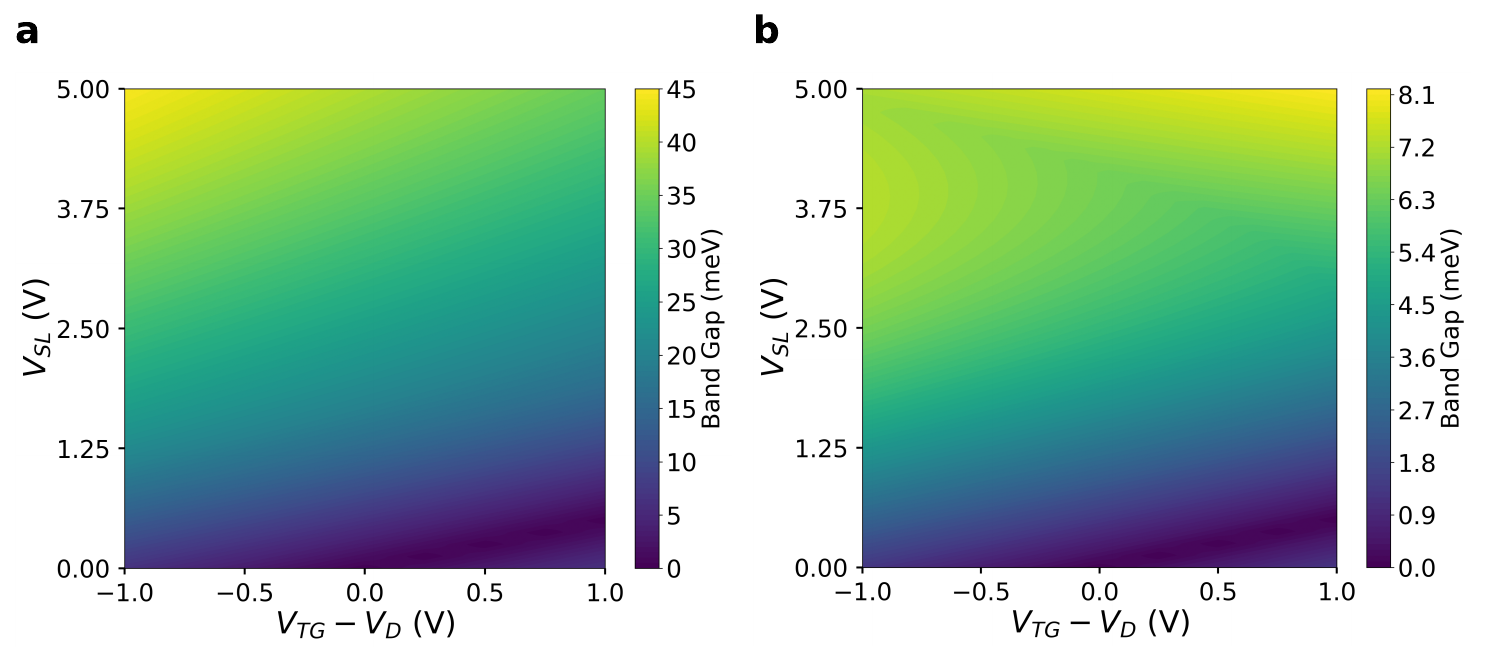}  \captionsetup{justification=raggedright,singlelinecheck=false}
    \caption{
        \justifying
        \textbf{Primary band gap plotted as a function of $V_{SL}$ and $V_{TG}-V_D$} for \textbf{a.} $\alpha = 0.55$ \textbf{b.} $\alpha = 0.51$
    }
    \label{band_gap}
\end{figure}

Figures~\ref{SM_fig1} and \ref{SM_fig2} present the evolution of the BLG band structure as a function of the superlattice gate voltage $V_{SL}$ for two representative choices of the screening parameter $\alpha$. Several systematic trends emerge. (i) For both values of $\alpha$, the gap at the charge-neutrality point increases as $V_{SL}$ is varied from $1$ to $5$~V, accompanied by a clear enhancement of particle--hole asymmetry. (ii) Reversing the sign of $V_{SL}$ inverts this asymmetry, as seen from the comparison between panels (e) and (f) in both figures. (iii) The band structure---and the size of the gap in particular---is strongly dependent on the choice of $\alpha$. A comparison between $\alpha = 0.51$ and $\alpha = 0.55$ shows that, even for a small difference in the value of $\alpha$, a substantial difference arises in the resulting gap. Fig.\ref{band_gap} shows how the primary (largest) band gap varies as a function of $V_{SL}$ and $V_{TG} - V_{D}$. At any $V_{SL}$, as $V_{TG}$ decreases, the primary band gap increases. For the remainder of this work, we use $\alpha = 0.55$, which yields a gap consistent with experimental estimates.  

The chemical potential is obtained from
\begin{equation}
    \mu = \frac{\hbar^{2}\pi n}{m^*},
\end{equation}
where $n$ is the carrier density and $m^{*} = 0.05\,m_{e}$ is the effective mass.

\subsection{Resistance Heat Maps}

\begin{figure}[bth]
    \centering
    \includegraphics[width=0.98\linewidth]{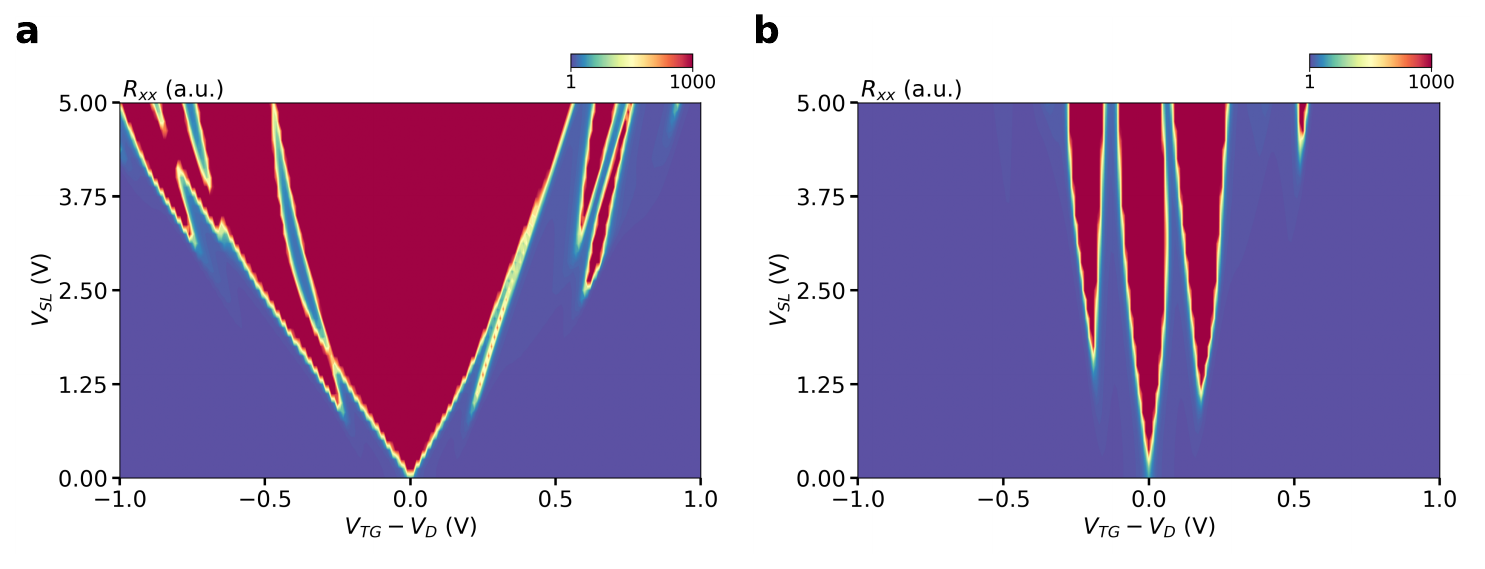}  \captionsetup{justification=raggedright,singlelinecheck=false}
    \caption{
        \justifying
        \textbf{Resistance heatmaps} for \textbf{a.} $\alpha = 0.55$ \textbf{b.} $\alpha = 0.51$
        }
        \label{resalpha}
\end{figure}

The extreme sensitivity of the band structure to the screening parameter $\alpha$ naturally leads to a corresponding sensitivity in the calculated resistivity. In Fig.~\ref{resalpha}, we compare the resistance heat maps obtained for $\alpha = 0.55$ and $\alpha = 0.51$. As evident from the figure, the resulting patterns differ substantially. Additional features also appear at larger $V_{SL}$, arising from the flattening of the bands and the concomitant opening of higher-order gaps. We also note that incorporating screening requires a self-consistent calculation---particularly in the flat-band regime, where screening effects become dominant and substantially increase the complexity of the problem. Such effects lie beyond the scope of the present work, and our theoretical results should therefore be regarded as providing qualitative support for the experimental evidence.

In constructing the resistance heat maps, the band structures obtained from the electrostatic model (as discussed in the previous subsection) were used as input for the conductivity calculations. While the previous analysis illustrated how the band structure evolves with $V_{SL}$, it is important to emphasize that it also varies with the top-gate voltage $V_{TG}$. In practice, both the band dispersion and the chemical potential evolve \emph{simultaneously} as $(V_{TG}, V_{SL})$ are tuned. 

\begin{figure}[bth]
    \centering
    \includegraphics[width=0.98\linewidth]{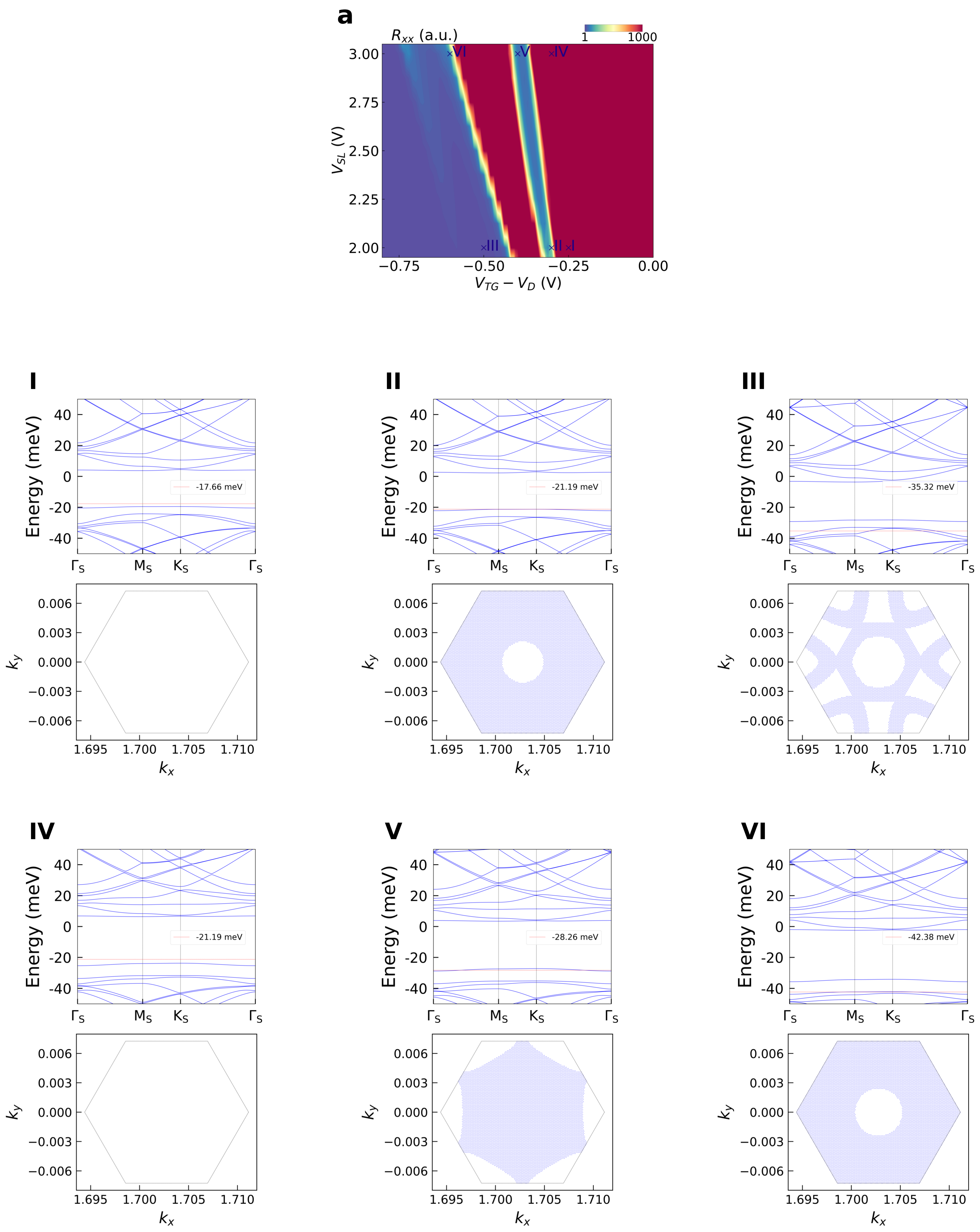}  \captionsetup{justification=raggedright,singlelinecheck=false}
    \caption{
        \justifying
        \textbf{Resistance heatmap and the corresponding band structures}
        \textbf{a.} Resistance heatmap for $\alpha = 0.55$. \textbf{I}--\textbf{III} shows the band structures and the Fermi surfaces corresponding to $V_{SL} = 2 \ V$, and $V_{TG} - V_{D} = -0.25, \ -0.3 \ \text{and} \ -0.5 \ V$, which are marked on the heatmap, respectively. Similarly, \textbf{IV}--\textbf{VI} corresponds to $V_{SL} = 3 \ V$ and $V_{TG} - V_{D} = -0.3, \ -0.4, \ \text{and} \ -0.6 \ V$, respectively.
    }
    \label{band_fs}
\end{figure}

To demonstrate this explicitly, we focus on a small region of the resistance heat map in Fig.~\ref{resalpha} and examine several representative gate-voltage combinations marked I--VI in Fig.~\ref{band_fs}a. For each point, we compute the corresponding band structure and determine the position of the chemical potential (indicated by the red dotted line). To identify the regions in momentum space that contribute most significantly to the conductivity, we also highlight the set of $k$ points satisfying $|E(k)-\mu| \le 5k_B T$, which depicts the Fermi contour together with its thermal broadening. All calculations are performed at $T = 2$~K.

A detailed inspection of these six points confirms that both the band structure and the chemical potential vary simultaneously as functions of $(V_{\mathrm{TG}}, V_{\mathrm{SL}})$, although their dominant roles differ: changes in $V_{SL}$ primarily reshape the band structure [cf. points I and IV], whereas changes in $V_{TG}$ primarily shift the chemical potential. We first consider points I--III, where $V_{SL}$ is fixed at 2~V and $V_{TG}$ is varied. At point I, the chemical potential lies within the primary band gap and intersects no band, resulting in negligible conductivity (high resistivity). As $V_{TG}$ is decreased to point II, the chemical potential enters the first available band, producing a finite carrier density and giving rise to the narrow low-resistance (blue) region that separates the two insulating domains in Fig.~\ref{band_fs}a. The associated Fermi surface contains multiple contributing $k$ points. A further decrease of $V_{TG}$ shifts the chemical potential into the second band gap, restoring a high-resistivity region in the heat map. Beyond this, the chemical potential drops into a dense manifold of bands, leading to a pronounced increase in conductivity. Points IV--VI correspond to $V_{SL}=3$~V and different values of $V_{TG}$. The qualitative behaviour mirrors that of the $V_{SL}=2$~V case, but the larger band gap results in the chemical potential entering the first conduction (or valence) band at a smaller value of $V_{TG}$. Finally, we note the strong sensitivity of the thermally broadened Fermi contour to small variations in gate voltage. For instance, points I and II differ by only $-0.05$~V in $V_{TG}$, yet exhibit markedly different Fermi-surface topologies.

\clearpage
\bibliographystyle{apsrev4-1}
\bibliography{References}
\end{document}